\documentclass[fleqn,usenatbib]{mnras}

\usepackage[T1]{fontenc}
\usepackage{ae,aecompl}

\usepackage{graphicx}	
\usepackage{amsmath}	
\usepackage{mathtools}	

\usepackage{hyperref}
\usepackage{xspace}
\usepackage{url}
\usepackage{longtable}
\usepackage[flushleft]{threeparttablex}
\usepackage{natbib,twoopt}

\def\simeq{
\mathrel{\raise.3ex\hbox{$\sim$}\mkern-14mu\lower0.4ex\hbox{$-$}}
}

\def\ltsima{$\; \buildrel < \over \sim \;$}
\def\simlt{\lower.5ex\hbox{\ltsima}}
\def\gtsima{$\; \buildrel > \over \sim \;$}
\def\simgt{\lower.5ex\hbox{\gtsima}}

\def\msun{{\rm M_{\odot}}}

\def\be{\begin{equation}}
\def\ee{\end{equation}}
\def\le{{L_{\rm Edd}}}

\def\del#1{{}}
\def\ltsima{$\; \buildrel < \over \sim \;$}
\def\simlt{\lower.5ex\hbox{\ltsima}}
\def\gtsima{$\; \buildrel > \over \sim \;$}
\def\simgt{\lower.5ex\hbox{\gtsima}}

\usepackage{newtxtext,newtxmath}

\title[Fossil AGN outflows]{Life after AGN switchoff: evolution and properties of fossil galactic outflows}

\author[K. Zubovas, G. Maskeli\={u}nas]{Kastytis Zubovas$^{1,2,\star}$, Gediminas Maskeli\={u}nas$^{2}$ \\
  $^{1}$Center for Physical Sciences and Technology, Saul\.{e}tekio al. 3, Vilnius LT-10257, Lithuania\\
  $^{2}$Astronomical Observatory, Vilnius University, Saul\.{e}tekio al. 3, Vilnius LT-10257, Lithuania\\
  $^{\star}$ {E-mail:~} {\rm kastytis.zubovas@ftmc.lt} }

\date{Accepted XXX. Received YYY; in original form ZZZ}

\pubyear{2019}

\begin{document}
\label{firstpage}
\pagerange{\pageref{firstpage}--\pageref{lastpage}}
\maketitle

\begin{abstract}
Galaxy-wide outflows driven by active galactic nuclei (AGN) are an important ingredient in galaxy evolution. Analytical calculations suggest that such outflows have significant inertia and can persist long after the AGN itself fades away. We use hydrodynamical simulations of outflows in idealised galaxy bulges to investigate the propagation of these `fossil' AGN outflows. We find that fossil outflows should be common in gas-poor galaxies but form only rarely in gas-rich ones; in general, fossil outflows should outnumber driven ones by a factor of a few in the local Universe, and possibly more at high redshift. When they do form, fossil outflows tend to be lopsided and detached from the nucleus, and colder than their driven counterparts, with a more prominent molecular phase. Spatially resolved and/or multiphase observations can help distinguish fossil AGN outflows from star formation-driven ones, which have similar integrated properties. We discuss a number of spatially-resolved observations of outflows, suggesting that most show evidence of fossil outflow existence, sometimes together with driven outflows on smaller scales. 
\end{abstract}




\section{Introduction} \label{sec:intro}

Active galactic nuclei (AGN) are well known to be an important element of galaxy evolution. Both semi-analytical \citep[e.g.,][]{Bower2006MNRAS, Croton2006MNRAS} and hydrodynamical simulations \citep[e.g.,][]{Sijacki2007MNRAS, Puchwein2013MNRAS, Dubois2014MNRAS, Vogelsberger2014MNRAS, Schaye2015MNRAS, Tremmel2019MNRAS} show that AGN feedback is necessary to regulate the star formation history and black hole mass function of massive galaxies. This feedback predominantly manifests as large-scale massive gaseous outflows that can reach velocities $>1000$~km~s$^{-1}$ and mass flow rates $>1000 \, \msun$~yr$^{-1}$ \citep[e.g.][]{Feruglio2010AA, Sturm2011ApJ, Rupke2011ApJ, Cicone2014AA, Rupke2017ApJ, Fiore2017AA, Fluetsch2019MNRAS, Lutz2020AA}. In many cases, outflow properties, such as the mass flow, momentum and energy rates, correlate well with AGN luminosity \citep{Cicone2014AA, Fluetsch2019MNRAS, Lutz2020AA}, suggesting a connection between the two phenomena. On the other hand, the scatter around this relationship is large \citep{Marasco2020AA, Zanchettin2021AA}, prompting some authors to question the simple connection picture.

The most successful physical model explaining the observed properties of large-scale outflows is the wind-driven feedback model \citep{King2003ApJ, King2010MNRASa, Zubovas2012ApJ, King2015ARAA}. In this model, a quasi-relativistic wind launched from the accretion disc around the supermassive black hole (SMBH) carries a significant fraction of the AGN bolometric luminosity as kinetic energy. As the wind shocks against the surrounding gas, it heats up to $T \sim 10^{10}$~K. The most efficient cooling process at this temperature is the inverse Compton effect, which is still inefficient if we treat the wind as a two-temperature plasma \citep{Faucher2012MNRAS}. The shocked interstellar medium (ISM) then expands adiabatically, resulting in an outflow with kinetic power $\dot{E}_{\rm out} \simeq 0.02-0.03 L_{\rm AGN}$, close to the average observational estimate. Importantly, the dynamical timescale of the outflow, defined as the ratio of outflow radius to its velocity, is of order a few times $10^5-10^6$~yr \citep{Zubovas2018MNRAS}, which is comparable to or longer than the typical duration of an AGN outflow \citep[$\sim 10^4-10^5$~yr; cf.]{King2015MNRAS, Schawinski2015MNRAS}. This means that when the AGN switches off, the outflow persists for a significant time before dissipating \citep{King2011MNRAS, Zubovas2020MNRAS, Zubovas2022MNRAS}. Such fossil outflows may comprise as much as $60\%$ of all observed outflows, based on semi-analytical estimates \citep{Zubovas2022MNRAS}.

In recent years, some observations have been interpreted as showing fossil outflows. In some cases, the evidence comes from spatially-resolved gas morphology that reveals outflowing clumps of gas at large distances from the nucleus. For example, in IRAS F08572+3915, there is an outflowing molecular gas blob $\sim 6$~kpc away from the nucleus, spatially detached from an ongoing outflow in the centre. \citep{Herrera-Camus2020A&A}; if it was launched by the AGN, it is more than $6$~Myr old. In PDS 456, a detached outflow component is visible $\sim 3$~kpc to the south of the nucleus \citep{Bischetti2019AA}; once again, if it was launched by the AGN, its age is $\sim 6$~Myr. In other systems, the fossil nature is evident due to the outflow momentum and energy rates being higher than the expectation $\dot{p} \sim 20 L_{\rm AGN}/c$ and $\dot{E} \sim 0.05 L_{\rm AGN}$ \citep{Audibert2019A&A, Fluetsch2019MNRAS, Davies2020MNRAS, Lutz2020AA}. This discrepancy can be alleviated if the AGN was much brighter in the past when the outflow was being inflated. However, the integrated properties of fossil outflows, such as velocities, mass, momentum and energy rates, can be indistinguishable from those of driven ones. Comparison of those properties with AGN luminosity may not reveal much either, since the AGN may have experienced multiple episodes of varying luminosity since the outflow started expanding, and a present-day agreement between outflow properties and AGN luminosity can be coincidental \citep{Zubovas2018MNRAS}. Numerical hydrodynamical simulations reveal that outflows retain much higher momentum and energy rates once the AGN fades \citep[e.g.,][]{Costa2018MNRASb}, but a thorough investigation of outflow evolution after AGN switchoff has so far not been carried out.

In this paper, we investigate the evolution of fossil AGN outflows using three-dimensional hydrodynamical SPH simulations of idealised galaxies. We run 20 simulations of AGN outflows propagating through turbulent spherical gas shells, designed to mimic galaxy bulges. We investigate the evolution of the AGN outflow during and after the AGN phase and identify the salient properties of fossil outflows: detachment from the nucleus of the galaxy, lopsidedness, low velocity and high mass. These properties are robust to changes in SMBH mass, galaxy gas fraction and AGN luminosity. We also determine the conditions necessary for fossil outflows to form and, when they form, the typical lifetime of such outflows. This allows us to infer the prevalence of fossil outflows and show that they should be more common in gas-poor than in gas-rich galaxies. We suggest that several currently known outflows are, in fact, fossils of earlier AGN episodes. We also discuss how fossil AGN outflows may be distinguished from outflows driven by star formation, and how star formation inside outflows may help identify fossils.

The paper is structured in the usual manner. In Section \ref{sec:physical_model}, we present the AGN wind-driven outflow model in more detail, focussing on the analytical predictions regarding fossil outflows. In Section \ref{sec:numerical_model}, we present the setup of our numerical simulations, as well as analytical predictions of fossil outflow propagation under such conditions. The results of the simulations are given in Section \ref{sec:results}, including the morphological, kinematical and thermodynamic evolution of the outflows. In Section \ref{sec:discuss}, we discuss our findings, including a direct comparison with several spatially resolved outflow observations, as well as a critical look at the assumptions made in our simulations and the way forward to making them more realistic and predictive. Finally, we summarize and conclude in Section \ref{sec:concl}.

\section{Physics of AGN wind-driven outflows} \label{sec:physical_model}

The AGN wind-driven outflow model, first proposed in \cite{King2003ApJ} and significantly developed in \cite{King2010MNRASa} and \cite{Zubovas2012ApJ}, is currently the most successful model in explaining the observed outflow properties and the origin of the $M-\sigma$ relation. For a thorough review, we refer the reader to \cite{King2015ARAA} and give here only a brief overview of the aspects relevant to our investigation.

A rapidly-accreting SMBH is surrounded by a geometrically thin, optically thick disc \citep{Shakura1973AA}. Radiation pressure, both in the continuum and especially in the spectral lines broadened due to gas motions, launches a wind with typical velocities $v_{\rm w} \sim 0.1c$ and kinetic power $\dot{E}_{\rm w} \sim 0.05 L_{\rm AGN}$ \citep{King2003MNRASb, Proga2000ApJ, Higginbottom2014ApJ}. The wind has a large opening angle \citep{Nardini2015Sci} and soon encounters the ISM around the SMBH. A strong shock develops which heats the wind plasma to $T_{\rm sh} > 10^{10}$~K. At this temperature, the only efficient cooling process is the inverse Compton effect. The electrons in the plasma cool rapidly, but most of the energy is in protons, so the cooling rate is governed by the electron-proton equilibration timescale \citep{Faucher2012MNRAS}. This leads to cooling being very inefficient outside of the central parsec. The shocked wind then expands as an approximately adiabatic bubble \citep{King2005ApJ, Zubovas2012ApJ}. An {\em energy-driven} outflow forms, with kinetic power $\dot{E}_{\rm out} \sim 0.02-0.03 L_{\rm AGN}$, with the rest of the wind kinetic power used to do work against gravity and $p$d$V$ work.

The morphology and kinematics of the outflow depend significantly on the ISM mass distribution. Globally, one may expect the gas density to be higher in the midplane of the galaxy than perpendicular to it. This makes it much easier for the outflow to expand along the galaxy (minor) axis, creating an hourglass-shaped outflow \citep{Zubovas2014MNRASb, Pillepich2021MNRAS} that may explain, for example, the presence of the Fermi bubbles in the Milky Way \citep{Su2010ApJ, Zubovas2012MNRASa}. Locally, dense clouds, filaments and similar structures produce obstacles for the outflow to overcome. In those directions, the outflow expands more slowly and washes around the obstacle. Depending on the size of the overdensity, the outflow may split into multiple lobes for a significant duration. The obstacle is pushed primarily by the wind momentum \citep{Zubovas2014MNRASb} and can evaporate in the hot bubble \citep{Cowie1977ApJ}.

The proximate cause of the bubble expansion is the extremely high pressure of the shocked wind:
\begin{equation}
\begin{split}
    P_{\rm w} &\simeq n_{\rm w} k_{\rm B} T_{\rm sh} \sim \frac{\dot{M}_{\rm w}}{4\pi R_{\rm out}^2 v_{\rm w} \mu m_{\rm p}} k_{\rm B} T_{\rm sh} \\&\sim 5 \times 10^{-10} \frac{\dot{M}_{\rm w}}{2.2 \, \msun {\rm yr}^{-1}} R_{\rm kpc}^{-2} \frac{T_{\rm sh}}{10^{10} {\rm K}} {\rm erg \, cm}^{-3}.
    \end{split}
\end{equation}
Here, we scaled the wind mass flow rate to the Eddington accretion rate of a $10^8 \, \msun$ SMBH. This pressure exceeds the typical ISM pressure of $P_{\rm ISM} \sim 10^4 k_{\rm B}$~K~cm$^{-3} \sim 10^{-12}$~erg~cm$^{-3}$ by more than two orders of magnitude. The dynamical timescale of the outflow is $t_{\rm dyn, out} \sim R_{\rm out} / v_{\rm out} \sim 1 R_{\rm kpc} v_8^{-1}$~Myr, where $v_8$ is outflow velocity scaled to $10^8$~cm~s$^{-1}$. When the AGN switches off, the shocked wind bubble is no longer supplied with energy, but its pressure decreases on the same dynamical timescale. It takes approximately ${\rm ln}\left(P_{\rm w}/P_{\rm ISM}\right) t_{\rm dyn,out} \sim 6 t_{\rm dyn,out} \sim 6$~Myr for the pressure to drop down to typical ISM values. The outflow keeps expanding throughout this time, even though its velocity gradually decreases. A slightly more detailed calculation shows that the outflow may persist for $\sim 10$~times longer than the AGN episode inflating it \citep{King2011MNRAS}.

The estimates above are, of course, applicable only to highly idealised systems. Absolute spherical symmetry and adiabaticity have been assumed in order to make the equations analytically tractable. There are good reasons to think that the derived timescale is an overestimate. Even if the assumptions were true, the outflow would become undetectable once its velocity drops below the typical velocity dispersion of the ISM. Lack of spherical symmetry results in the mixing of the outflowing material with the ISM, slowing the outflow down and further decreasing the detectability. Additionally, detection of a non-spherical outflow depends on its orientation with respect to the line of sight. Finally, radiative cooling of the shocked wind bubble, especially when considering the density increase due to evaporating clouds, is going to reduce the bubble pressure and cause the outflow to stall earlier. Testing these complications, however, requires numerical simulations, which we present in this paper.

\section{Numerical model} \label{sec:numerical_model}

\begin{table*}
	\centering
	\caption{Initial conditions of the simulations}
	\label{table:initial_conditions}
	\begin{tabular}{ccccccc} 
		\hline
		Simulation name & SMBH mass [$\msun$] & Bulge gas mass [$\msun$] & Particle mass [$\msun$] & Bulge radius [kpc] & Gas fraction & AGN luminosity [erg s$^{-1}$] \\
		\hline
		M7L05 & $10^7$ & $4.1 \times 10^8$ & $410$ & $0.64$ & $0.1$ & $0.65 \times 10^{45}$\\
		M7L07 & $10^7$ & $4.1 \times 10^8$ & $410$ & $0.64$ & $0.1$ & $0.91 \times 10^{45}$\\
		M7L10 & $10^7$ & $4.1 \times 10^8$ & $410$ & $0.64$ & $0.1$ & $1.3 \times 10^{45}$\\
		M7L12 & $10^7$ & $4.1 \times 10^8$ & $410$ & $0.64$ & $0.1$ & $1.56 \times 10^{45}$\\
		M7-control & $10^7$ & $4.1 \times 10^8$ & $410$ & $0.64$ & $0.1$ & $0$\\
		M8L05 & $10^8$ & $3.7 \times 10^9$ & $3700$ & $2.57$ & $0.1$ & $0.65 \times 10^{46}$\\
		M8L07 & $10^8$ & $3.7 \times 10^9$ & $3700$ & $2.57$ & $0.1$ & $0.91 \times 10^{46}$\\
		M8L10 & $10^8$ & $3.7 \times 10^9$ & $3700$ & $2.57$ & $0.1$ & $1.3 \times 10^{46}$\\
		M8L12 & $10^8$ & $3.7 \times 10^9$ & $3700$ & $2.57$ & $0.1$ & $1.56 \times 10^{46}$\\
		M8-control & $10^8$ & $3.7 \times 10^9$ & $3700$ & $2.57$ & $0.1$ & $0$\\
		\hline
		M7fg002L01 & $10^7$ & $8.2 \times 10^7$ & $82$ & $0.64$ & $0.02$ & $0.13 \times 10^{45}$\\
		M7fg002L014 & $10^7$ & $8.2 \times 10^7$ & $82$ & $0.64$ & $0.02$ & $0.182 \times 10^{45}$\\
		M7fg002L02 & $10^7$ & $8.2 \times 10^7$ & $82$ & $0.64$ & $0.02$ & $0.26 \times 10^{45}$\\
		M7fg002L024 & $10^7$ & $8.2 \times 10^7$ & $82$ & $0.64$ & $0.02$ & $0.364 \times 10^{45}$\\
		M7fg002-control & $10^7$ & $8.2 \times 10^7$ & $82$ & $0.64$ & $0.02$ & $0$\\
		M8fg002L01 & $10^8$ & $7.4 \times 10^8$ & $740$ & $2.57$ & $0.02$ & $0.13 \times 10^{46}$\\
		M8fg002L014 & $10^8$ & $7.4 \times 10^8$ & $740$ & $2.57$ & $0.02$ & $0.182 \times 10^{46}$\\
		M8fg002L02 & $10^8$ & $7.4 \times 10^8$ & $740$ & $2.57$ & $0.02$ & $0.26 \times 10^{46}$\\
		M8fg002L024 & $10^8$ & $7.4 \times 10^8$ & $740$ & $2.57$ & $0.02$ & $0.364 \times 10^{46}$\\
		M8fg002-control & $10^8$ & $7.4 \times 10^8$ & $740$ & $2.57$ & $0.02$ & $0$\\
		\hline
	\end{tabular}
\end{table*}

\subsection{Simulation setup} \label{sec:sim_setup}

We run the simulations using Gadget-3, a modified version of the publicly available Gadget-2 \citep{Springel2005MNRAS}. Our version uses the SPHS modification \citep{Read2010MNRAS} with a Wendland $C^2$ kernel \citep{Wendland95} with 100 neighbour particles. Gas particles use an adaptive softening length with the minimum possible value set to $0.1$~pc, much smaller than any of the structures investigated in this paper. AGN feedback is tracked using the `virtual particle' method described in \citet{Nayakshin2009MNRASb}. Briefly, the AGN emits tracer particles that travel in straight lines at a velocity $v_{\rm w} = 0.1c$ until they enter the smoothing kernel of an SPH particle. Each tracer particle carries momentum and energy that it passes on to the SPH particles it encounters over several timesteps. This method has significant advantages over the more common spherical energy injection into the gas surrounding the AGN, mainly because our method allows outflows to become realistically non-spherical when encountering a non-uniform gas distribution \citep{Zubovas2016MNRASa}.

Gas heating and cooling are tracked using a composite prescription. For gas at temperatures above $T_{\rm thresh} = 10^4$~K, we use the prescription from \citet[][hereafter SOCS]{Sazonov2005MNRAS}, which accounts for Compton heating and/or cooling, bremsstrahlung and metal line cooling in optically thin gas exposed to an AGN radiation field. While the AGN is inactive, the prescription acts as purely a cooling function, i.e. it contains no other heating terms. For temperatures below $T_{\rm thresh}$, we use the cooling function from \citet{Mashchenko2008Sci} instead. This function does not have a heating term to account for the AGN radiation, but we implicitly assume that the cold (and, usually, dense) gas is efficiently shielded from the radiation. The code tracks the ionization fraction and effective mean physical particle mass $\mu$ within each SPH particle following the prescription of \citet{Katz1996ApJS}. Within this prescription, the value of $\mu$ varies from $\mu_{\rm max} \simeq 1.22$ at $T \ll 10^4$~K to $\mu_{\rm min} = 0.63$ at $T \gg 10^4$~K, with a rather sharp transition between the two around $T = 10^4$~K. We also impose a density-dependent temperature floor, which is selected so that the Jeans mass of the gas never drops below the resolvable mass in the simulation, which is $m_{\rm res} = 100 m_{\rm SPH}$, with $m_{\rm SPH}$ being the mass of a single particle:
\begin{equation}
    T_{\rm floor} = \rho^{1/3}\frac{\mu m_{\rm p}G}{\pi k_{\rm B}}\left(m_{\rm res}\right)^{2/3}.
\end{equation}
Here, $m_{\rm p}$ is the proton mass and $k_{\rm B}$ is the Boltzmann constant. In order to mitigate the use of computing resources for tracking very dense gas, we convert gas with particle density $n > 100$~cm$^{-3}$ that is at the temperature floor into star particles. This is done stochastically, with each gas particle having a probability of $10\%$ to be converted over its dynamical timescale. We do not intend this approach to produce realistic star formation rates that can be compared to real data, however, comparisons between simulations and qualitative analysis of fragmentation trends are still possible.

\subsection{Initial conditions} \label{sec:init_cond}

The initial conditions in each simulation are chosen to represent an idealised form of a galaxy bulge. They consist of an SMBH placed at the origin of the coordinate system, surrounded by a shell of gas starting at an inner radius $R_{\rm in} = 0.1$~kpc. The shell contains $N_{\rm gas} \simeq 10^6$ particles\footnote{ This number is motivated by our available computational results; we present a resolution study in Appendix \ref{app:resolution}.}. The gas in the shell is given a convergence-free ($\nabla \cdot \vec{v} = 0$) turbulent velocity field based on the prescription described in \citet{Dubinski1995ApJ} and \citet{Hobbs2005MNRAS}. Turbulence is not driven during the course of the simulation.

We run 20 simulations in total, split into four groups depending on the SMBH mass and gas density. Half the simulations have SMBH masses of $10^7 \, \msun$ and half have $10^8 \, \msun$; we label them M7 and M8, respectively. The SMBH mass also determines three other parameters of the simulations: the total bulge mass, the outer radius of the gas shell and the strength of the background gravitational potential.

We choose the galaxy parameters based on observationally-derived typical values of present-day galaxies. The bulge mass is set by the $M_{\rm BH} - M_{\rm bulge}$ relation \citep{McConnell2013ApJ}:
\begin{equation} \label{eq:mbh-mbulge}
    \log\left(\frac{M_{\rm b}}{10^{11}~ \msun} \right) = \frac{\log\left(\frac{M_{\rm BH}}{\msun}\right) - 8.46}{1.05};
\end{equation}
for the two simulation groups, this gives
\begin{equation}
    M_{\rm b, M7} = 4.1\times10^9 \, \msun; \quad M_{\rm b,M8} = 3.7\times10^{10} \, \msun.
\end{equation}
In half the simulations, the gas mass is set to $10\%$ of the total bulge mass. This is roughly appropriate for very gas-rich, cluster/group central galaxies in the local Universe \citep{Guo2017ApJ}. These simulations do not have a specific label. In the other half, the gas mass is five times lower, appropriate for local field galaxies \citep{Guo2020ApJ}; these simulations are labeled `fg002'.

Similarly, we calculate the bulge velocity dispersion from the $M-\sigma$ relation \citep{McConnell2013ApJ}:
\begin{equation} \label{eq:mbh-sigma}
    \log\left(\frac{\sigma}{200 {\rm km s}^{-1}} \right) = \frac{\log\left(\frac{M_{\rm BH}}{\msun}\right) - 8.32}{5.64};
\end{equation}
the values for the two simulation groups are
\begin{equation}
    \sigma_{\rm M7} = 117 \, {\rm km\, s}^{-1}; \quad \sigma_{\rm M8} = 176 \, {\rm km\, s}^{-1}.
\end{equation}
We set up a static isothermal gravitational potential in our simulations that provides a constant circular velocity equal to $v_{\rm circ} = \sqrt{2}\sigma$. In addition to this external potential, the gas is affected by its self gravity and the gravity of the SMBH; however, the gravitational potential of the SMBH is dominant only in the central few parsecs, well inside the sink radius (see below). The gravitational softening length for gas particles is set to be equal to the smoothing length, and for the non-SPH particles it is set to $0.1$~pc, effectively making them point sources of gravity.

Finally, we use the two values to determine the radius of the bulge:
\begin{equation} \label{eq:rbulge}
    R_{\rm bulge} = \frac{G M_{\rm tot, b}}{2 \sigma^2};
\end{equation}
the values are
\begin{equation}
    R_{\rm b, M7} = 0.64 \, {\rm kpc}; \quad R_{\rm b,M8} = 2.57 \, {\rm kpc}.
\end{equation}
We set this radius to be the outer radius of the gas shell.

Within each of the four groups defined by SMBH mass and gas density, the five simulations differ in the AGN luminosity. One simulation is a control run, with $L_{\rm AGN} = 0$ throughout. In the other four, the AGN turns on at $t = 1$~Myr after the start of the simulation and stays on for $t_{\rm q} = 1$~Myr, after which the AGN switches off forever. The AGN luminosity is set to $L_{\rm AGN} = l L_{\rm Edd}$, where $L_{\rm Edd} \simeq 1.3 \times 10^{45} M_{\rm BH} / \left(10^7 \, \msun\right)$~erg~s$^{-1}$ is the Eddington luminosity. In the higher density simulations, the Eddington factor $l$ is set to $l = 0.5$, $0.7$, $1.0$ and $1.2$; we label these simulations L05, L07, L10 and L12, respectively. In the `fg002' simulations, the values of $l$ are also five times lower: $l = 0.1$, $0.14$, $0.2$ and $0.24$; we label them L01, L014, L02 and L024, respectively. The simulation parameters are summarized in Table \ref{table:initial_conditions}.

The first million years of each simulation are intended to allow the gas distribution to relax and develop an uneven density structure. For this reason, we set the sink boundary of the SMBH particle to $R_{\rm in}$ for the first million years, so that the inner edge of the shell stays constant. After the relaxation period, the sink boundary is reduced to $R_{\rm sink} = 0.01$~kpc. Neither radius is intended as any physical boundary of gas capture by the SMBH; this is merely a solution that helps us save computational resources by refraining from simulating the regions very close to the SMBH. Keeping the sink boundary large during relaxation has the added benefit of preventing the formation of unphysically dense clumps in the central regions that would preclude any outflow from forming. Since turbulent velocity in SPH simulations decays on a dynamical timescale \citep{Hobbs2005MNRAS}, we want the inner edge of the distribution to have a dynamical timescale $t_{\rm dyn} \sim 1$~Myr. Given our setup, $t_{\rm dyn} \sim R_{\rm in}/\sigma \sim 0.6-0.8$~Myr, which is close to the requirement.

We run each simulation for $10$~Myr physical time, making snapshots every $50$~kyr. Subsequent analysis is performed using Python scripts, including pyGadgetReader \citep{pygadgetreader}. All plots are made using the matplotlib package\footnote{https://matplotlib.org/}.


\begin{figure*}
	\includegraphics[width=0.33\textwidth]{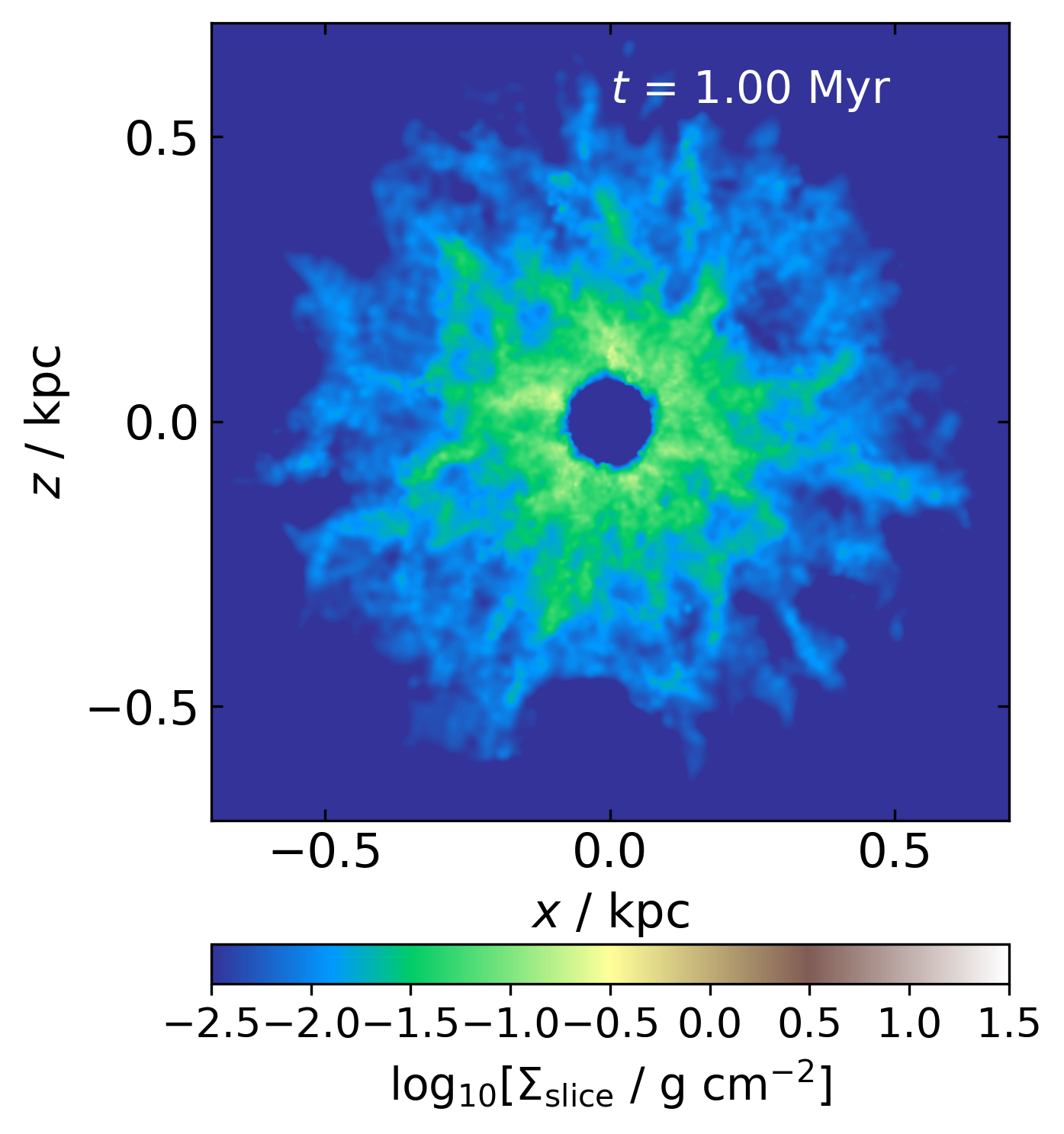}
	\includegraphics[width=0.33\textwidth]{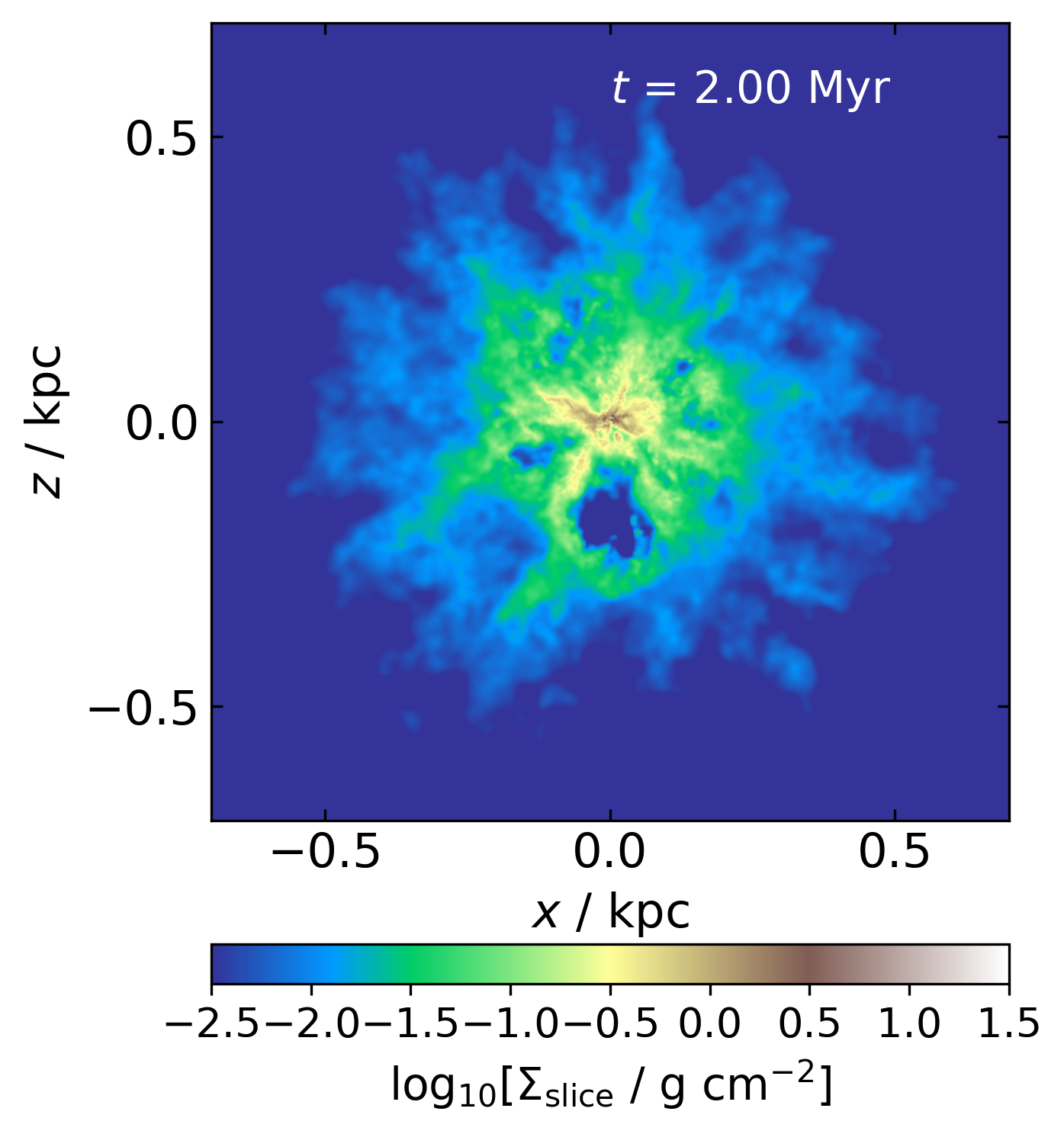}
	\includegraphics[width=0.33\textwidth]{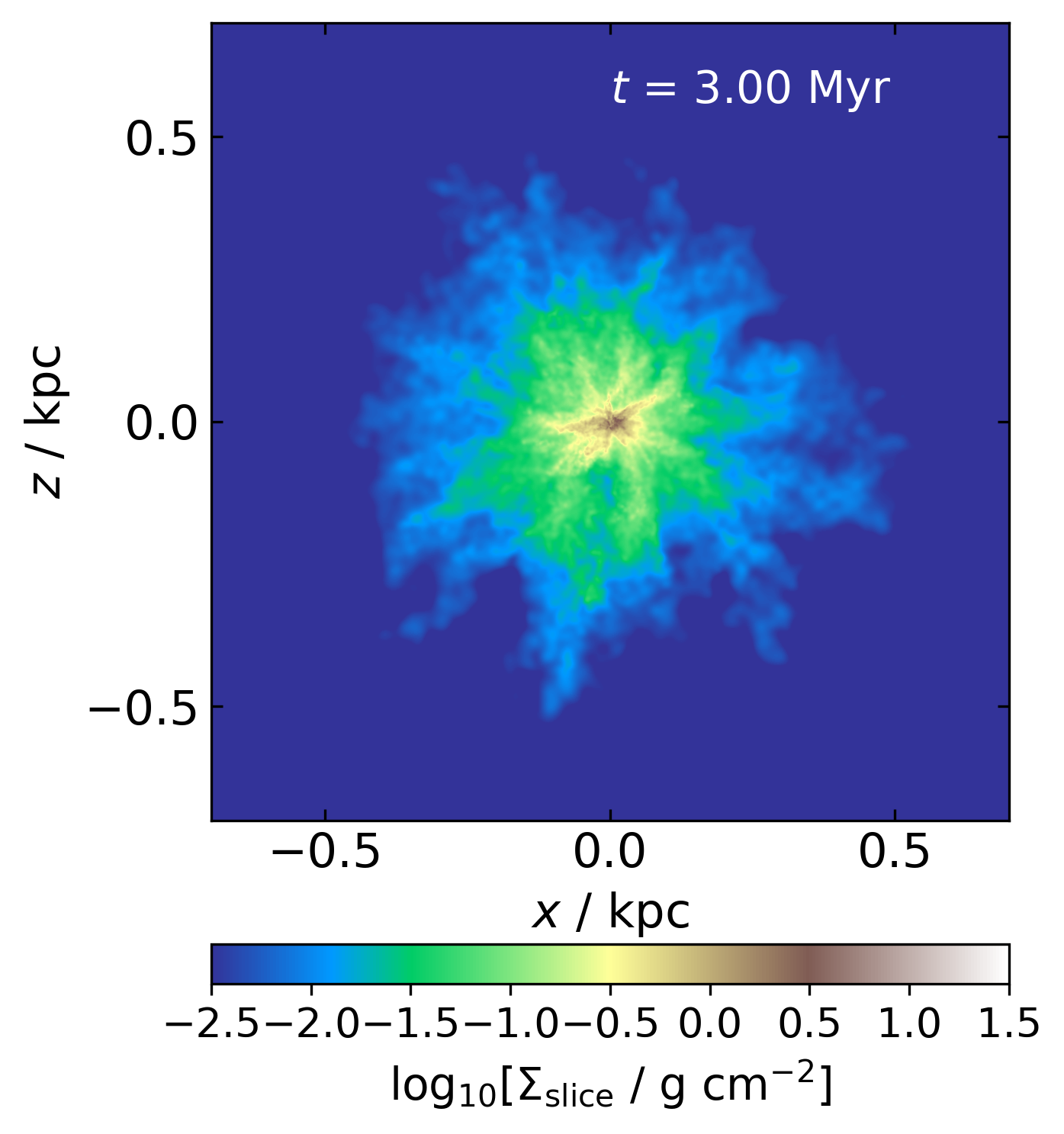}
    \caption{Density evolution of the M7L07 simulation, showing no prominent fossil outflow. {\em Left}: density slice at $t = 1$~Myr, just before the AGN switches on. {\em Middle}: density slice at $t = 2$~Myr, when the AGN switches off. {\em Right}: density slice at $t = 3$~Myr.}
    \label{fig:evolution_M7L07}
\end{figure*}

\begin{figure*}
	\includegraphics[width=0.33\textwidth]{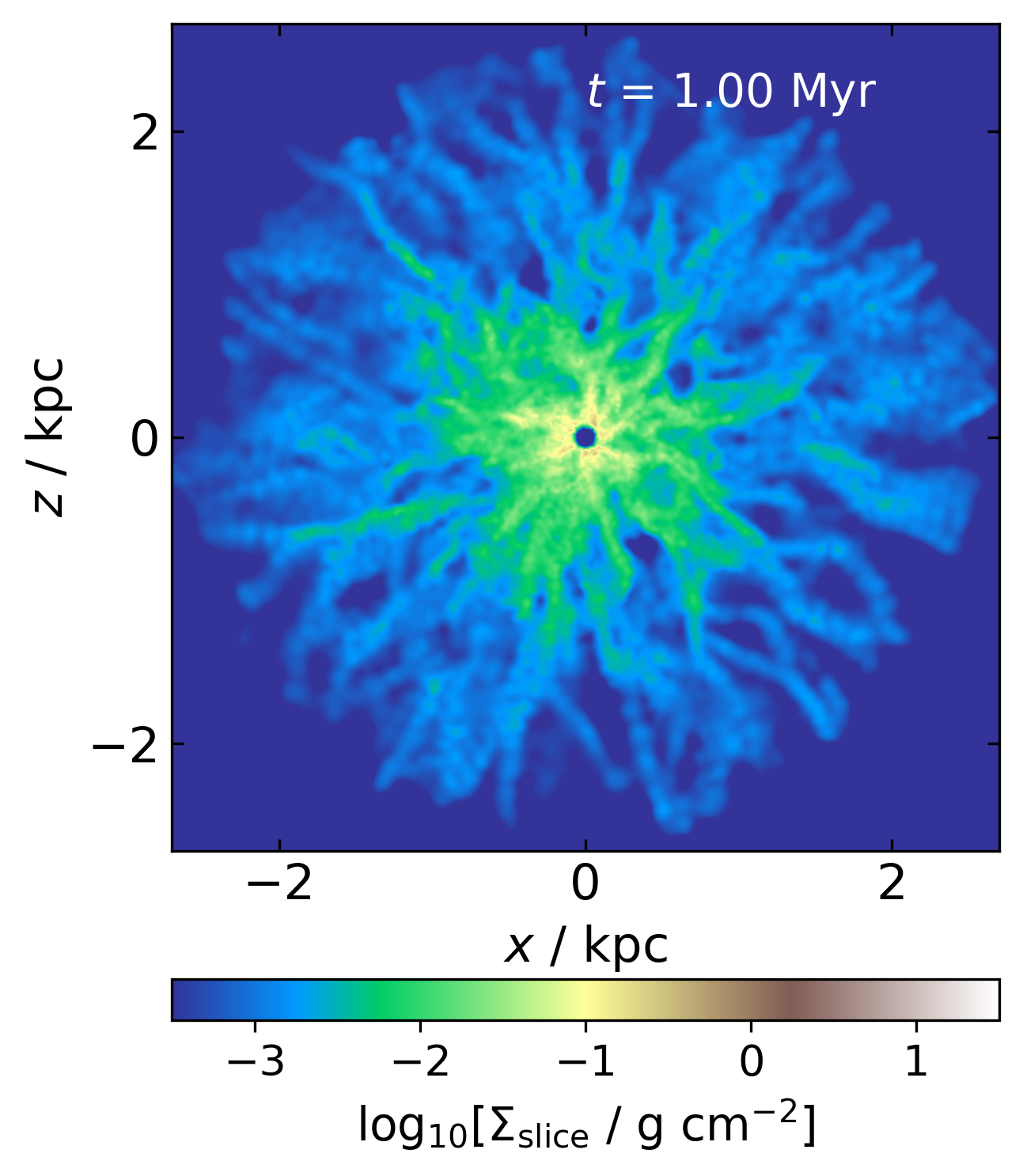}
	\includegraphics[width=0.33\textwidth]{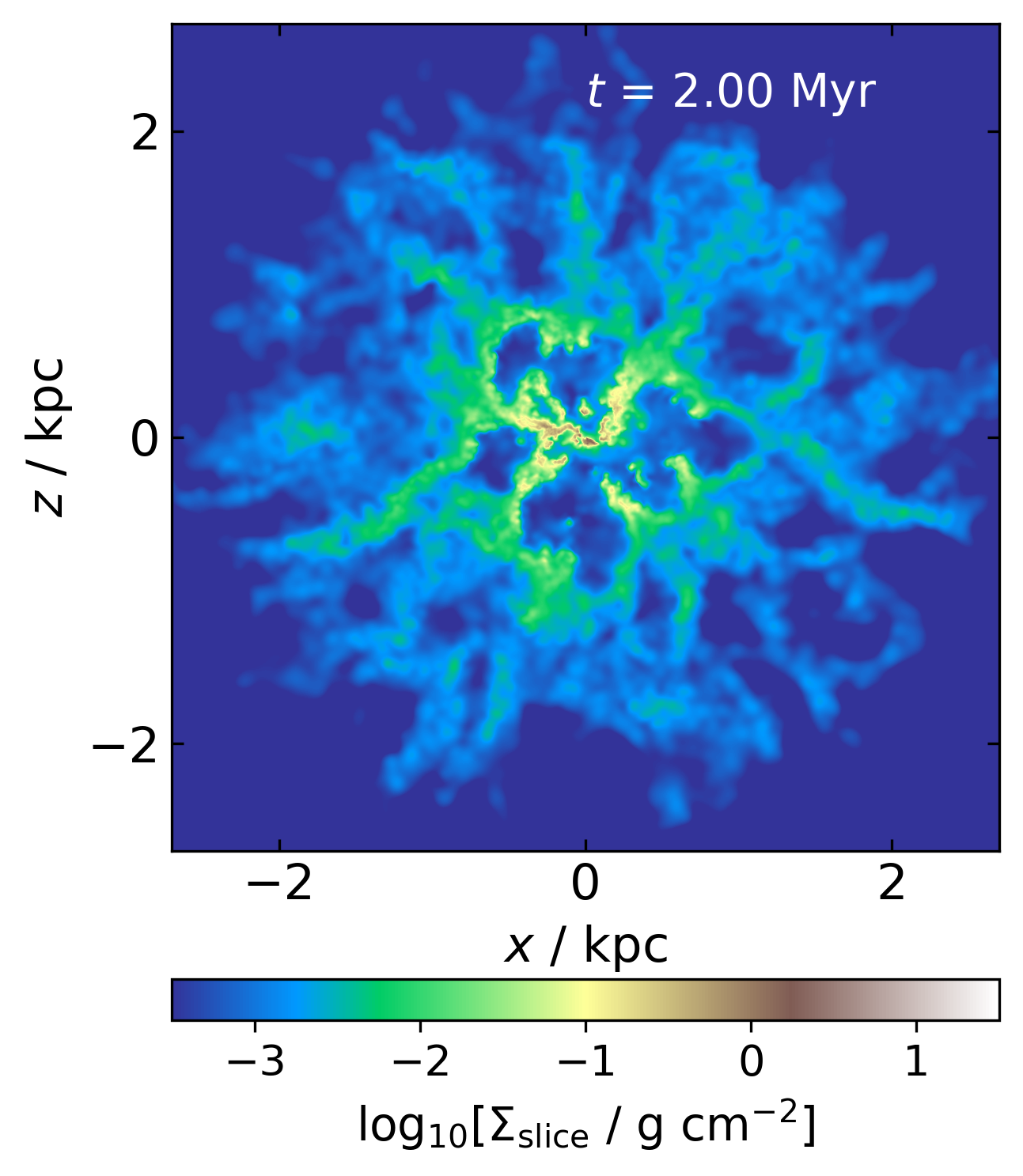}
	\includegraphics[width=0.33\textwidth]{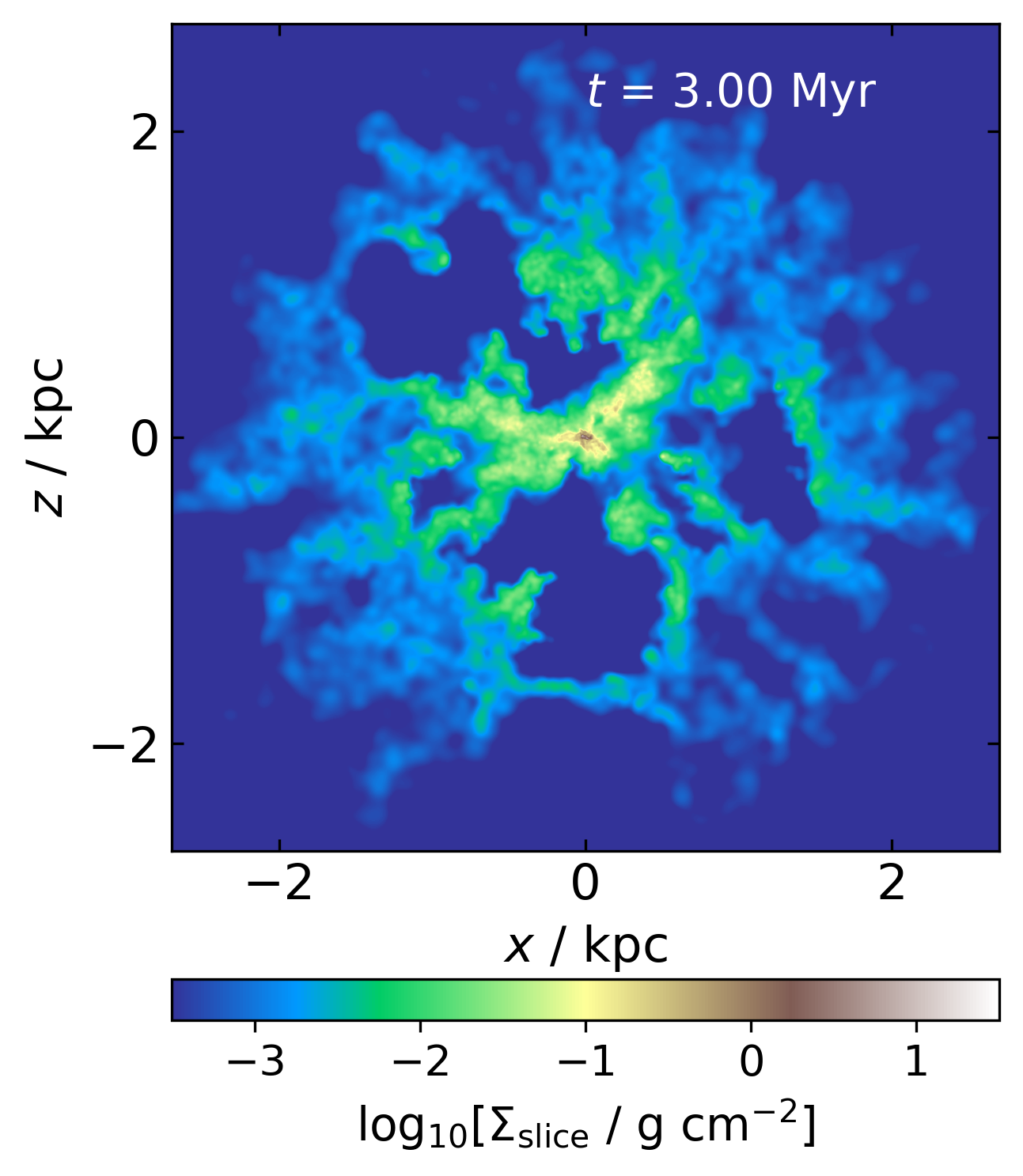}
    \caption{Density evolution of the M8L10 simulation, showing a prominent multi-lobed fossil outflow. Panels show the same times as in Figure \ref{fig:evolution_M7L07}. Note the different spatial scales between the two figures.}
    \label{fig:evolution_M8L10}
\end{figure*}

\begin{figure*}
	\includegraphics[width=0.33\textwidth]{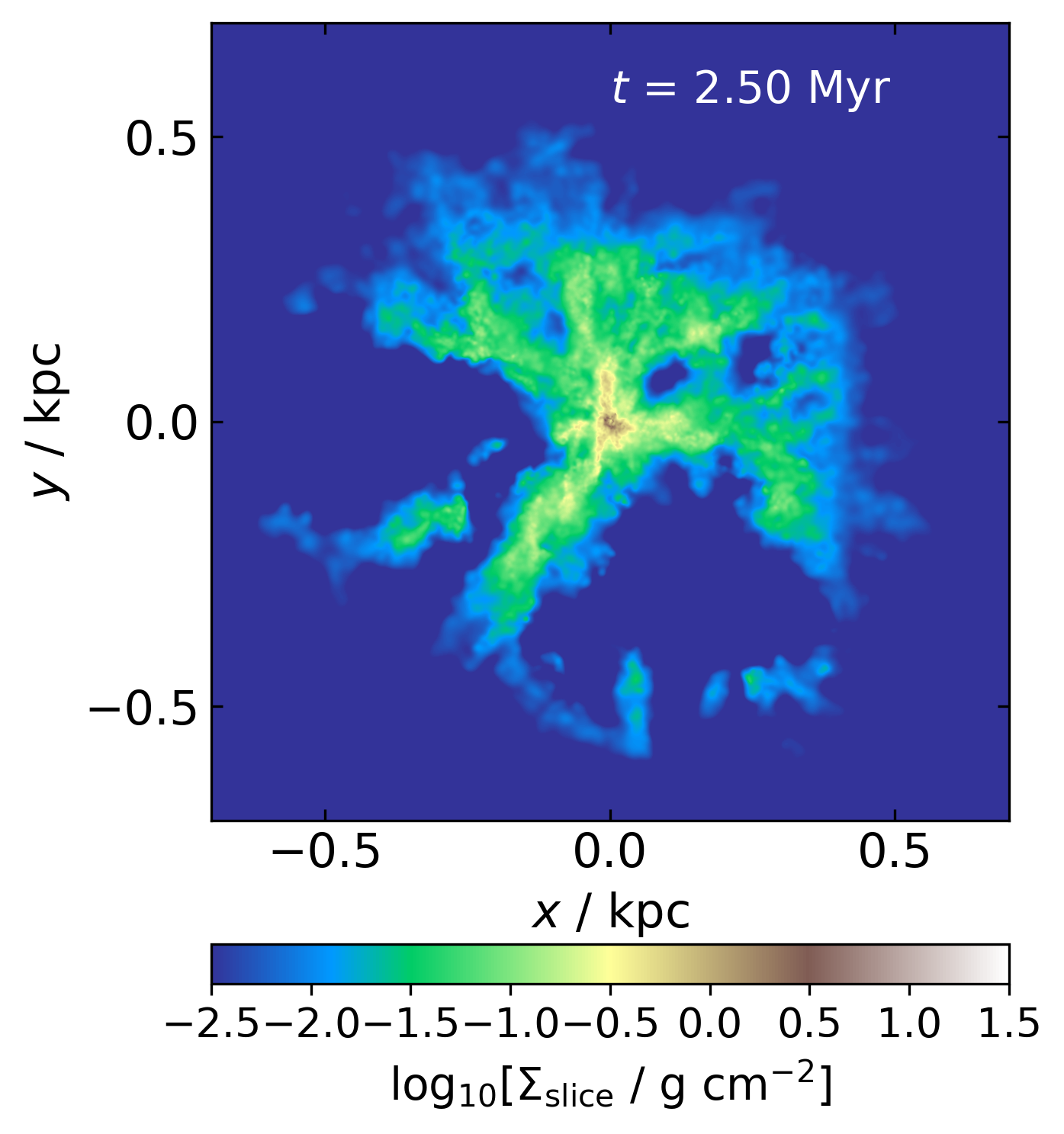}
	\includegraphics[width=0.33\textwidth]{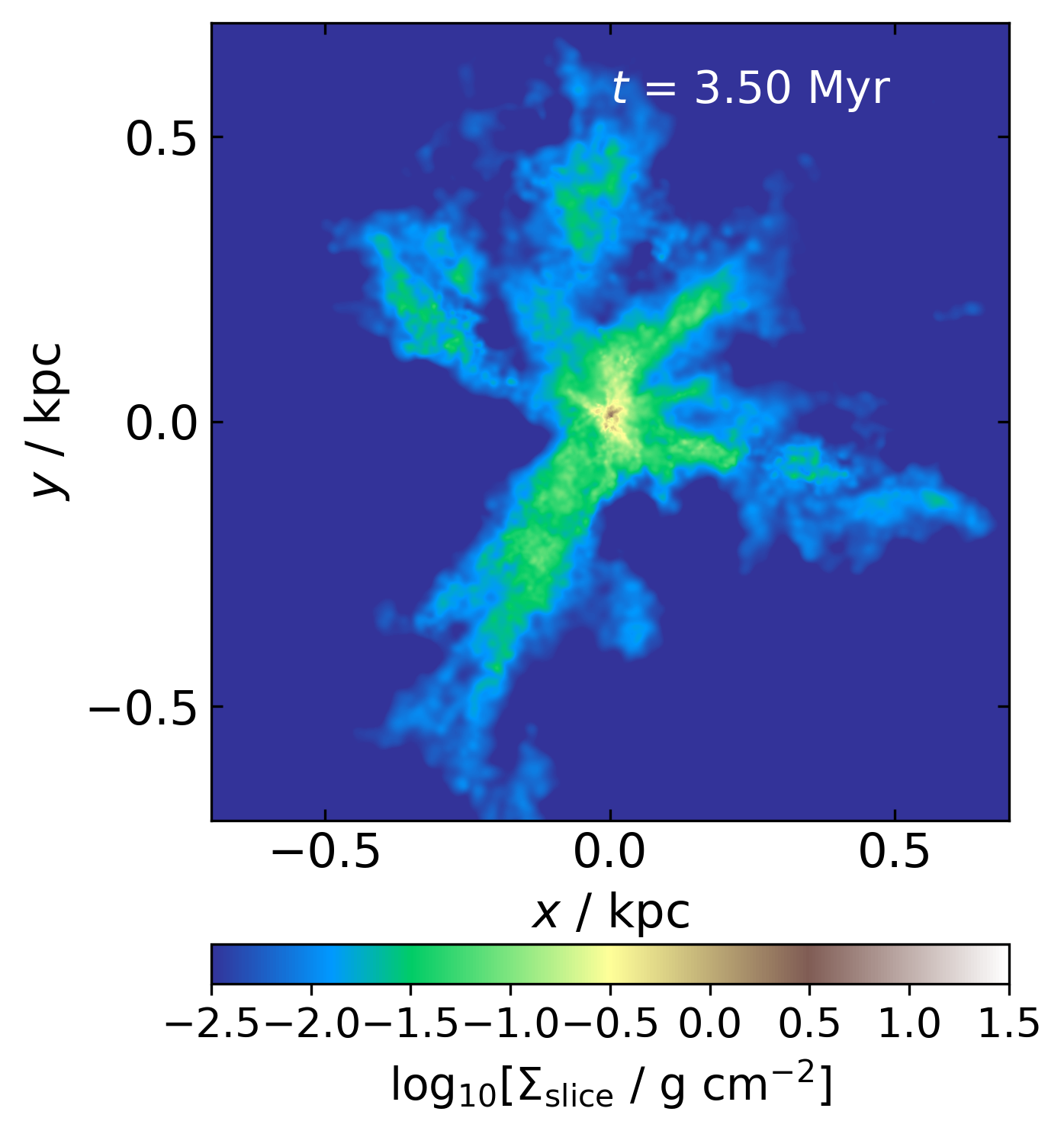}
	\includegraphics[width=0.307\textwidth]{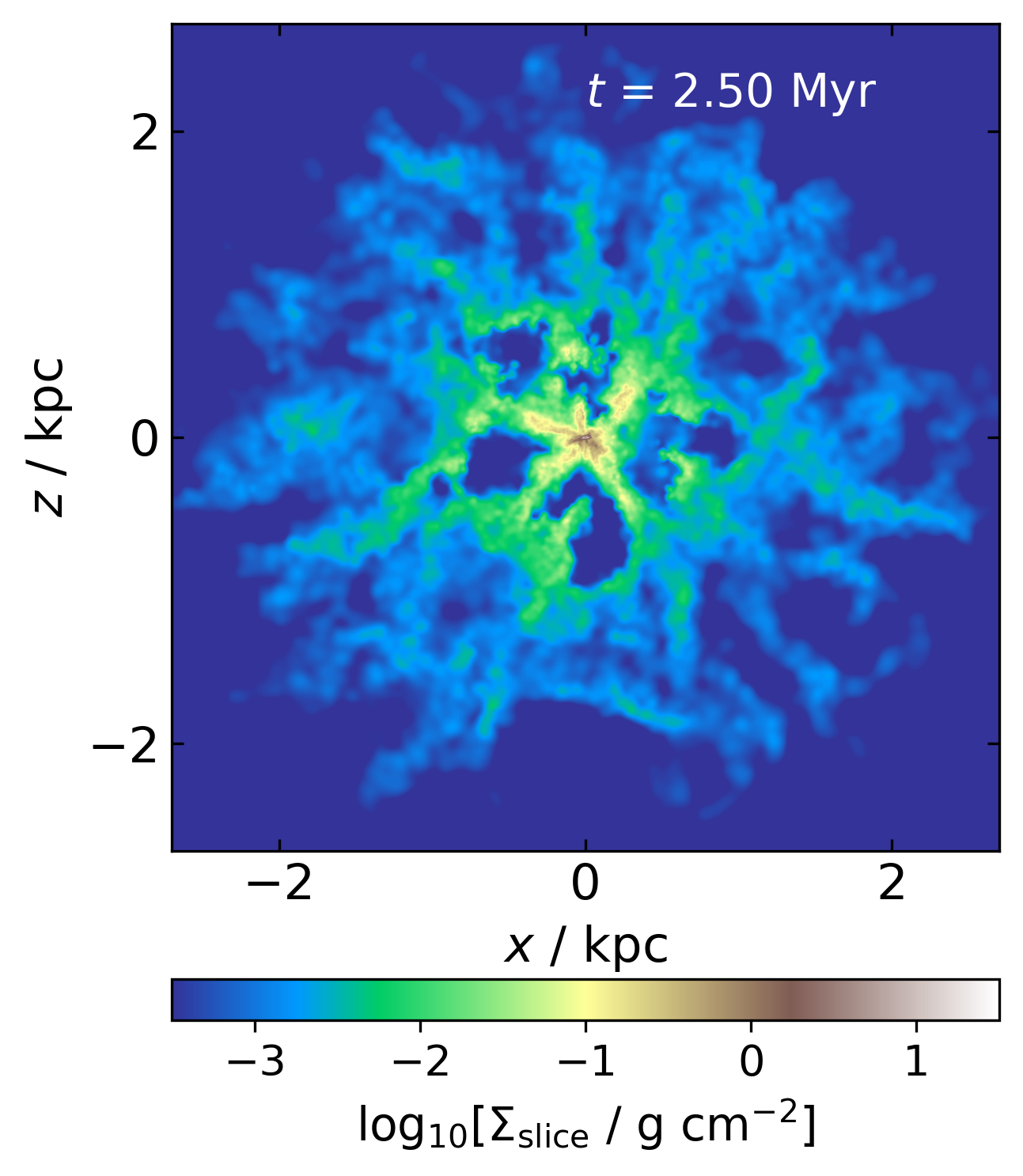}
    \caption{Density maps of three other fossil outflows, showcasing interesting behaviour. {\em Left}: in M7L10, one bubble breaks out of the bulge soon after the AGN switches off, followed $\sim 0.3$~Myr later by another one to the bottom right. {\em Middle}: in M7L12, two bubbles break out of the nucleus and compress the gas between, leading to a very asymmetric distribution. {\em Right}: in M8L07, the bubbles stall within $0.5$~Myr, but maintain clear shapes.}
    \label{fig:other_density_maps}
\end{figure*}

\begin{figure*}
	\includegraphics[width=0.33\textwidth]{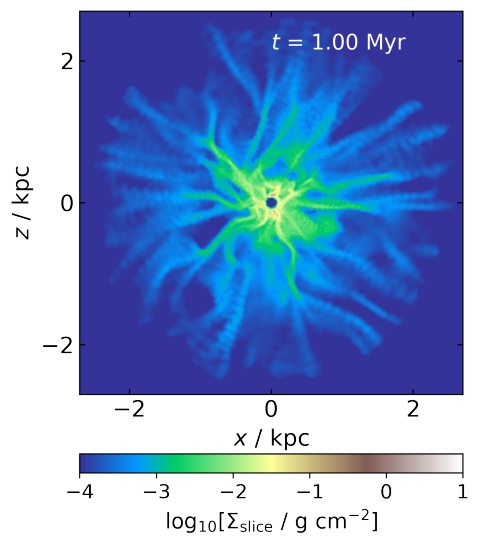}
	\includegraphics[width=0.33\textwidth]{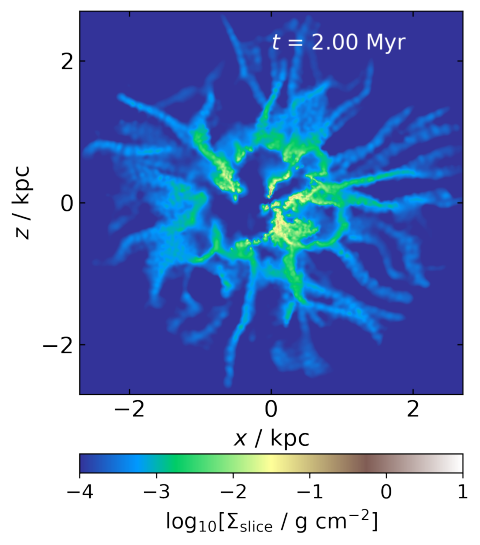}
	\includegraphics[width=0.33\textwidth]{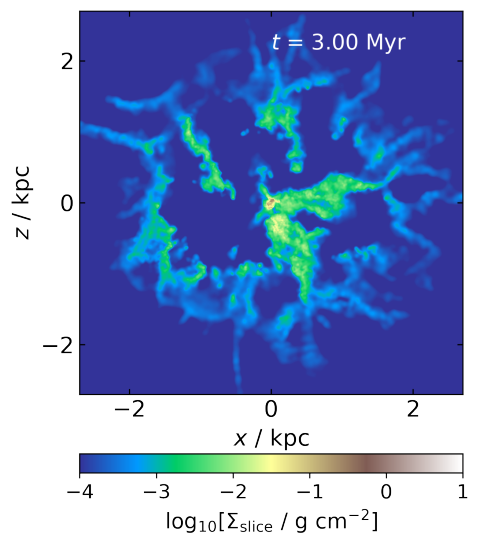}
    \caption{Density evolution of the M8fg002L02 simulation. Panels show same times as in Figure \ref{fig:evolution_M8L10}.}
    \label{fig:evolution_M8fg002L02}
\end{figure*}

\subsection{Analytical expectations} \label{sec:analytical}

In an idealised spherically symmetric non-turbulent gas distribution, the energy-driven outflow would expand spherically and would attain a constant velocity\footnote{Formally, in a spherically symmetric distribution with our chosen bulge and SMBH parameters, an AGN shining at $L \geq L_{\rm Edd}$ would drive a large-scale outflow purely by its momentum input. In our simulations, AGN wind momentum is transferred to the gas but is not enough to push it away because of two reasons. First of all, most of the gas close to the SMBH has a significant negative radial velocity, which slows down and can even prevent a momentum-driven outflow from forming \citep{Nayakshin2010MNRAS}. Secondly, most of the gas that is resilient to feedback is much denser than the average and so resists the momentum push effectively.}. We can determine this velocity following the derivation in \citet[see also Section \ref{sec:physical_model}]{King2011MNRAS}, but allowing for the different relationship between $M_{\rm BH}$ and $\sigma$:
\begin{equation}
    \frac{\eta}{2}L_{\rm AGN} = \frac{3 f_{\rm g} \sigma^2}{G} \dot{R}^3 + \frac{10 f_{\rm g} \sigma^4}{G} \dot{R}.
\end{equation}
Rearranging gives
\begin{equation} \label{eq:dotr_deriv}
    \frac{\eta G}{2 f_{\rm g} \sigma^2} \frac{4 \pi G M_{\rm BH} c l}{\kappa} = 3 \dot{R}^3 + 10 \sigma^2 \dot{R}.
\end{equation}
Assuming that the outflow velocity is significantly higher than $\sigma$ allows us to neglect the second term on the right to get
\begin{equation}
    \dot{R}_{\rm exp} \simeq \left(\frac{2 \pi \eta G^2 M_{\rm BH} c l}{3 f_{\rm g} \kappa \sigma^2}\right)^{1/3}.
\end{equation}
Putting in the numerical values gives the velocity estimates for the two groups of simulations:
\begin{equation} \label{eq:dotr_M7}
    \dot{R}_{\rm exp, M7} \simeq 465 \left(l/f_{0.1}\right)^{1/3} \, {\rm km\, s}^{-1}
\end{equation}
and
\begin{equation} \label{eq:dotr_M8}
    \dot{R}_{\rm exp, M8} \simeq 765 \left(l/f_{0.1}\right)^{1/3} \, {\rm km\, s}^{-1}.
\end{equation}
Here, $f_{0.1} \equiv f_{\rm g}/0.1$. In the `fg002` simulations, $f_{\rm g} = 0.02$, but the values of $l$ are also five times lower, so the expected outflow velocity in the corresponding simulations is the same.

After $1$~Myr of AGN activity, the outflows should reach radii $R_{\rm off} \simeq \dot{R}_{\rm exp} t_{\rm q}$, i.e. $R_{\rm off, M7} \simeq 475 \left(l/f_{0.1}\right)^{1/3}$~pc and $R_{\rm off, M8} \simeq 780 \left(l/f_{0.1}\right)^{1/3}$~pc; in both cases, this is smaller than the bulge radius. The stalling timescale and radius, following equations (22) and (28) of \citet{King2011MNRAS}, are
\begin{equation}
    t_{\rm stall, M7} \simeq 7.9 \left(l/f_{0.1}\right)^{2/3} \,{\rm Myr}; \quad R_{\rm stall, M7} \simeq 1.8 \left(l/f_{0.1}\right)^{2/3} \,{\rm kpc}
\end{equation}
and
\begin{equation}
    t_{\rm stall, M8} \simeq 9.4 \left(l/f_{0.1}\right)^{2/3} \,{\rm Myr}; \quad R_{\rm stall, M8} \simeq 3.2 \left(l/f_{0.1}\right)^{2/3} \,{\rm kpc}.
\end{equation}
All of these estimates are upper limits. The velocity values would be reduced if we accounted for the neglected $10 \sigma^2 \dot{R}$ factor in eq. \ref{eq:dotr_deriv}. The stalling timescale calculation relies on the implicit assumption that the outflow persists until its velocity drops to zero, while in reality, it becomes indistinguishable from galactic gas once $\dot{R} < \sigma$. The combination of both effects leads to an overestimate of the stalling radius as well.

On the other hand, turbulence changes the density distribution and allows some outflows to expand through low-density channels \citep[see also Section \ref{sec:physical_model}]{Zubovas2014MNRASb}, so the outflow velocity, stalling timescale and radius may be higher than given by the analytical estimates.

\section{Results} \label{sec:results}

\subsection{Morphological evolution} \label{sec:morphology}

We first describe the qualitative evolution of the simulated systems, using the higher-density simulations as a base. In general, they can be divided into two groups: those with prominent fossil outflow bubbles lasting at least $10^5$~yr after the AGN switches off, and those without. The first group includes simulations M7L10, M7L12, M8L07, M8L10 and M8L12; the second includes M7L05, M7L07 and M8L05.

Figures \ref{fig:evolution_M7L07} and \ref{fig:evolution_M8L10} showcase the evolution of simulations M7L07 and M8L10, respectively. The panels, from left to right, show density maps at $t = 1, 2$ and $3$~Myr. The projection is in the XZ plane and only gas with $-70$~pc~$< y < 70$~pc is shown; since our simulations are broadly spherically symmetric, cuts along any other plane passing through the centre would look qualitatively the same.

At $t = 1$~Myr, both simulations have developed an uneven density distribution, with denser filaments and more diffuse voids in between. Due to the higher characteristic velocities, the density contrasts in the M8L10 simulation are higher than in the lower-mass counterpart. The hole in the centre exists due to the imposed sink radius of the SMBH particle. As the AGN switches on, the sink radius is reduced by a factor 10 and gas begins falling inward. Simultaneously, the AGN wind pushes the gas outward, developing outflow bubbles.

In M7L07, the AGN wind is generally too weak to push the gas against the combined action of ambient pressure, gravitational potential and inertia. As a result, while some material is accelerated to positive radial velocities, only one significant outflow bubble remains by $t = 2$~Myr (Fig. \ref{fig:evolution_M7L07}, middle panel, below the centre). The vast majority of material continues pouring down on to the SMBH and quickly swamps the outflow once the AGN switches off. The outflow bubbles disappear within $< 0.5$~Myr of quiescence, and by $t = 3$~Myr, there is no evidence of fossil outflows.

In contrast, the AGN in M8L10 is powerful enough to create significant outflow bubbles in all directions. By $t = 2$~Myr, the largest bubbles reach a radius $R_{\rm out} \sim 1$~kpc, i.e. the maximum velocity of diffuse gas is $v_{\rm out,diff} \sim 1000$~km~s$^{-1}$ (see also Section \ref{sec:out_kinematics} below). This velocity is somewhat higher than analytically expected (eq. \ref{eq:dotr_M8}), but this is reasonable since the analytical estimate is based on the assumption of gas density being the same in all directions. Nevertheless, there is some gas that keeps falling toward the centre, forming dense filaments resilient to feedback. This is a common feature seen in many numerical simulations of feedback, including idealised \citep[e.g.,][]{MacLow1988ApJ, Zubovas2014MNRASb}, isolated galaxy- and/or cluster-scale \citep[e.g.,][]{Sijacki2008MNRAS, Gabor2014MNRAS} and cosmological \citep[e.g.][]{Costa2018MNRAS, Pillepich2021MNRAS}. After AGN switchoff, the four outflow lobes visible in the middle panel of Figure \ref{fig:evolution_M8L10} continue expanding and clearing out large volumes of gas. Each lobe has an opening angle of $\sim 60-70^\circ$, but this value is very sensitive to our adopted turbulent velocity spectrum. By $t = 3$~Myr, two lobes have expanded to $R > 1.5$~kpc (Fig. \ref{fig:evolution_M8L10}, right); there is essentially no gas inside them. Parts of the top-left bubble escape the bulge entirely and only slow down to a halt past the $5$~Myr mark. Remnants of that bubble are still visible by the time the simulation finishes at $t = 10$~Myr, although all the gas is falling back toward the centre. The other two bubbles stall and fall back or dissipate much more rapidly.

In the other high-density simulations, the shape of the fossil outflows is varied. For example, the evolution of M7L10 is intermediate between those of M7L07 and M8L10. Its outflow initially expands in all directions, similarly to the one in M8L10. When the AGN episode ends, however, the outflow becomes much more lopsided much more rapidly. One bubble breaks out of the bulge almost immediately after the AGN switchoff, expanding in the negative X direction (Figure \ref{fig:other_density_maps}, left panel); $\sim 0.3$~Myr later, another bubble breaks out to the bottom right. By this time, most of the material of the earlier bubble has left the bulge and become very dilute, even though the outflowing mass is still large (see Section \ref{sec:out_mass} below). It is unlikely that both bubbles will be detected at the same time. The reason for the difference between the morphological evolution of the outflow in M7L10 and M8L10 is twofold. First of all, the gravitational potential is shallower in the M7 simulations; secondly, the average gas density is lower. Both effects make it easier for the hot gas to push the bubbles outward. The expanding bubbles compress some gas laterally and smother the smaller bubbles, creating a more lopsided appearance than in M8L10. In M7L12, the same effect is even more pronounced: two bubbles quickly break out of the initial gas distribution and expand away to infinity, compressing the gas in between them into relatively narrow filaments (Figure \ref{fig:other_density_maps}, middle panel). This change in gas morphology persists until the end of the simulation, suggesting that any further AGN episodes would produce outflows in similar directions to the original ones, while the rest of the gas can easily fall toward the SMBH. Conversely, in M8L07, the bubbles stall within $0.5$~Myr after the AGN episode ends, but retain their clear shapes even as they fall back inward (Figure \ref{fig:other_density_maps}, right panel).

Morphologically, the low-density simulations show very similar behaviour to the high-density ones at early times after the AGN switches off (see Fig. \ref{fig:evolution_M8fg002L02}). By $t=2$~Myr, the outflow bubble has a radius of $\sim 1$~kpc, just as in the high-density simulation. By $t=3$~Myr, the highest radius reached by the outflow is also similar to that in the high-density simulation, but the outflowing material subtends a larger solid angle. At later times, the evolution of the simulations becomes progressively more different, because the low-density gas cools down more slowly and can expand for longer (see Section \ref{sec:thermodynamics}). This trend is seen when comparing all the corresponding low- and high-density simulations, although the presence or absence of fossil outflows is not affected by the density. In addition, the low-density fossil outflows appear more asymmetric by $t = 3$~Myr compared with the high-density ones, although this is a somewhat subjective distinction.

\subsection{Outflowing mass} \label{sec:out_mass}

\begin{figure}
	\includegraphics[width=\columnwidth]{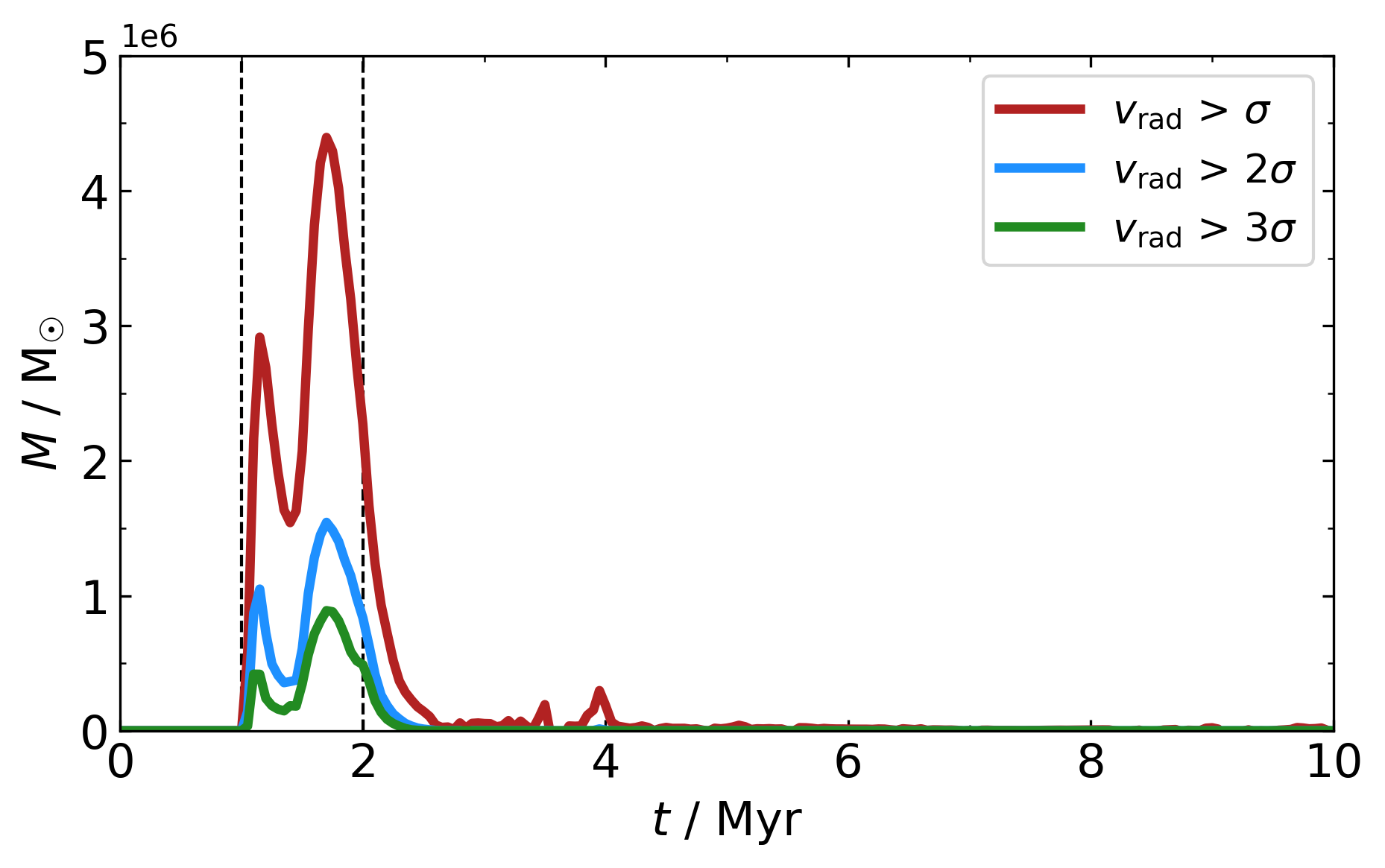}
	\includegraphics[width=\columnwidth]{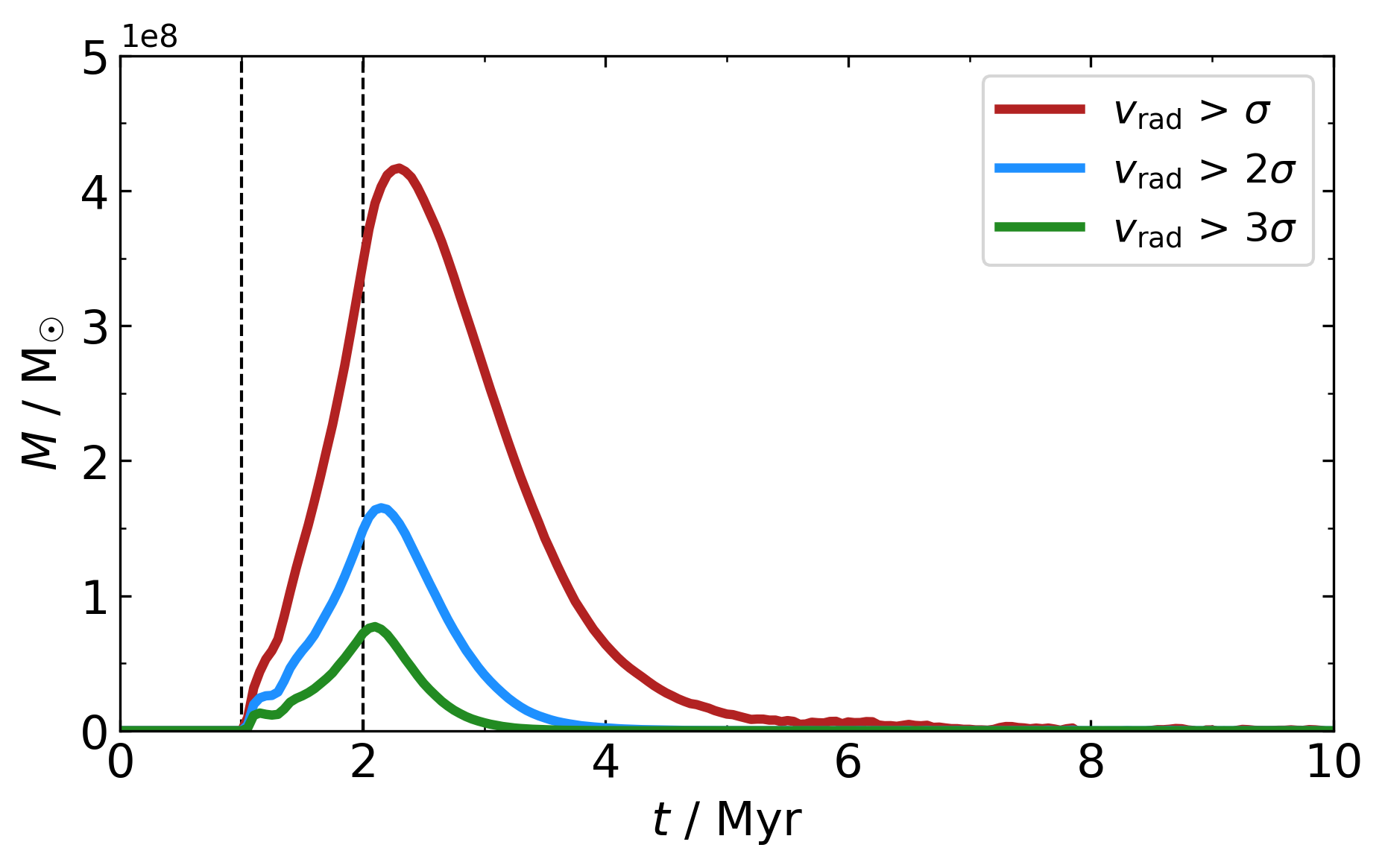}
    \caption{Total outflowing mass. Solid lines show the difference in mass moving with radial velocity above $\sigma$ (brown), $2\sigma$ (blue) and $3 \sigma$ (green) in the M7L07 (top) and M8L10 (bottom) simulations and the corresponding control simulations, against time. Vertical dashed lines show the start and end times of the AGN episode. Note that the velocity dispersions between the two simulations are different.}
    \label{fig:mass_time}
\end{figure}

\begin{figure}
	\includegraphics[width=\columnwidth]{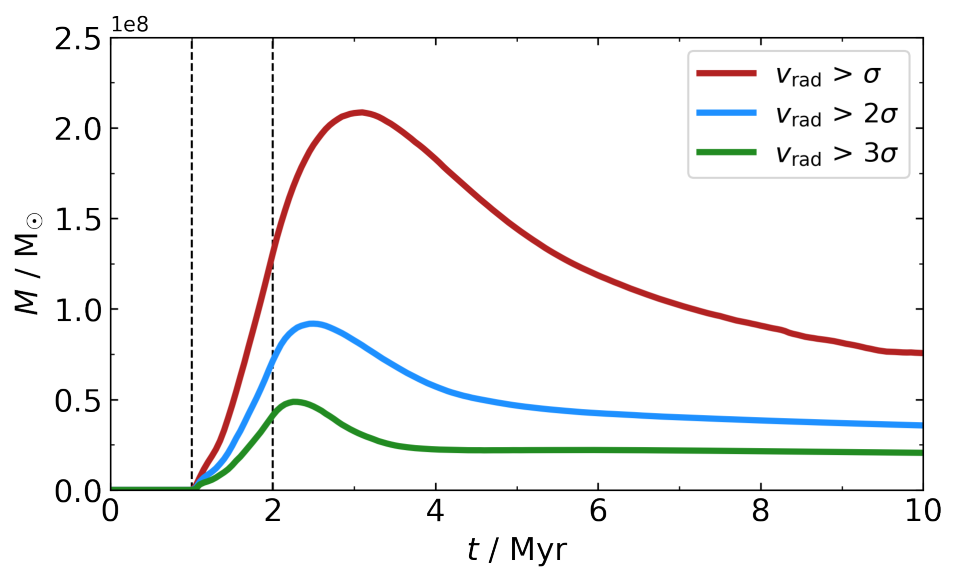}
    \caption{Same as Figure \ref{fig:mass_time}, but for the M8fg002L02 simulation.}
    \label{fig:mass_time_fg002}
\end{figure}

Figure \ref{fig:mass_time} shows the mass of the outflowing gas in the M7L07 (top) and M8L10 (bottom) simulations. Each line shows the difference between gas mass in the simulation with an AGN and its corresponding control simulation, in order to remove the presence of ``false outflows'' at early times when the initial gas distribution is relaxing. Brown, blue and green lines show, respectively, the mass of gas moving with radial velocities greater than $\sigma$, $2\sigma$ and $3\sigma$ as a function of time. For the first 1 Myr, the simulations evolve identically to their controls, so the outflow mass is zero. AGN feedback immediately creates some outflowing gas, but in the M7L07 simulation, its total mass is not very significant. The maximum mass of gas with $v_{\rm rad} > \sigma$ - that could be interpreted as ``outflowing'' when observed - is at most $4.5 \times 10^6 \, \msun$, i.e. $\sim 1.1\%$ of the total gas mass in the simulation. By $t = 2$~Myr, this mass is already decreasing and drops to essentially zero before $t = 2.5$~Myr. The mass of gas surpassing higher thresholds is $<1.5 \times 10^6 \, \msun$ at all times.

In M8L10, the situation is radically different. The mass of outflowing material begins to rise immediately when the AGN switches on and keeps rising beyond $t = 2$~Myr. This happens because the outflow bubbles are filled with extremely hot gas that expands adiabatically, further accelerating the dense gas on the bubble edges. The maximum mass of gas with $v_{\rm rad} > \sigma$ is $\sim 4.2 \times 10^8 \, \msun$, $\sim 11.4 \%$ of all the gas in the simulation. A mass $> 1.5 \times 10^8 \, \msun$ reaches $v_{\rm rad} > 2\sigma$ and $\sim 0.7 \times 10^8 \, \msun$ reaches $v_{\rm rad} > 3 \sigma$; all three peaks are reached at almost the same time. Once the peak is reached, the outflowing mass keeps decreasing at a similar or slightly lower rate than it was growing. By $t = 4$~Myr, there is no longer any gas expanding with $v > 2\sigma$, and the mass of gas with $v > \sigma$ has dropped to $0.5 \times 10^8 \, \msun$, i.e. $< 1.4\%$ of the total gas mass. By this time, the outflow bubbles are unlikely to be detected kinematically, although a spatially-resolved gas density map would still reveal significant cavities in the galaxy's bulge \citep[cf.,e.g.,][]{Rosario2019ApJ, Shimizu2019MNRAS, Feruglio2020ApJ}.

Fig. \ref{fig:mass_time_fg002} shows the outflowing mass evolution in the low-density simulation M8fg002L02. It is rather different from M8L10, which is remarkable considering the similarity in their morphological evolution. The total outflowing mass in M8fg002L02 peaks at $t \simeq 3$~Myr, more than $0.7$~Myr later than in M8L10 (Fig. \ref{fig:mass_time}, bottom), although the higher velocity components peak earlier, similarly to those in M8L10. The total outflowing mass reaches $\sim 2.2 \times 10^8 \, \msun$. This is $\sim 30\%$ of the total gas mass in the simulation. Remarkably, it is also approximately half of the value in M8L10, despite the gas density (and total mass) being five times lower in this simulation. The decay of outflowing mass is much slower than the rise. Some of the material that escapes the galaxy bulge entirely exceeds the escape velocity of the system, and even by the end of the simulation, at $10$~Myr, there is $\sim 8 \times 10^7 \, \msun$ of material ($\sim 11\%$ of all gas) expanding with $v_{\rm rad} > \sigma$. A similar trend is seen in the other low-density simulations as well: if the fossil outflow exists at all, it has a much higher chance of breaking out of the potential well in the low-density simulations than in the high-density ones.

\begin{figure}
	\includegraphics[width=\columnwidth]{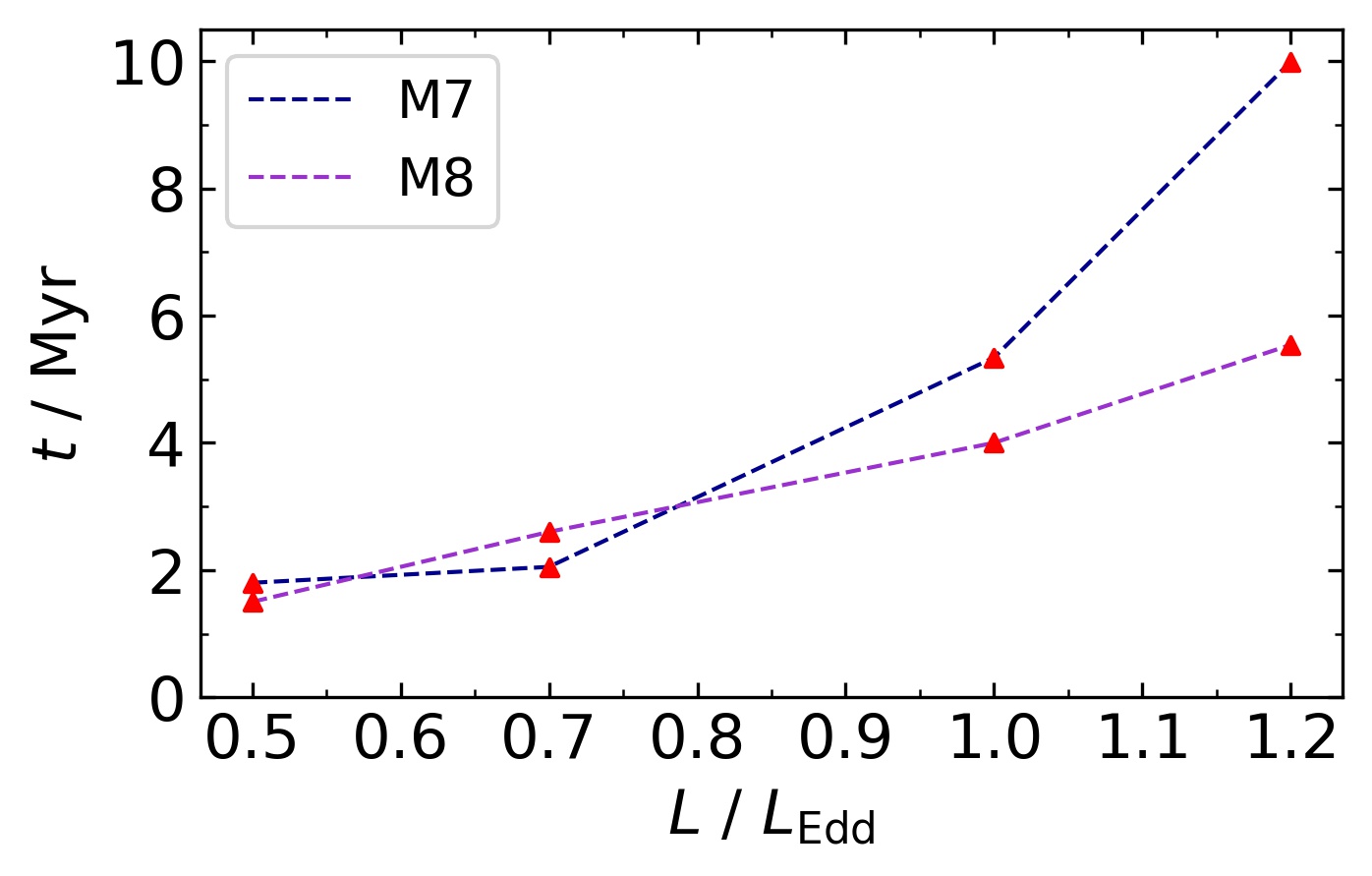}
    \caption{Time at which the mass of material with $v_{\rm rad} > \sigma$ falls below the value at $t = 1$~Myr, as function of AGN luminosity, for the M7 (dark blue) and M8 (purple) simulations. The uncertainty is $\pm 0.025$~Myr, i.e. half the time between two simulation snapshots.}
    \label{fig:fossil_duration}
\end{figure}

We use the outflowing gas mass to define the duration of the fossil phase in the high-density simulations. Assuming that the gas velocity distribution at $t = 1$~Myr is representative of the average state in an inactive galaxy, we say that the fossil outflow disappears once the total mass of gas with $v_{\rm rad} > \sigma$ becomes lower than it was at $t = 1$~Myr. In simulations M7L07, M8L10 and M8fg002L02, this mass is $\sim 2.5\times10^6 \, \msun$, $\sim 8\times10^7 \, \msun$ and $\sim 2\times10^7 \, \msun$, respectively. We plot the times when that happens in each of the high-density simulations in Figure \ref{fig:fossil_duration}. In three of the simulations - M7L05, M7L07 and M8L05, the outflow dissipates at or before $t = 2$~Myr, i.e. before the AGN switches off. In M8L07, the fossil outflow exists for $< 0.5$~Myr (see the end of the previous section and right panel of Figure \ref{fig:other_density_maps}). However, in the other four simulations, the fossil phase lasts for at least $2$~Myr after the AGN switches off. The trends with luminosity are similar for the two simulation groups, however, in the M7 simulations, the fossil outflow persists for longer when it forms. In fact, in the M7L12 simulation, due to the breakout of outflow bubbles from the initial gas shell, the mass of material accelerated to escape velocity from the system is larger than the mass moving with $v_{\rm rad} > \sigma$ at $t=1$~Myr. As a result, the fossil outflow, according to the definition we use here, does not disappear at all. This is also the case in all the low-density simulations where fossil outflows exist. Of course, in reality, at some point, the outflowing material becomes too dilute to be detected.

\subsection{Outflow kinematics} \label{sec:out_kinematics}

\begin{figure*}
	\includegraphics[width=0.33\textwidth]{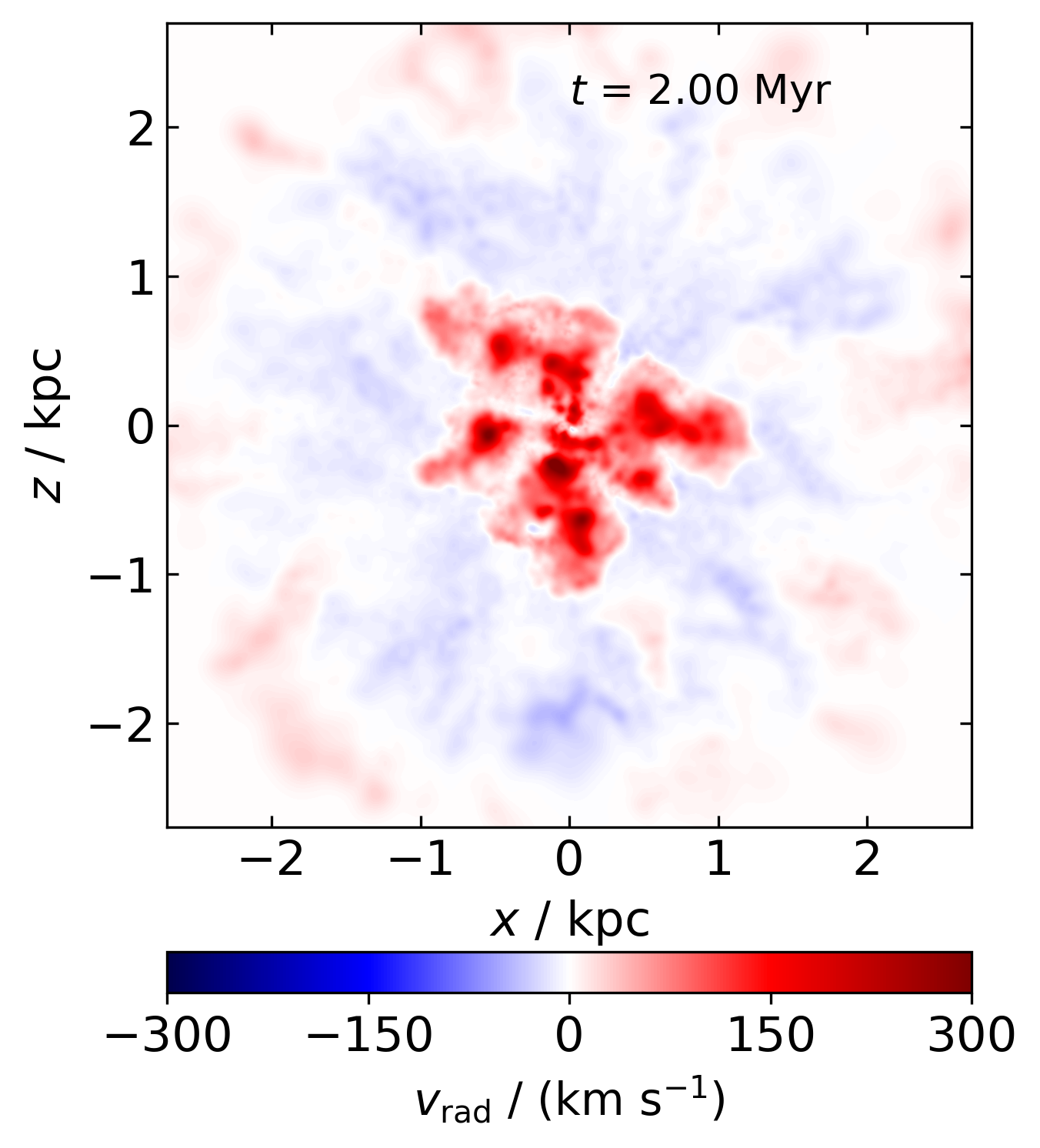}
	\includegraphics[width=0.33\textwidth]{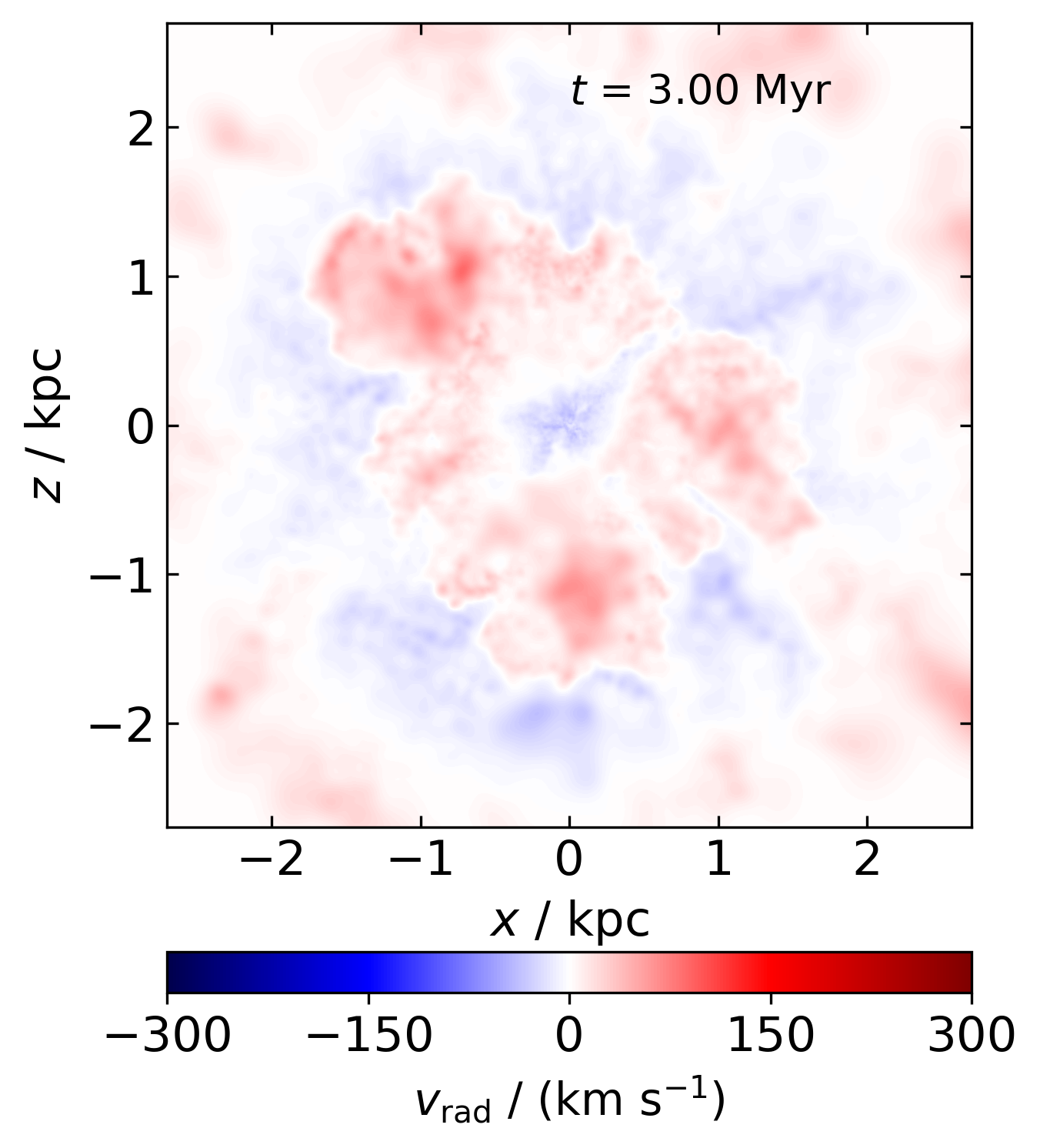}
	\includegraphics[width=0.33\textwidth]{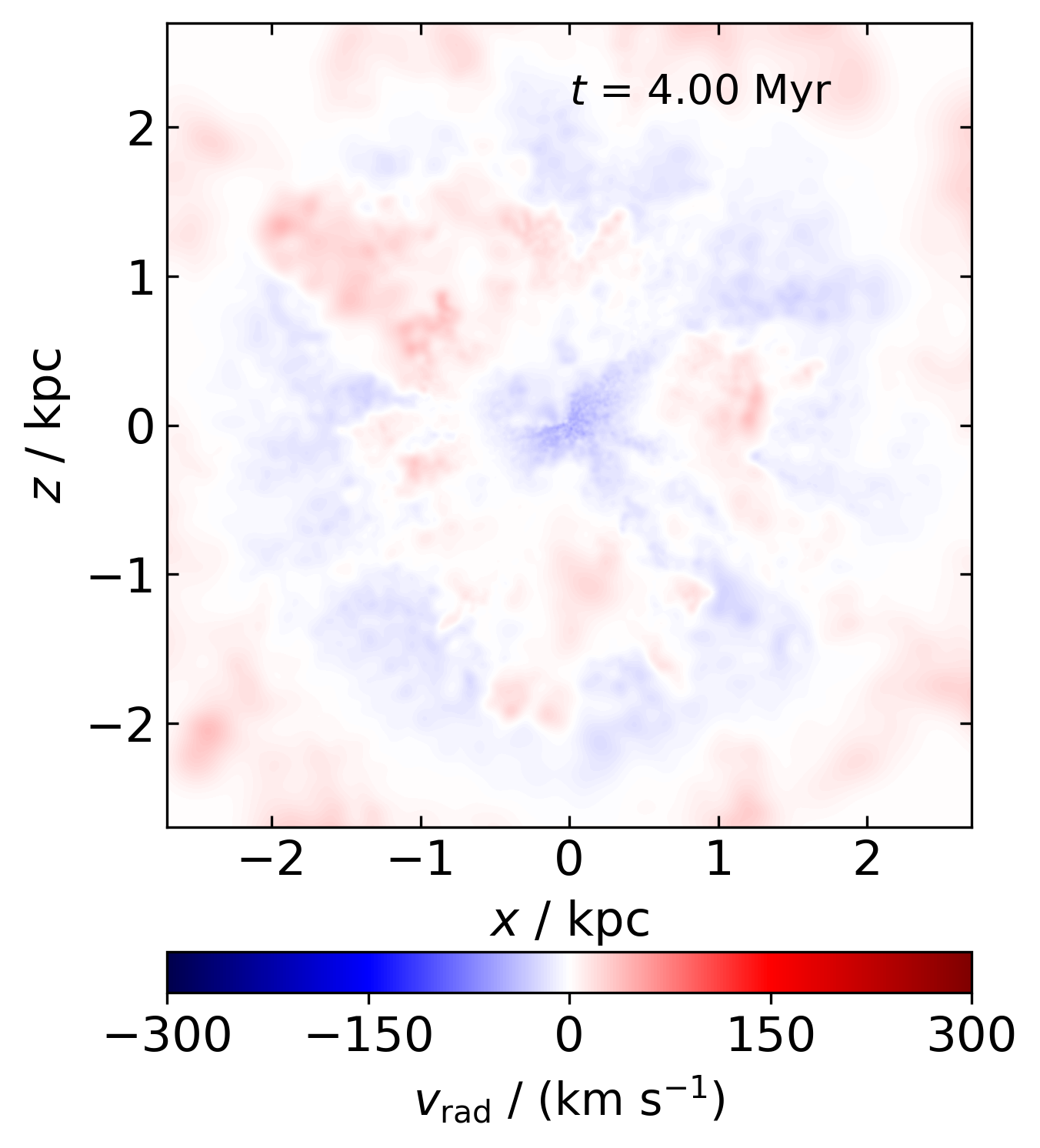}
    \caption{Radial velocities of gas in the M8L10 simulation. Panels show snapshots at $t = 2$, 3 and 4 Myr. Outflowing gas is shown in red, inflowing in blue.}
    \label{fig:velocity_maps}
\end{figure*}

\begin{figure*}
	\includegraphics[width=0.33\textwidth]{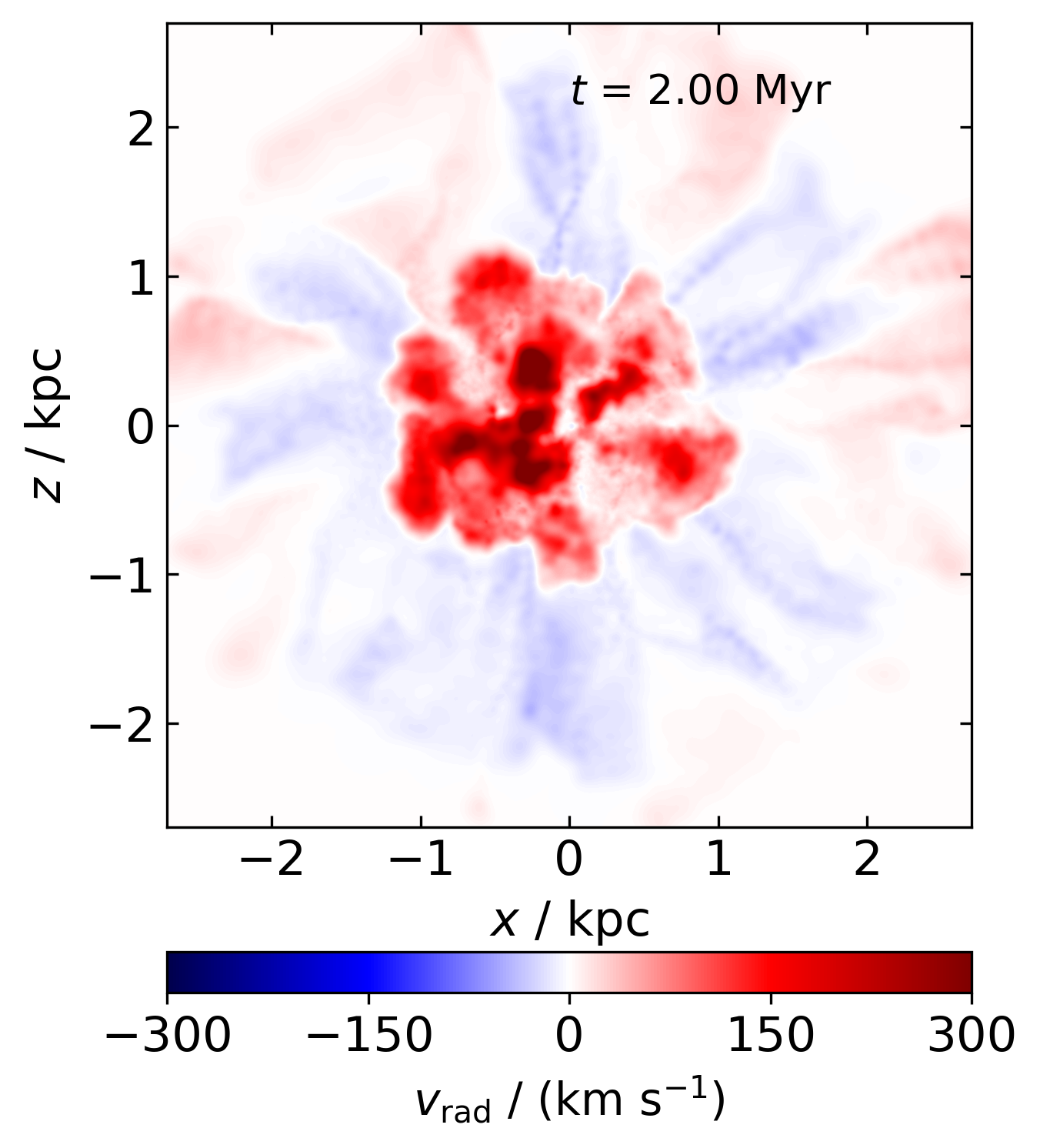}
	\includegraphics[width=0.33\textwidth]{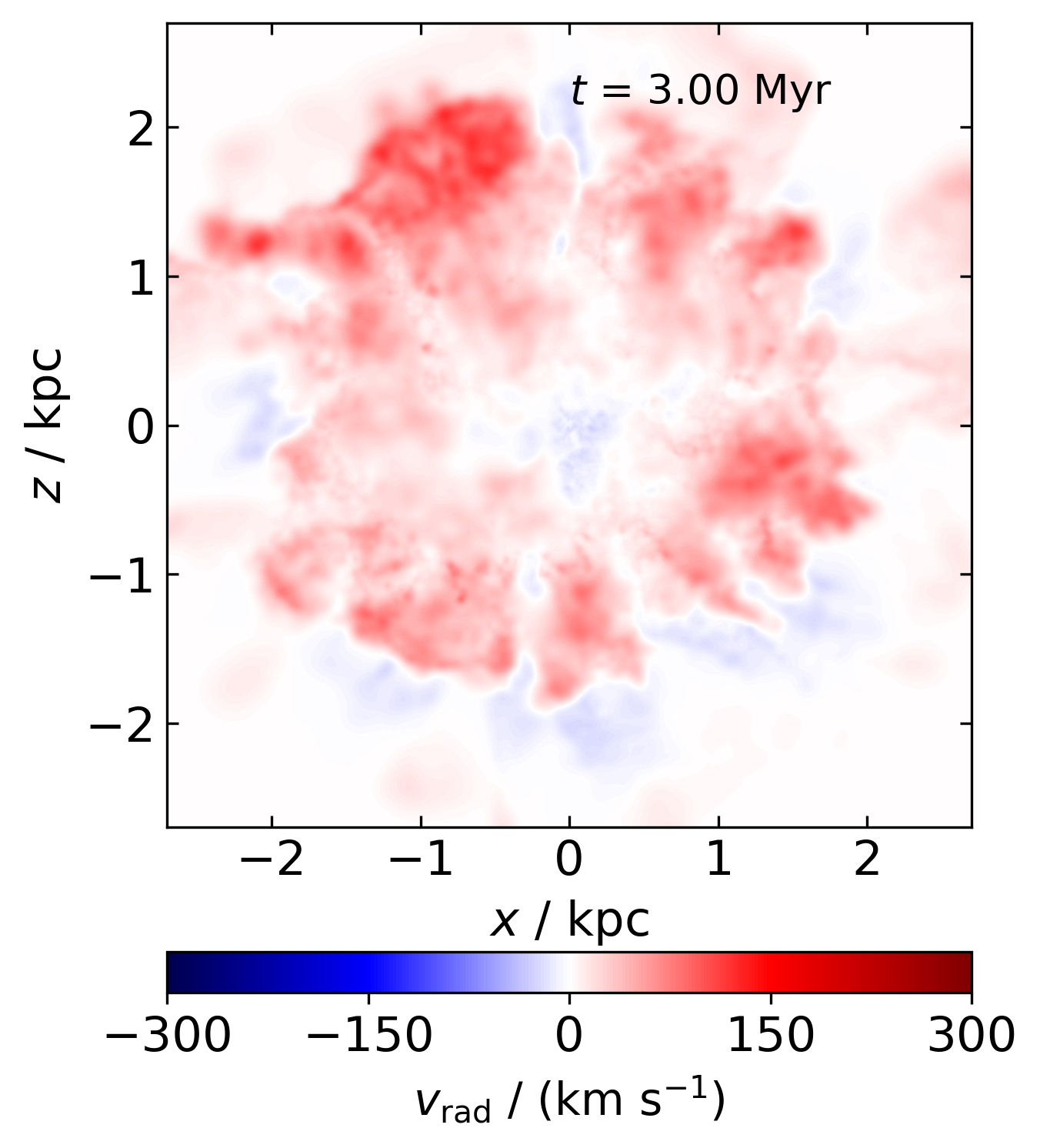}
	\includegraphics[width=0.33\textwidth]{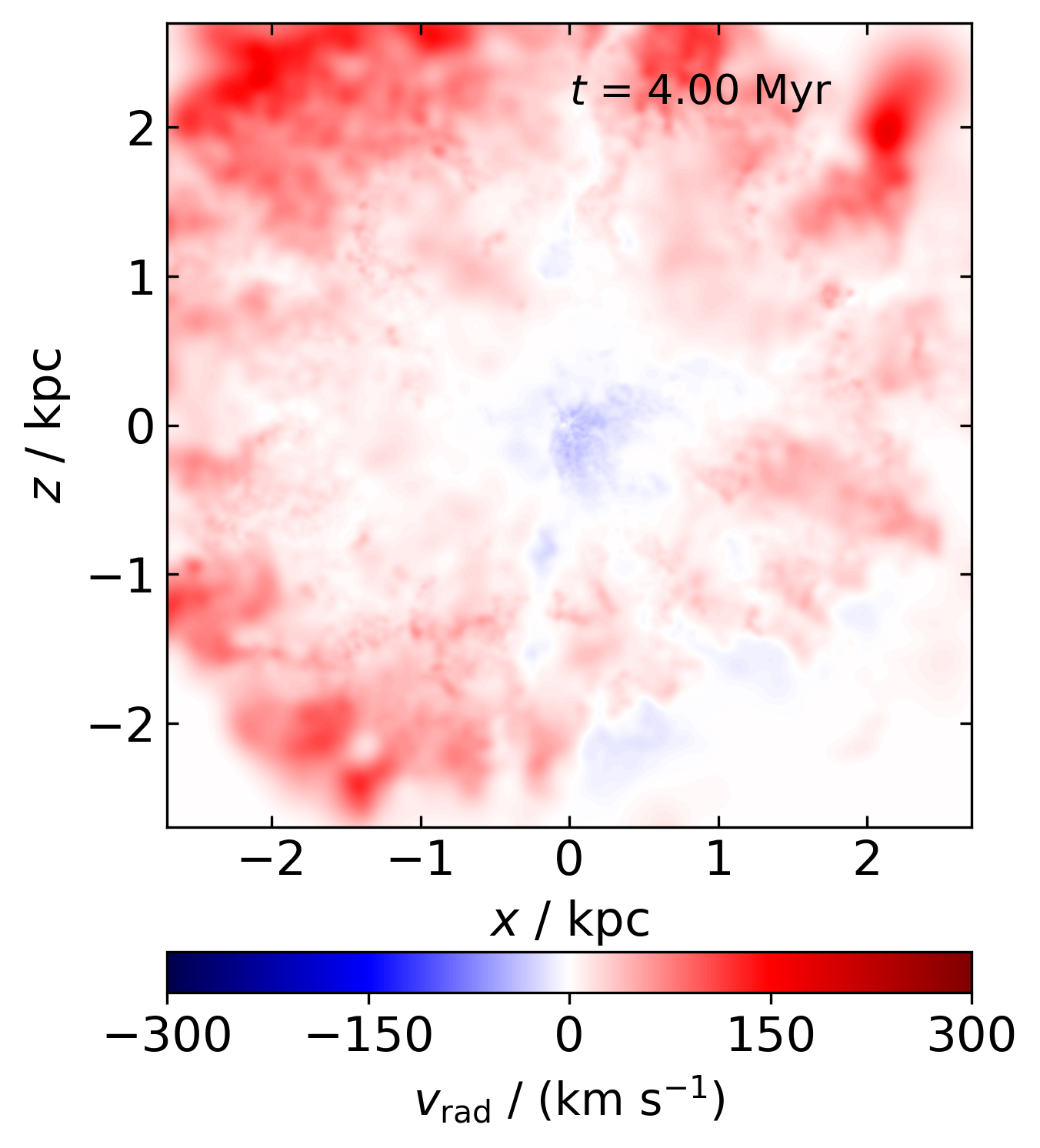}
    \caption{Same as Figure \ref{fig:velocity_maps}, but for the M8fg002L02 simulation.}
    \label{fig:velocity_maps_fg002}
\end{figure*}

\begin{figure}
	\includegraphics[width=0.88\columnwidth]{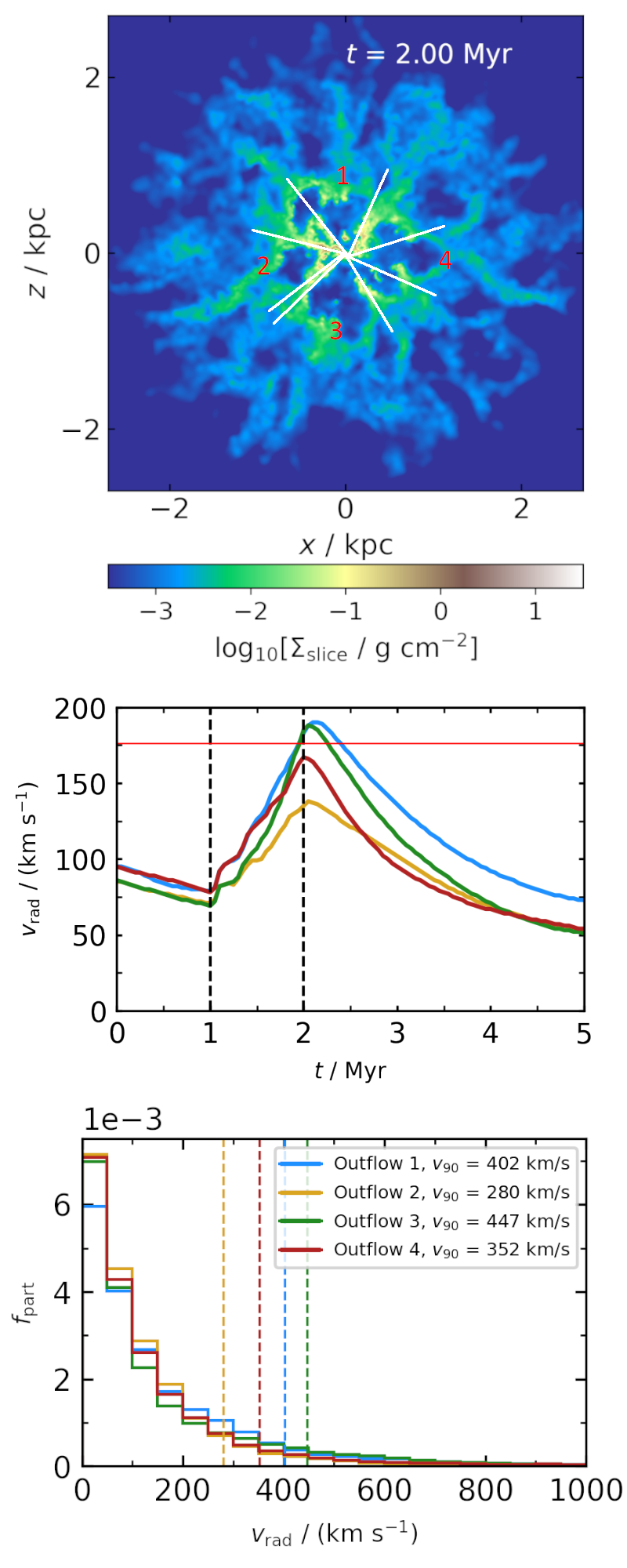}
    \caption{Properties of the four outflow bubbles in simulation M8L10. Top: identification of the four conical outflowing regions at $t = 2$~Myr. Middle: evolution of mean radial velocity with time of each region; red horizontal line shows $v_{\rm rad} = \sigma$, vertical dashed lines show the start and end of the AGN episode. Bottom: histograms of gas radial velocities in each region at $t = 2$~Myr, with vertical dashed lines showing the 90th percentile velocities.}
    \label{fig:four_bubbles}
\end{figure}

We now concentrate on one simulation with significant fossil outflows, M8L10, and investigate the main differences between the driven and fossil outflow phases, as well as among the four outflow bubbles visible at $t = 2$~Myr (see Figure \ref{fig:evolution_M8L10}, middle panel). In Figure \ref{fig:velocity_maps}, we show maps of gas radial velocities at $t = 2$, 3 and 4 Myr. Initially, just as the AGN episode is ending, the outflowing gas is concentrated in the central $\sim 1$~kpc, in a multi-lobed structure. The fossil outflow expands and slows down over time, while also detaching from the centre of the galaxy. By $t = 3$~Myr, i.e. 1 Myr after the AGN episode, there is an irregular-shaped, roughly 0.5-kpc-sized region in the centre of the simulation where gas is flowing toward the SMBH. By $t = 4$~Myr, the size of the region has increased to $\sim 1$~kpc. Its existence confirms two conclusions from numerous earlier works \citep[e.g.,][]{GarciaB2009AA, Davies2014ApJ, Feruglio2020ApJ, Dominguez2020AA, Costa2020MNRAS}: that AGN outflows are ``leaky'' and that an individual AGN episode cannot shut down SMBH accretion in a gas-rich galaxy. The weight of dense material is too large for the AGN wind momentum to remove it directly, and the hot outflow bubble preferentially escapes along the low-density channels, leading to formation/retention of infalling filaments. While the AGN is active, these filaments remain dense and narrow, since any material that expands away from them is quickly accelerated by the AGN wind. When the AGN fades, the infalling material can fill the region evacuated by the outflowing gas. In addition, the hot outflowing bubbles are filled with supervirial gas and so rise due to buoyancy. This further facilitates the lateral expansion of the inflowing filaments, so the multi-lobe structure of the outflow becomes more pronounced, until it breaks up into separate outflow bubbles, even if such bubbles were not immediately obvious initially. This structural difference between filled outflow structures connected to the nucleus while they are driven and hollow distinct bubbles detached from the nucleus during the coasting phase may help distinguish between driven and fossil outflows.

The low-density simulations again show qualitatively similar behaviour to the high-density ones, but the quantitative details are different. If we compare the radial velocity maps in Figures \ref{fig:velocity_maps} and \ref{fig:velocity_maps_fg002}, we see that at $t = 2$~Myr, both outflows look very similar, with radii $R \sim 1$~kpc. The velocity distributions are also similar, although the low-density simulation has a somewhat higher median velocity ($\sim 130$~km~s$^{-1}$ versus $\sim 85$~km~s$^{-1}$ in M8L10). The low-density simulation also has a more substantial tail of velocities $> 500$~km~s$^{-1}$. By $t = 3$~Myr, the low-density bubble is clearly both larger and faster than the high-density counterpart; this difference is even stronger at $t = 4$~Myr. This difference is mainly the result of differences in hot gas evolution: in M8L10, it cools down radiatively, while in M8fg002L10, it is almost completely adiabatic and so can transfer more energy to the surrounding colder material (see Section \ref{sec:thermodynamics}). However, the region of infalling gas is present in the centre of the low-density bubble as well as in the high-density one, and the velocity field is strongly asymmetric, with the highest velocities achieved in the top-left direction, mimicking the clearly larger bubbles in that direction seen in the density map (Figure \ref{fig:evolution_M8fg002L02}).

In order to understand what local conditions correlate with the expansion of a bubble as a fossil outflow as opposed to its rapid stalling and collapse, we analyse the evolution of four outflow bubbles at $t = 2$~Myr in simulation M8L10. Since each bubble expands approximately radially, we identify them with cones extending from the nucleus, as depicted in Figure \ref{fig:four_bubbles}, top panel. Bubbles 1 and 3 produce fossils that remain visible for $> 2$~Myr, while 2 and 4 collapse within $\sim 1-1.5$~Myr after the AGN switches off. In the middle panel of Figure \ref{fig:four_bubbles}, we show the evolution of the average radial velocity of all gas particles in each region with time, while in the bottom panel, we show the histogram of gas radial velocities at $t = 2$~Myr. Vertical dashed lines show $v_{90}$, the 90th percentile radial velocity, with values given in the legend. Two important aspects differentiate regions 1 and 3 from regions 2 and 4. First of all, the mean radial gas velocity at AGN switchoff in regions 1 and 3 is $v_{\rm rad} \sim 190$~km~s$^{-1} > \sigma$, while in regions 2 and 4 it is $v_{\rm rad,2} \sim 135$~km~s$^{-1}$ and $v_{\rm rad,4} \sim 170$~km~s$^{-1}$, both $< \sigma$. Furthermore, while the four regions have roughly similar distributions of radial velocity, the 90th percentile velocities are markedly different. In regions 1 and 3, these velocities are $\simgt 2.2\sigma$, while in 2 and 4, they are $< 2\sigma$. 

It is interesting to consider what these conditions represent physically. Velocity dispersion is a measure of the depth of the gravitational potential. Having $v_{\rm rad} > \sigma$ in a conical region suggests that the kinetic energy associated with radial motions in that region is high enough to overcome gravity, even when including the gas that is not currently part of the outflow. Adding to this the additional push provided by the residual heat in the shocked gas, the outflow is capable of expanding for a significant period of time.

Qualitatively similar results are seen in other simulations with prominent fossil outflows. These results suggest that the radial velocity of the outflow at AGN switchoff is important in determining its evolution during the fossil stage. In particular, we tentatively identify $v_{\rm 90} > 2 \sigma$ and $\left\langle v_{\rm rad} \right\rangle > \sigma$ as necessary conditions in order for the outflow to produce a fossil after AGN switchoff.

\subsection{Radial profiles}

\begin{figure}
	\includegraphics[width=\columnwidth]{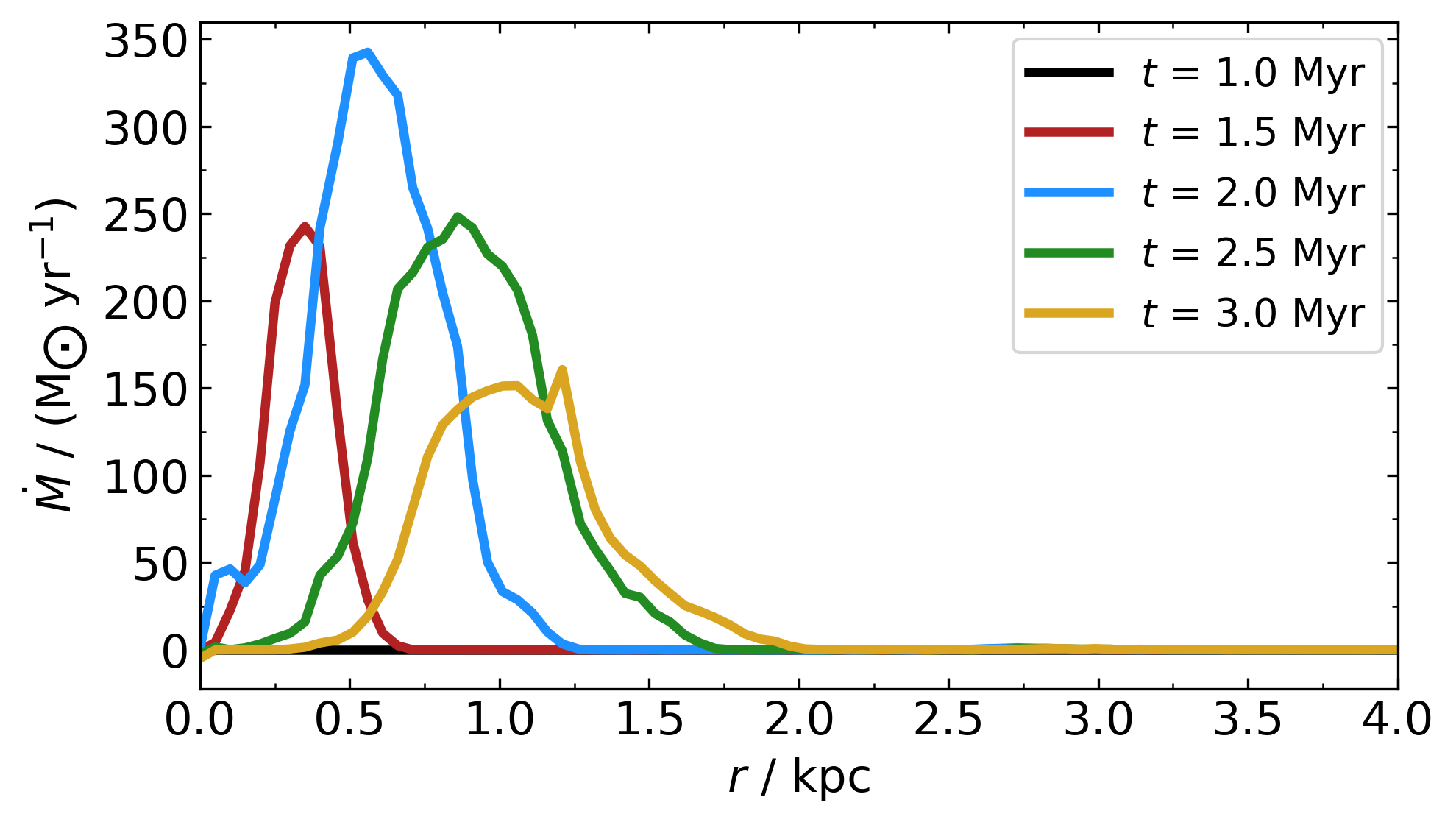}
	\includegraphics[width=\columnwidth]{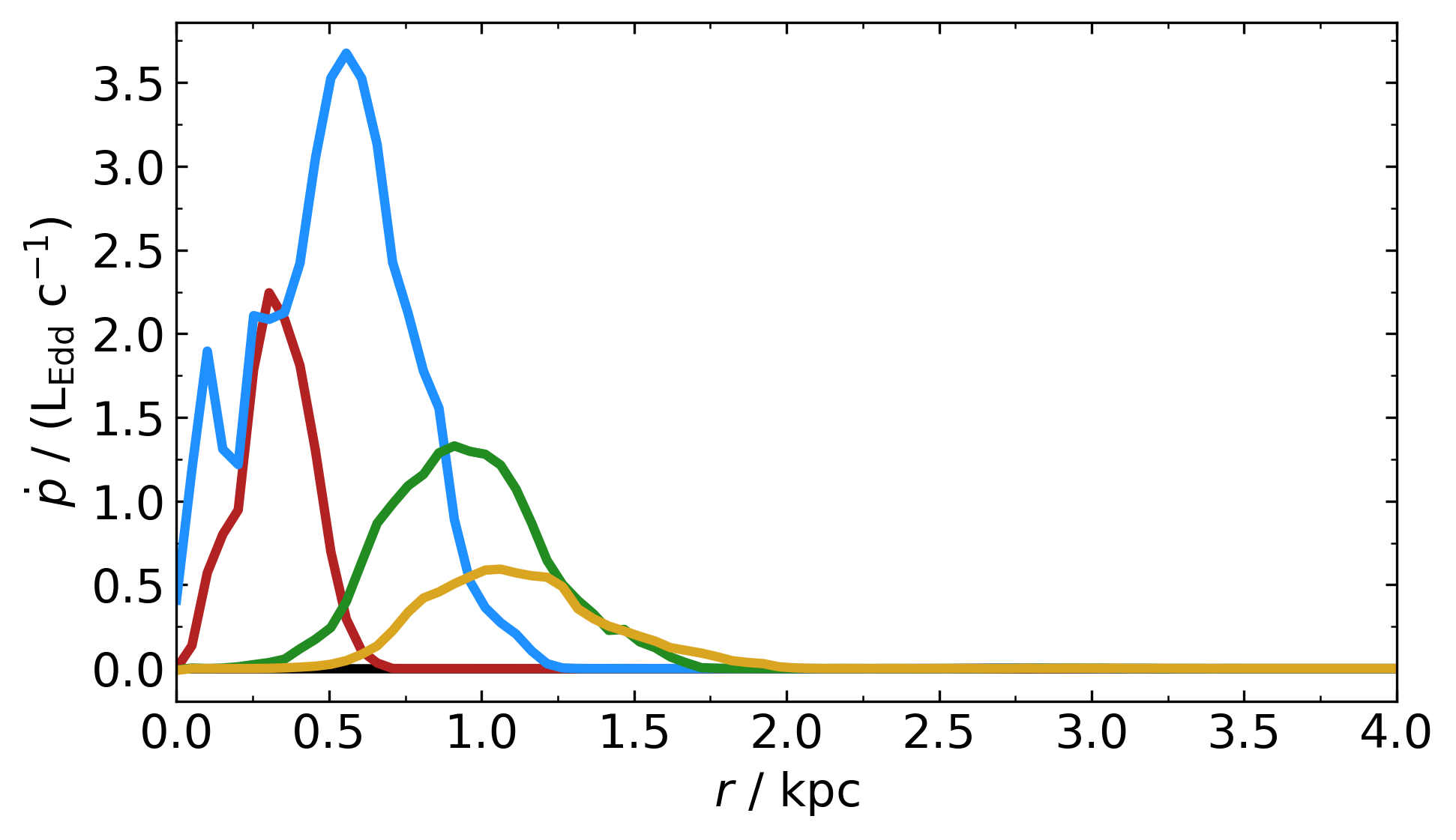}
	\includegraphics[width=\columnwidth]{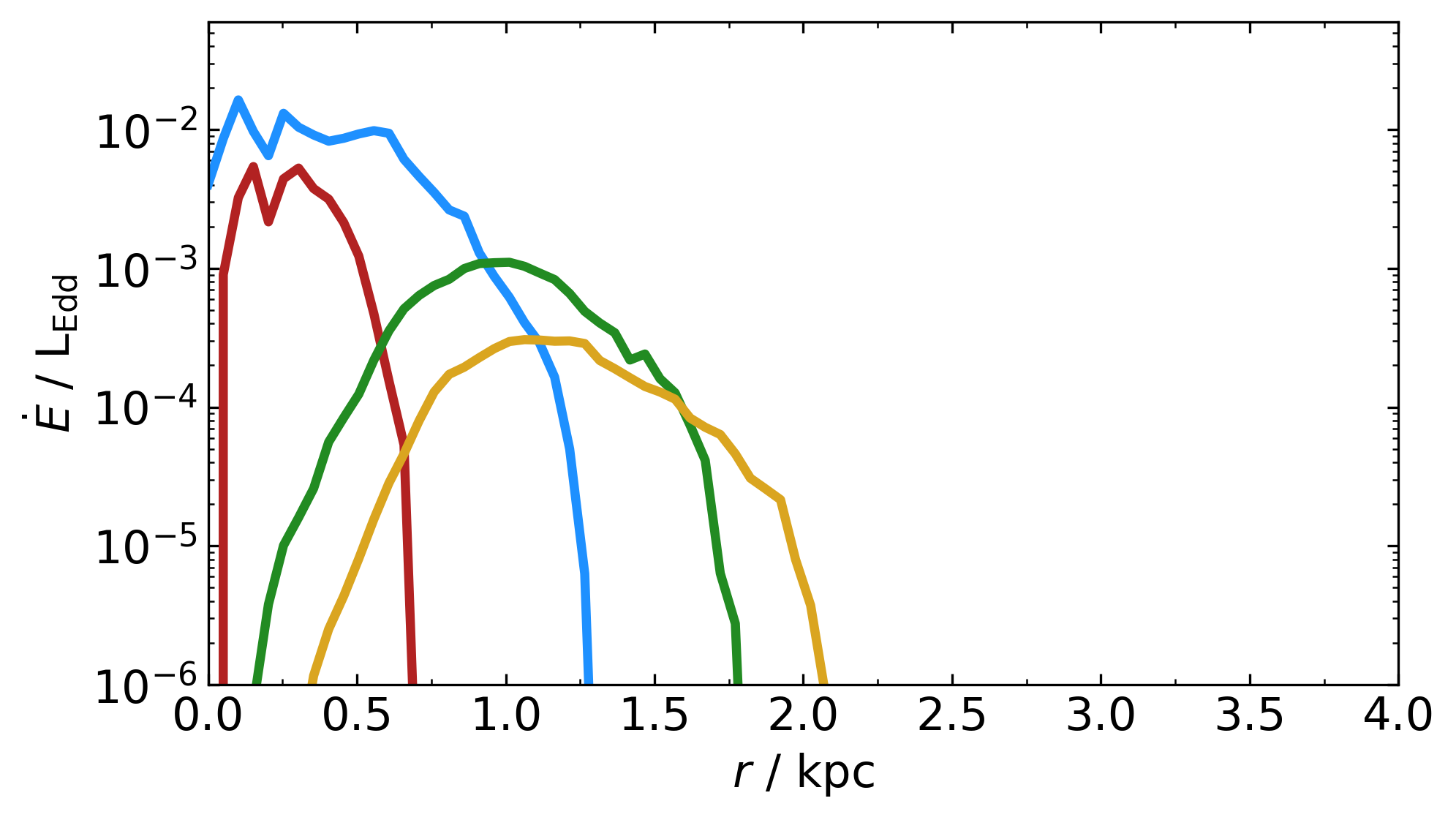}
    \caption{Radial profiles of mass (top), momentum (middle) and energy (bottom) flow rates in the simulation M8L10. Mass flow rates given in $\msun$~yr$^{-1}$, momentum flow rates in units of $L_{\rm Edd}/c$, energy flow rates in units of $L_{\rm Edd}$. Each line represents the profile at different times, as given in the legend.}
    \label{fig:mdot_pdot_edot}
\end{figure}

\begin{figure*}
	\includegraphics[width=\textwidth]{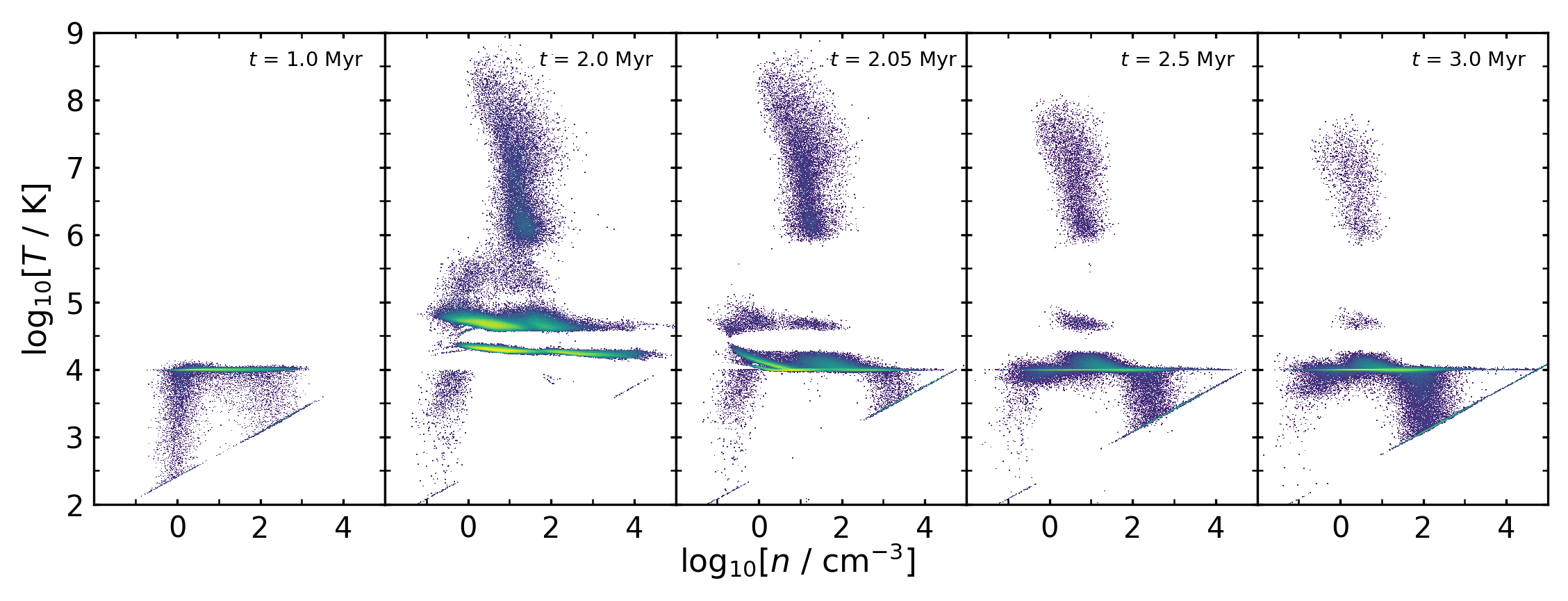}
    \caption{Phase diagrams of gas in the M8L10 simulation. Panels show snapshots at $t = 1$, 2, 2.05, 2.5 and 3~Myr. Colour represents gas mass in each pixel.}
    \label{fig:phase_diagrams}
\end{figure*}

\begin{figure*}
	\includegraphics[width=\textwidth]{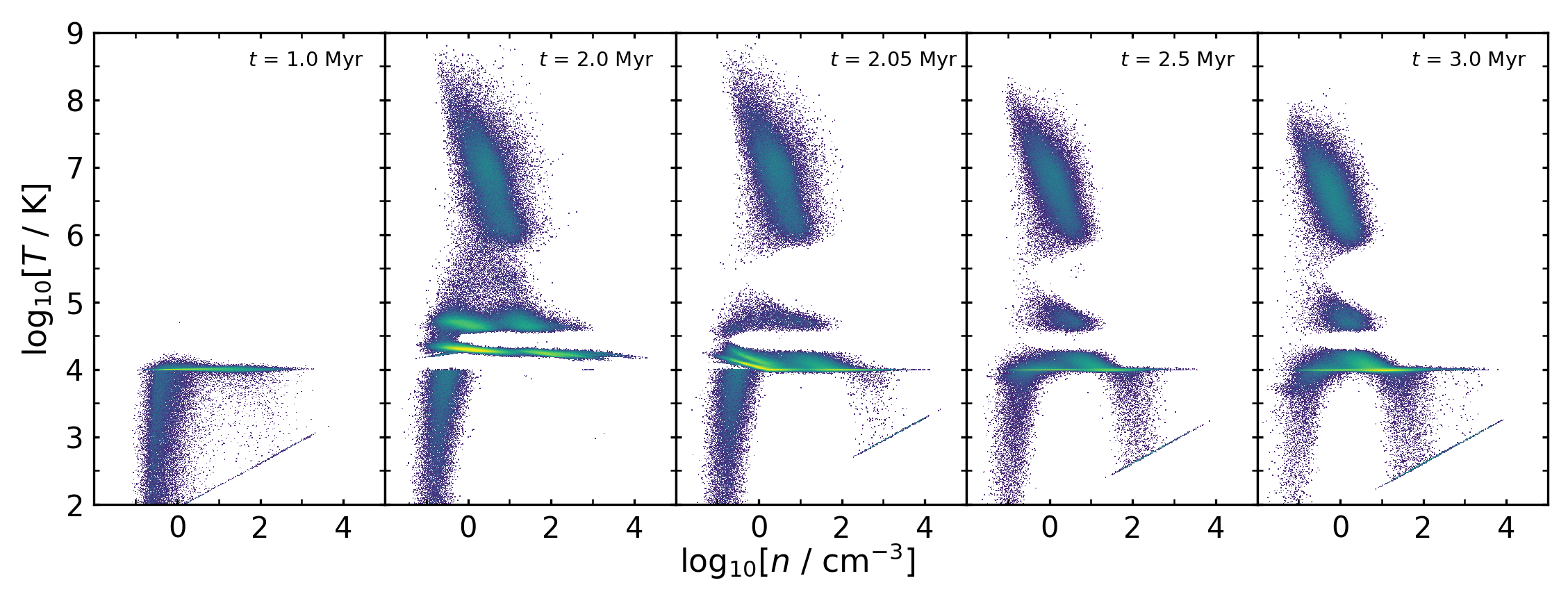}
    \caption{Same as Figure \ref{fig:phase_diagrams}, but for the M8fg002L02 simulation.}
    \label{fig:phase_diagrams_lowfg}
\end{figure*}

\begin{figure}
	\includegraphics[width=\columnwidth]{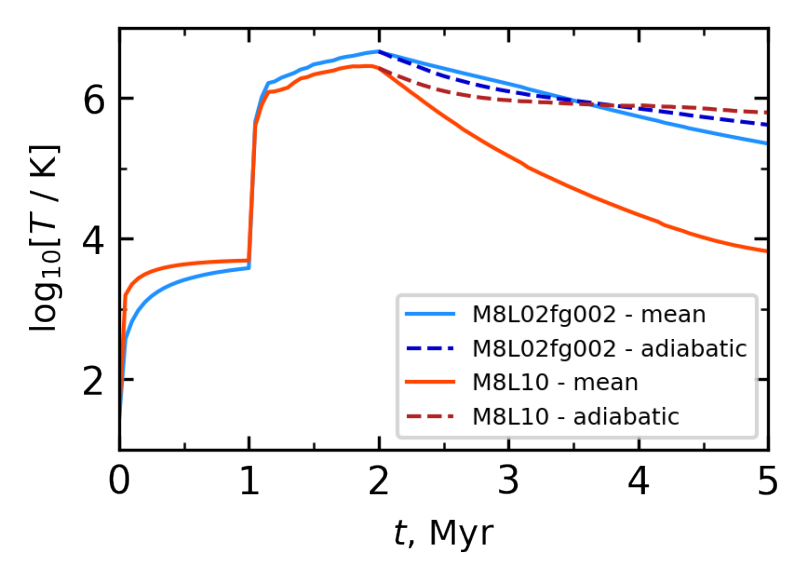}
    \caption{Mean gas temperature over time (solid lines) in the simulations M8L10 (blue) and M8fg002L02 (red). Darker dashed lines show the predicted temperature evolution assuming purely adiabatic cooling of the outflowing gas.}
    \label{fig:mean_temps}
\end{figure}

The trend of fossil outflow detachment from the nucleus can also be seen when looking at the azimuthally averaged properties of the outflowing gas at different times. We define the mass flow rate, momentum rate and energy rate through a thin spherical shell as follows:
\begin{equation}
\dot{M}_{\rm out}(r)\Delta r=\sum_{i}{m_{i}v_{{\rm rad, }i}},
\end{equation}
\begin{equation}
\dot{p}_{\rm out}(r)\Delta r=\sum_{i}{m_{i}v_{{\rm rad}, i}^{2}},
\end{equation}
and
\begin{equation}
\dot{E}_{\rm k, out}(r)\Delta r=\sum_{i}{\frac{1}{2}m_{i}v_{{\rm rad}, i}^{3}},
\end{equation}
where $m_{i}$ and $v_{{\rm rad, }i}$ are the mass and radial velocity of the $i$th particle within the spherical shell, which extends from $r$ to $r+\Delta r$. In Figure \ref{fig:mdot_pdot_edot}, we show these three quantities (top to bottom panels, respectively) as a function of radius at different times. Similarly to Figures \ref{fig:mass_time} and \ref{fig:mass_time_fg002}, we actually plot the difference between M8L10 and the M8-control simulations. For the first 1 Myr, the evolution of M8L10 is identical to control and there is no outflow. By $t = 1.5$~Myr (red lines), there is a clear outflow extending from the centre out to $R \sim 0.6$~kpc, peaking around $R = 0.3$~kpc. The very central region is already mostly devoid of gas, but whatever gas remains has very high velocity, so the momentum and energy rates in the central $\sim 0.1$~kpc are still substantial\footnote{It should be noted that the total $\dot{p}$ and $\dot{E}$ of the outflow are obtained by integrating, respectively, $\dot{M}$ and $\dot{p}/2$ across the whole radial range. The values seen in the radial plots here are not directly comparable to the integrated quantities derived from observations.}. The situation persists until the end of the AGN episode ($t = 2$~Myr, blue lines): the outflow has expanded to $R \sim 1.2$~kpc with a peak around $R = 0.6$~kpc in the mass and momentum flow rate profiles. The energy rate, however, is almost flat from the centre to $0.6$~kpc. 

Once the AGN switches off, the outflow continues to move outward and retains a peak around the middle of its radial extent (green and yellow lines, $t = 2.5$ and 3~Myr, respectively). The central regions become completely devoid of outflowing gas - this is visible in all three radial plots. It is also worth noting that while the peak value of the mass flow rate is the same at $t = 2.5$~Myr as at $t = 1.5$~Myr, the momentum and energy flow rates are significantly lower. Later on, the mass flow rate decreases, but the momentum and energy flow rates drop even more substantially. This happens because initially, although the outflow velocity is decreasing, this is compensated by the increase in total swept-up mass, leading to a modest decrease of the mass flow rate. Momentum and energy rates depend more strongly on velocity and so decrease more rapidly.

Overall, the radial profiles of our simulation results suggest two main properties of fossil outflows. First of all, they should be detached from the centre of the galaxy, separated by a region of inflowing (or, at least, non-outflowing) turbulent gas. Secondly, since they have gathered a lot of gas from the central regions, they should have large masses and mass flow rates, although their velocities are low. Both of these properties are qualitative in nature, i.e. there is no definite threshold when a massive but slow and/or detached outflow can be definitely said to be fossil. However, the more clear these properties are, the more likely is the fossil nature of the outflow. We discuss this in the context of several observed outflows in Section \ref{sec:real_comparison}.

\subsection{Outflow thermodynamics} \label{sec:thermodynamics}

The evolution of gas density and temperature during and after the AGN episode offers additional insights into the difference between driven and fossil outflow states. In Figure \ref{fig:phase_diagrams} we show the phase diagrams of gas in the M8L10 simulation at $t = 1$~Myr (just before the AGN switches on), $t = 2$~Myr (the end of the AGN episode), $t = 2.05$~Myr (just after the end of the episode), $t = 2.5$~Myr and $t = 3$~Myr. Initially, the balance between turbulent shock heating and radiative cooling for most of the gas is reached at $T \simeq 10^{4}$~K, as the cooling rate is much higher for higher temperatures. Once the AGN switches on, most of the gas gets very rapidly heated to two temperatures of approximate heating-cooling balance: $T \simeq 10^{4.2}$~K and $T \simeq 10^{4.7}$~K. This balance is reached throughout the simulation volume, except for the densest regions, where the upper branch hardly exists. Two equilibrium temperatures exist because of the particular combination of AGN luminosity and gas density, which leads to most of the gas having an ionization parameter $\xi \equiv L_{\rm AGN}/\left(n R^2\right) \sim 200$, where $n$ is the gas number density. Around this ionization value, the SOCS prescription leads to net heating of the gas, due to the Compton process, at temperatures $T \simlt 1.5\times10^4$~K and at $3\times10^4$~K~$\simlt T \simlt 4\times10^4$~K. As a result, gas that enters these temperature ranges quickly gets heated to their higher ends, i.e. the two equilibrium temperatures. Gas with temperatures outside these ranges cools down to the same equilibrium temperatures. Gradually, shock heating of the outflowing material fills a high-temperature region of the phase space, with temperatures $T > 10^6$~K. By $t = 2$~Myr, more than $99\%$ of the gas is above $10^4$~K, with $\sim 6.5\%$ above $10^6$~K. We note that the almost complete lack of dense cold gas at this time should not be taken as a definite prediction of our simulation; rather, it is an artifact of our adopted cooling prescription, where we do not account for gas self-shielding against the AGN radiation field.

Once the AGN switches off, the radiatively heated gas cools down very rapidly. By $t = 2.05$~Myr, $\sim 70\%$ of the gas remains at $T > 10^4$~K, while only $\sim 10\%$ above $10^{4.5}$~K. Most of the remaining hot gas is the shock-heated material, which has a much longer cooling time; it comprises $\sim 5.8\%$ of all gas at this time, a comparatively minor decrease compared to $t = 2$~Myr. The total mass of the hot gas keeps gradually decreasing and falls down to $\sim 1.2\%$ by $t = 3$~Myr. The hot gas also dominates the thermal energy budget, so the mean gas temperature only drops by approximately an order of magnitude, from $T_2 \sim 10^{6.4}$~K to $T_3 \sim 10^{5.2}$~K, in 1 Myr (Figure \ref{fig:mean_temps}, orange solid line). This gradual temperature decrease continues for as long as the fossil outflow persists, with another approximately order-of-magnitude decrease to $T_4 \sim 10^{4.4}$~K by $t = 4$~Myr. Even so, this is much faster than adiabatic cooling (orange dashed line in Figure \ref{fig:mean_temps}). We calculate the adiabatic cooling prediction as $T_{\rm ad}\left(t\right) = T_2\left(V\left(t\right)/V_2\right)^{-2/3} \simeq T_2 \left(R\left(t\right)/R_2\right)^{-2}$, where $V_2$ and $R_2$ are the outflow volume and radius at $t=2$~Myr, respectively. We take the 95th percentile of the radial coordinates of outflowing gas as the bubble radius estimate. This calculation implicitly assumes that the bubble shapes do not change during the fossil phase, which is not strictly correct, but reasonable for illustrative purposes. By $t=3$~Myr, adiabatic cooling would only decrease the gas temperature to $10^6$~K, while by $t=4$~Myr, the predicted temperature is $10^{5.9}$~K, a decrease by merely a factor of three in 2 Myr. The rest of the cooling - a factor of 32 drop in mean temperature - occurs due to radiative losses. Some of these losses occur in the material outside the outflow bubble, which was also heated by the AGN, but the majority of energy losses take place within the shocked gas.

Radiative cooling of the fossil outflow is one of the reasons why our fossil outflows stall faster, and expand less, than predicted analytically (Section \ref{sec:analytical}). Analytical estimates assume that the shocked wind is perfectly adiabatic and transfers its energy only to the outflowing gas. We discuss the realism of that assumption, as well as of the cooling rates in our simulations, in Section \ref{sec:discussion_cooling}.

In the low-density simulation M8fg002L02, the qualitative evolution of the gas temperature is similar, but there is an important quantitative difference. Looking at Figure \ref{fig:phase_diagrams_lowfg}, we see significantly more gas heated to $T > 10^6$~K by $t = 2$~Myr: $\sim 17.7\%$ of particles are in this regime. The mean gas temperature, however, is only slightly higher, $T_2 \sim 10^{6.6}$~K (Figure \ref{fig:mean_temps}). The hot gas remains hot for much longer as well; its total mass actually increases after AGN switchoff, with $\sim 20.7\%$ of particles having $T > 10^6$~K by $t = 3$~Myr. This occurs because some of the hottest gas ($T > 10^8$~K) spreads its energy to neighbouring colder particles. As a result, the average gas temperature decreases much more slowly, only dropping to $T_4 \sim 10^{5.7}$~K, i.e. a factor $\sim 8$, by $t = 4$~Myr (blue solid line in Figure \ref{fig:mean_temps}). The adiabatic cooling prediction (blue dashed line) gives a temperature of $10^{5.85}$ at $t=4$~Myr, only a factor $1.4$ higher than the actual temperature. We see that the temperature decrease can be explained almost entirely by adiabatic cooling, with radiative losses being negligible. This difference explains the much longer persistence of the low-density fossil outflow.



\section{Discussion} \label{sec:discuss}

\subsection{Salient properties of fossil outflows} \label{sec:salient_properties}

Our simulation results suggest certain salient properties of fossil outflows that should help distinguish them from driven ones. In particular, these are:
\begin{itemize}
    \item Asymmetry: fossil outflows often consist of one or several lobes to one side of the nucleus, while driven outflows have more spherical symmetry. In real galaxies, the driven outflow probably has a bicone shape due to being confined in the plane of the galaxy by dense gas discs/rings; even so, the two outflow cones encounter somewhat different ambient conditions and one of them is likely to stall, collapse and/or dissipate earlier than the other, leading to a lopsided shape \citep[for a similar effect, see][]{Gabor2014MNRAS}. However, it should be noted that projection effects may lead to bicone outflows appearing one-sided \citep[e.g.,][]{Venturi2017FrASS}.
    \item Detachment: fossil outflows are spatially detached from the nucleus, with a region of inflowing/turbulent gas in between the nucleus and the outflowing bubble. Detachment by itself may be difficult to ascertain, as some common outflow tracers, e.g. CO emission, are produced in the radiative layer behind the outer shock and would naturally appear detached from the nucleus independently of the full outflow morphology. However, a combination of detachment plus inflow/turbulence close to the centre should be seen as stronger evidence for the existence of a fossil.
    \item Mass and velocity: outflow momentum and energy rates decrease more rapidly than the mass flow rate during the fossil stage, implying rather high total gas masses in the fossil outflows; however, this is not a qualitative difference, so drawing a distinction would be difficult.
    \item Multiphase structure: gas at moderately high temperatures $10^4$~K $< T < 10^6$~K cools down much faster than the fossil outflow dissipates, so the ratio of ionized to atomic/molecular gas in a fossil outflow should be lower than in a driven one; as above, this is not a qualitative difference. Counter-intuitively, this behaviour may lead to high-luminosity AGN having low ionized-to-molecular outflowing gas mass ratios, as observed \citep{Fluetsch2021MNRAS}. To understand this, consider that fossil outflows exist either in systems without an AGN, which would not be selected for analysis of `AGN outflows', or in systems where the AGN has only recently turned on and is merely coincident with an outflow inflated by an earlier episode. These AGN are the most likely to have high Eddington ratios and hence high luminosities.
    \item Age: the estimated age of a fossil outflow, $t_{\rm age} = R_{\rm out}/v_{\rm out}$, is longer than the duration of the AGN episode that inflated it, which is likely to be $t_{\rm ep} \simlt 10^5$~yr \citep{Schawinski2015MNRAS, King2015MNRAS}. The estimate $t_{\rm age}$ is also the upper limit of the real age, because the velocity has been higher in the past. 
    \item Weakness: fossil outflows are weaker than driven ones in terms of both momentum and energy rates and, when accompanied by a recently-reignited AGN, momentum- and energy-loading factors.
\end{itemize}

None of these properties is sufficient, by itself, to identify a particular outflow as a fossil. However, the better that an outflow aligns with these properties, the higher the chance it has been coasting for some time.

\subsection{Comparison with observed outflows} \label{sec:real_comparison}

We now look at a few observations of spatially-resolved AGN outflows and consider how their properties line up with the expectations of our simulations (see Section \ref{sec:salient_properties}, above). This is not intended to be an exhaustive list; we only wish to showcase several interesting systems where fossil outflows appear to be evident.

{\em Mrk 231}: this galaxy has a very thoroughly investigated AGN-driven outflow. Its energy and momentum rates agree very well with the simple energy-driven model with an implicit assumption that the AGN has been shining throughout the evolution of the outflow \citep{Cicone2012AA}. The radial structure of the outflow shows spherical symmetry and a clear connection to the nucleus \citep{Feruglio2015AA}. All of these properties agree with each other and show that the outflow is {\em not} a fossil. The low ionized mass fraction can be explained as a result of high gas density rather than prolonged cooling.

{\em PDS 456}, conversely, has a notoriously weak outflow, with energy rates lower than even the momentum-driven model would predict \citep{Bischetti2019AA}. The CO map of the outflow \citep[Fig. 3a in][]{Bischetti2019AA} shows that the ``extended'' outflow is clearly detached from the nucleus. It is also highly lopsided: the two outflow bubbles are located to the east and south of the nucleus. Both properties suggest that it is a fossil; this conclusion helps explain the weakness of the outflow component as well. The system hosts a powerful ultra-fast outflow (AGN wind) in the nucleus \citep{Nardini2015Sci, Matzeu2022arXiv}, but it is likely that the AGN has just recently restarted after a period of quiescence \citep{Zubovas2020MNRAS}. An ionized gas outflow has been discovered in the system recently, with a spatial extent coincident with that of the molecular one, and the outflow rate similar to, or slightly higher than, molecular (Travascio et al., in prep.). The high ionized outflow rate disagrees with the conclusion that fossil outflows should be strongly dominated by the molecular component and suggests more complex outflow dynamics.

{\em NGC 5728} has a very weak outflow. Its scale is rather small, well below a kiloparsec, but there are signs of detachment - the bulk of the outflow, in terms of mass, momentum and energy flow rates, is located $\sim 200$~pc away from the nucleus \citep[Fig. 15 of][]{Shimizu2019MNRAS}. Fig. 7 of the same paper shows that the northwestern part of the outflow is clearly detached, but the southeastern is connected to the nucleus.

{\em XID 2028} has a very powerful outflow with $\dot{E}_{\rm out} \sim 0.04 L_{\rm bol}$ \citep{Perna2015AA}, however the molecular component is extremely weak, with $\dot{E}_{\rm out, mol} \sim 2.5\times10^{-3} L_{\rm bol}$ \citep{Brusa2018AA}. In the latter paper, we see in Fig. 9, left panel, that the blue wing of the molecular outflow has a spatially detached component with a distance from the nucleus of $\sim 1'' \sim 8.5$~kpc. No counterpart is visible in the red wing. The connected component, visible in both blue and red wings, has a radius of $\sim 0.5'' \sim 4.2$~kpc. The detached lopsided component may represent a weak fossil outflow. An ionized outflow exists as well, but it is $\sim 20$ times weaker \citep{Fluetsch2021MNRAS}, in agreement with the expectation for a fossil.

{\em Mrk 509} has an outflow that can be spatially separated into two regions: region A is connected to the nucleus, while region B is clearly spatially detached \citep[][Fig. 6]{Zanchettin2021AA}. Both regions are also clearly offset to the west of the nucleus. Region B has lower $\dot{M}$, $\dot{p}$ and $\dot{E}$ than region A (by factors 6, 6 and 10 respectively). Even though both regions have rather low energy loading factors, region B can be identified as a noticeably weaker fossil.

{\em zC400528}: this galaxy has a rather weak outflow, with momentum and energy rates consistent with continuous momentum driving \citep{Herrera-Camus2019ApJ}. Fig. 7 of that paper, however, shows one detached component of the outflow (component 3); in general, the whole outflow is very lopsided: it extends only in one direction from the nucleus both in the plane of the sky (northwest) and radially (all components have a positive velocity relative to the nucleus). All of these properties suggest it is a fossil, perhaps comprised of remnants of several previous outflows.

{\em IRAS F08572+3915} has an outflow component offset by about $R = 6$~kpc to the north from the nucleus, redshifted by $v \sim 900$~km~s$^{-1}$ \citep[Fig. 2 in][]{Herrera-Camus2020A&A}. These authors interpret it as a fossil remnant of an earlier episode of nuclear activity. Using the total gas mass in the component as given in that paper, $M_{\rm out} \sim 5\times10^7\,\msun$, we estimate the mass flow rate as $M_{\rm out} v/R \sim 7.6 \, \msun$~yr$^{-1}$, which leads to an energy rate of $\dot{E} \sim 2\times10^{42}$~erg~s$^{-1}$, almost two orders of magnitude lower than the main outflow. The low mass flow rate can be explained by the small solid angle of the outflow. An ionized outflow exists in the system, but its mass rate is more than three orders of magnitude lower than the molecular one. The lopsidedness, detachment, weakness and minuscule ionized component all agree with the interpretation of this being a fossil.

{\em ESO428-G14} has a very weak outflow in the central $\sim 700$~pc \citep{Feruglio2020ApJ}. The inset in Fig. 6 in that paper shows that both the red and blue wings of the outflow are observed to the northwest of the nucleus, hinting at some lopsidedness. This may mean that the outflow is a fossil remnant of a more spherical one, but the evidence, in this case, is much weaker than in the previous ones.

{\em IRAS F11119+3257} has a large-scale outflow that has been interpreted both as energy-conserving \citep{Tombesi2015Natur} and as significantly weaker \citep{Veilleux2017ApJ, Nardini2018MNRAS}. Figure 5 in \citet{Veilleux2017ApJ} shows both blue and red wings are offset to the west of the nucleus, suggesting that the outflow is at least somewhat lopsided and hence could be a (recent) fossil. This lopsidedness may also explain why the outflow has rather low $\dot{M} \sim 100 \msun$~yr$^{-1}$, despite having a high velocity $v \sim 1000$~km~s$^{-1}$: if only a few of the sightlines from the nucleus contain outflowing material, the total solid angle is small.

Finally, {\em NGC 6240} is a dual AGN, with the outflow located in between the two nuclei, with the blue wing very close to the southern nucleus and the red wing approximately in the middle between them \citep[Fig. 1, panels d and e in][]{Cicone2018ApJ}. This lopsidedness (and, in the case of the red wing, detachment) suggests the outflow may be a fossil. On the other hand, the $\dot{p}$ and $\dot{E}$ values of the outflow agree with expectations of energy driving, calling the fossil interpretation into question. However, the current luminosity of the AGN, $L_{\rm AGN} \sim 1.1\times10^{45}$~erg~s$^{-1}$ \citep{Cicone2018ApJ}, may be only a small fraction of the Eddington luminosity. The SMBH mass in the southern nucleus is estimated to be $M_{\rm BH} \sim \left(7-8\right) \times10^8\msun$ \citep{Medling2011ApJ, Kollatschny2020AA}, which corresponds to an Eddington luminosity $L_{\rm Edd} \sim \left(0.9-1\right) \times 10^{47}$~erg~s$^{-1}$, i.e. $L_{\rm AGN}/L_{\rm Edd} \sim 0.01$. In that case, the outflow is in fact very weak and the agreement with the energy-driving expectation based on the current AGN luminosity is merely a coincidence.

All of the outflows discussed above are detected in systems with ongoing AGN episodes, but this is a result of selection. We think it is very likely that a large number of fossil outflows can be found in currently quiescent galaxies. Some of them may have been discovered and interpreted as driven by star formation (see Section \ref{sec:sf_comp} below).

\subsection{Conditions for the formation of fossil outflows} \label{sec:conditions}

In all our simulation groups, fossil outflows appear when the condition $l/f_{\rm g} > 7$ is satisfied; this corresponds to $l > 0.7$ for the high-density simulations and $l > 0.14$ for the low-density ones. This can be easily translated to a condition
\begin{equation}\label{eq:condition_1}
    L_{\rm AGN} > 0.07 \frac{f_{\rm g}}{0.01} L_{\rm Edd}.
\end{equation}
One should keep in mind that we define gas fraction as the ratio between gas density and total density, including dark matter. Observationally, gas fraction is usually defined as the ratio of gas mass to stellar mass or the ratio of gas mass to total baryonic (gas + stellar) mass \citep[e.g.,][]{Zhang2009MNRAS, Combes2013A&A}. In order to recast the above condition in terms of this type of gas fraction (which we denote $f_{\rm gas,obs}$ here), we can consider the stellar-to-halo mass ratio $f_*$. This relation is non-monotonous with halo mass and peaks at $M_{\rm halo} \sim 10^{12} \, \msun$ \citep{Behroozi2013ApJ, Girelli2020A&A}, which corresponds to $M_{\rm BH} \sim 10^7 \, \msun$ \citep{Ferrarese2002ApJ, Bandara2009ApJ}. The peak value of the ratio is $f_{\rm *, max} \sim 0.03$, while its value is $> 0.003$ within the halo mass range $3\times 10^{10} \, \msun < M_{\rm h} < 10^{14}\,\msun$, essentially encompassing all large galaxies \citep{Behroozi2013ApJ, Girelli2020A&A}. The observational gas fraction tends to decrease with increasing stellar mass, albeit with a large scatter. At $M_* \sim 3\times 10^{10} \,\msun$ (the peak of $f_*$), the gas fraction is $f_{\rm gas,obs} \sim 0.1$ \citep{Combes2013A&A}. If we now combine those values, we find
\begin{equation}
    f_{\rm g} \equiv \frac{M_{\rm gas}}{M_{\rm halo}} = \frac{M_{\rm gas}}{M_*} \frac{M_*}{M_{\rm halo}} = f_{\rm gas,obs} f_*.
\end{equation}
So we can recast the condition for the formation of fossil outflows (eq. \ref{eq:condition_1}) as
\begin{equation}\label{eq:condition_2}
    L_{\rm AGN} > 0.02 \frac{f_{\rm gas,obs}}{0.1} \frac{f_*}{0.03} L_{\rm Edd}.
\end{equation}
This suggests that in `typical' low-redshift galaxies, essentially all AGN that reach a radiatively efficient accretion state characterised by $L_{\rm AGN} > 0.01 L_{\rm Edd}$ \citep{Heckman2014ARAA} should produce fossil outflows. At high redshift, where the gas fraction is typically higher, while the ratio $f_*$ is similar to the low-redshift value, the condition becomes more stringent (see also Sections \ref{sec:prevalence} and \ref{sec:high-z} below).

Additionally, we note that the formation of fossil outflows does not require clearing the galaxy bulge. In all our simulations, the AGN switches off before the outflow expands to the size of the bulge. In low-density simulations, the energy in the hot gas bubble is enough to push the outflow out of the bulge into essentially a vacuum. On the other hand, in the high-density simulations, especially those with the higher SMBH mass, the fossil outflow evolution occurs entirely within the bulge.

\subsection{Prevalence of fossil outflows} \label{sec:prevalence}

By comparing the lifetime of the fossil outflow phase with the duration of the AGN episode, we can estimate the prevalence of fossil outflows in two ways. First of all, and more directly, we can estimate the ratio of galaxies with fossil outflows to the number of AGN host galaxies. Secondly, we can use the information on AGN duty cycles to estimate the fraction of all galaxies that can be expected to host fossil outflows.

To start with, as an illustrative example, let us consider the simple analytical estimates from \citet{King2011MNRAS}. They suggest that the outflow takes an order of magnitude longer to stall than the AGN episode that inflated it (see also Section \ref{sec:analytical}). If this were the case, we would expect the number of galaxies hosting fossil outflows to be an order of magnitude larger than the number of AGN hosts. Given that the AGN duty cycle is of order a few percent \citep{Wang2006ApJ}, several tens of percent of galaxies should contain fossil outflows.

Our results paint a somewhat more modest picture that depends strongly on the gas density. In gas-rich systems, fossil outflows appear after high-luminosity episodes that reach $L \simgt 0.7 L_{\rm Edd}$ and can persist $2-3$ times longer than the AGN episode itself (cf. Fig. \ref{fig:fossil_duration}). In extreme cases, when $L > L_{\rm Edd}$, the fossil may last $4-10$ times longer than the AGN episode. This still suggests that fossil outflows are more common than the very high luminosity Eddington episodes. However, such episodes are themselves rare, with a duty cycle $<10^{-4\ldots-3}$ in the local Universe depending on SMBH mass \citep{Schulze2010A&A}; the required Eddington ratio is more than $1\sigma$ above the mean value for even the brightest local AGN populations \citep{Lusso2012MNRAS}, meaning they comprise $<15\%$ of the brightest and $<2\%$ of all AGN. Therefore, the fraction of gas-rich galaxies with fossil outflows is also small, $< 10^{-2}$. On the other hand, the numbers may be higher at high redshift (see Section \ref{sec:high-z} below). It is also worth noting that finding a fossil outflow in a gas-rich system automatically suggests that it has experienced a nearly-Eddington-limited AGN episode within the past few Myr.

In gas-poor systems, more typical of galaxies today, fossil outflows persist for much longer, essentially becoming undetectable due to gas dilution rather than stalling. This consequence becomes more pronounced the more diffuse the gas in the galaxy, primarily because the hot gas inside the outflow is more adiabatic in lower-density systems and can push the outflow out of the bulge, leading to much easier expansion. The detection of the fossil outflow then becomes a question of telescope sensitivity and signal-to-noise ratio. Optimistically, we will assume that the outflow may be detectable for up to 10 times longer than its driving episode.

In these systems, fossil outflows form after most AGN episodes (see Section \ref{sec:conditions}, above). Assuming that the actual fraction is a half and multiplying by the lifetime ratio, we see that the fraction of galaxies with fossil outflows can be a factor $\sim 5$ higher than the AGN duty cycle, i.e. $f_{\rm fossil} \sim 5 \times 10^{-3}\ldots 5\times10^{-2}$, depending on the SMBH mass. In order to make a better estimate, a full population synthesis model is required, but this is beyond the scope of this paper.

Overall, our results suggest that fossil outflows should be more common than outflows in AGN host galaxies. In addition, there should be an anti-correlation between galaxy gas fraction and prevalence of fossil outflows. Although unambiguously fossil outflows may be rare \citep[e.g.,][identify only 5 fossil outflows in a sample of 45 ($11\%$)]{Fluetsch2019MNRAS}, a careful examination of the available data, as well as future dedicated search campaigns, may easily reveal a plethora of candidates (see also Sections \ref{sec:real_comparison} and \ref{sec:sf_comp}). This development should significantly improve our understanding of the interplay between AGN and their host galaxies, especially the effect that AGN outflows have on their galaxies on timescales longer than that of a single activity episode. We caution, however, that the low densities of the fastest-moving material within fossil outflows may make them difficult to detect, especially as they slow down and the shocks driven into the surrounding medium become progressively weaker.

We note that recently, \citet{Zubovas2022MNRAS} also estimated the prevalence of fossil outflows using a semi-analytical model that allows for more realistic AGN luminosity variations with time. They found that fossil outflows should outnumber driven ones by about $1.6$ to one. Despite the significant difference in model setup and the large number of uncertainties involved, our result agrees with that from \citet{Zubovas2022MNRAS}.

\subsection{Fossil outflows at high redshift} \label{sec:high-z}

In the simulations presented in this paper, we set up the initial conditions to be broadly reminiscent of present-day galaxy bulges. AGN outflows have also been observed at high redshift, around $z\sim2$ \citep{Harrison2016MNRAS, Bischetti2017AA, Circosta2018AA, ForsterSchreiber2019ApJ} and even as far as $z\sim6$ \citep{Bischetti2019AAb}. Whether these outflows leave behind fossils depends on how the typical properties of their galaxies differ from those in the local Universe.

The ratio of bulge mass to SMBH mass is typically smaller at high redshift, by as much as 1-2 orders of magnitude \citep{Bischetti2021AA}. Furthermore, the average Eddington ratio of AGN increases with redshift by up to a factor of a few (\citealt{Lusso2012MNRAS, Martocchia2017AA, Delvecchio2020ApJ, Bischetti2021AA}; however, see \citealt{Suh2015ApJ} for an interpretation that the apparent increase in Eddington ratio is due to selection effects). Both effects suggest that an AGN of a given luminosity at $z = 2$ would produce a more powerful and long-lasting outflow than the same AGN at $z = 0$. 

On the other hand, the bulges are more compact at $z \sim 2$ than those of the same mass at $z = 0$, by a factor of a few \citep{Tacchella2018ApJ}, making the gravitational potential deeper. Gas fractions also tend to be higher, with typical $z \sim 2$ galaxies having $f_{\rm g} \sim 0.1$,  similar to group/cluster centrals in the Local Universe \citep{Tacconi2010Natur, Zhang2021AA}. These two effects lead to more difficult outflow expansion.

Determining which effects are the most important and the variation of typical AGN (fossil) outflow properties with redshift is beyond the scope of the present paper. However, based on the evidence presented above, it appears that the bulge-to-SMBH mass ratio experiences the greatest evolution of all the relevant parameters, so we expect that high-redshift galaxies should produce fossil outflows more often than local ones. Given that the AGN duty cycle is also higher at high redshift \citep{Delvecchio2020ApJ}, it is possible that fossil outflows are present in essentially all galaxies at $z \sim 2$.

\subsection{Star formation inside outflows}

\begin{figure}
	\includegraphics[width=\columnwidth]{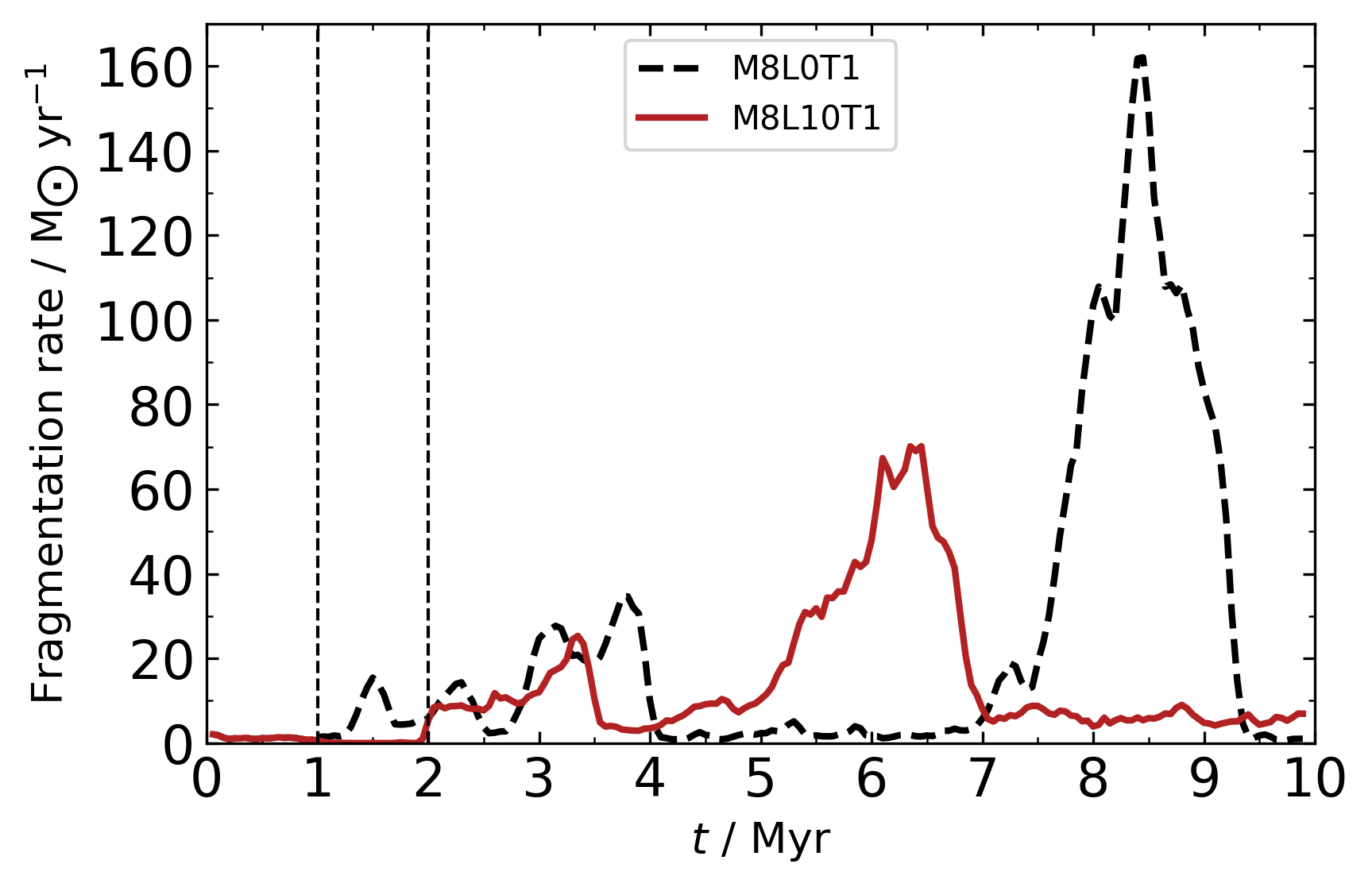}
	\includegraphics[width=\columnwidth]{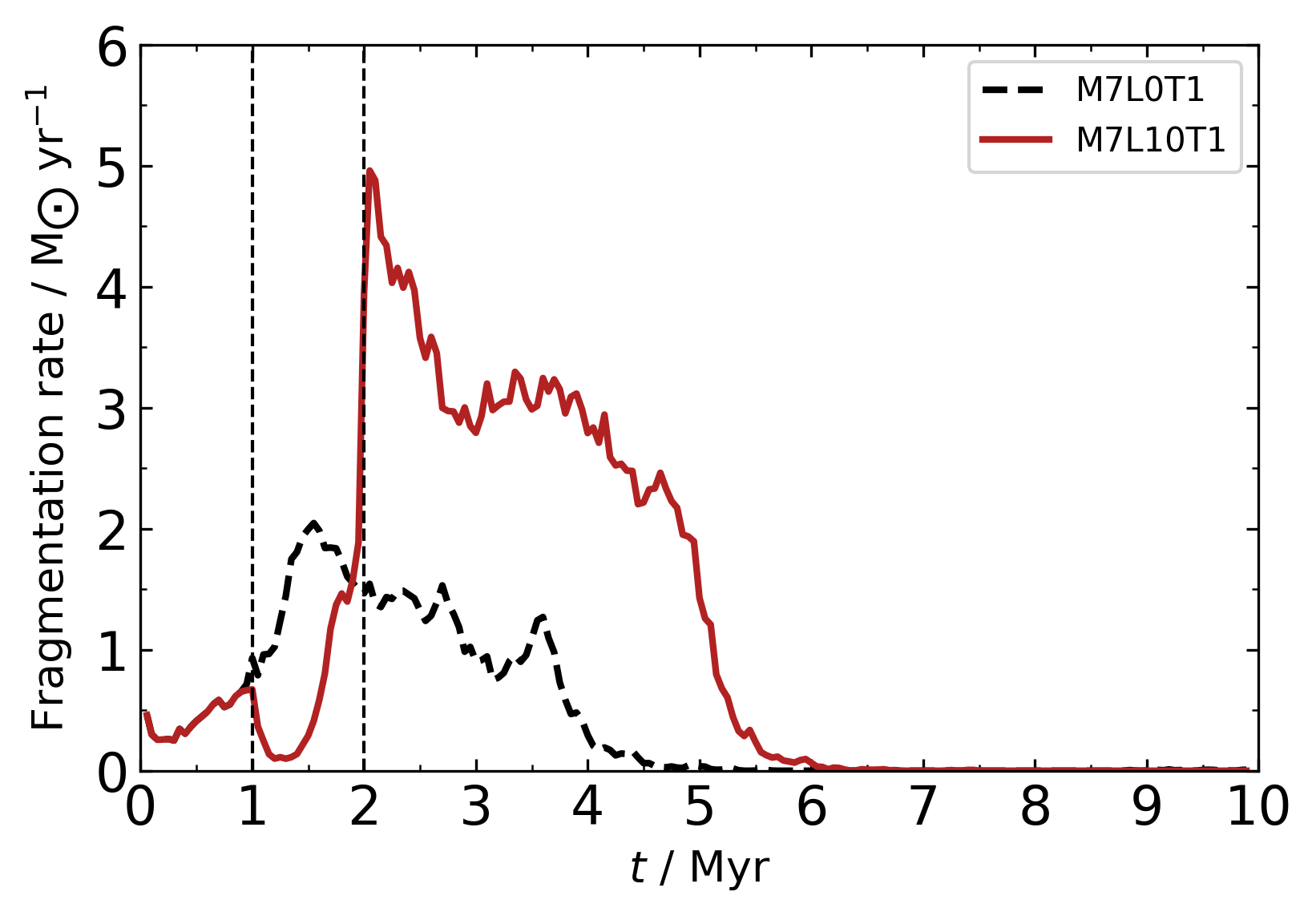}
    \caption{Time dependence of the fragmentation rate (solid lines) in simulations M8L10 (top) and M7L10 (bottom) against corresponding control simulations (black dashed lines).}
    \label{fig:fragmentation_rates}
\end{figure}

Gas in massive outflows is susceptible to fragmentation and star formation \citep{Silk2005MNRAS, Nayakshin2012MNRASb, Zubovas2013MNRASb}; numerous galactic outflows show evidence of this ongoing process \citep{Maiolino2017Natur, Gallagher2019MNRAS}. In principle, this process should be even stronger for a short period of time after the AGN switches off. The AGN radiation field disappears first, so the ionization and heating rates of gas are reduced significantly. On the other hand, the hot shocked wind bubble provides pressure that diminishes only on the outflow dynamical timescale, maintaining high gas density.

We looked for any specific trends of gas fragmentation in our simulations. We track gas fragmentation by considering the total mass of gas converted to star particles (see Section \ref{sec:sim_setup}); the results would be essentially unchanged if we added the total mass of gas on the temperature floor. Due to the simplistic numerical prescription of fragmentation, we cannot predict exact star formation rates, but we can compare the fragmentation rate histories in simulations with AGN to those in control simulations. Figure \ref{fig:fragmentation_rates} shows these comparisons for simulations M8L10 (top) and M7L10 (bottom). In both cases, fragmentation is suppressed during the AGN episode (marked with vertical dashed lines), as the AGN radiation heats the gas up and prevents fragmentation. Note that this isn't necessarily a real effect, but may instead be a result of the lack of gas self-shielding in our adopted cooling function. After the AGN switches off, the evolution of the two simulations is wildly different. M8L10 has fragmentation rates very similar to those of the control simulation for the duration of the fossil outflow, i.e. until $t \sim 4$~Myr. Only later, around $5$~Myr, a significant difference emerges, when the compressed formerly outflowing gas collapses to the centre and reaches the threshold of fragmentation. In the control simulation, a similar peak of fragmentation is visible at a later time, but it actually results from most of the gas falling into the centre after the turbulent motions die down; we expect a similar occurrence in the M8L10 simulation at an even later time. Conversely, in M7L10, the fragmentation rate increases significantly above the control value as soon as the AGN switches off. This happens because the fossil outflow bubbles expand in only a few directions and compress the gas between them; this gas is susceptible to fragmentation as soon as it is able to cool down. These significant differences preclude us from making any quantitative conclusions regarding star formation in fossil outflows. We echo the conclusion of \citet{Zubovas2017MNRAS}, who found that the interplay between AGN and star formation is very complex, with many separate effects cancelling each other out on global scales and obfuscating simple trends of either enhancement or reduction of the star formation rate.

That said, the kinematics of newly formed stars may reveal the presence of fossil AGN outflows. The outflowing gas slows down both due to doing work against gravity and sweeping up ever more material as it expands. Stars, on the other hand, are only affected by gravity. Therefore, as the outflow stalls, stars that formed within it should overtake the gas and stream out on radial trajectories. This is the opposite effect of that found in \citet{Zubovas2013MNRAS}: there, stars forming within an AGN outflow would create radial streams behind the outflow, as the gas expands at a roughly constant velocity while the stars decelerate due to gravity. Once the AGN switches off and outflow is no longer driven, the behaviour flips. Radial streams of (very) young stars ahead of cold gas clumps may be an important sign of fossil AGN outflows.

Only the high-density simulations show significant fragmentation at all; this means that in most fossil outflows, fragmentation and star formation are not relevant. This result agrees well with the observations of our own Galaxy, where the Fermi bubbles \citep{Su2010ApJ} are almost certainly fossils of a past AGN outflow \citep{Zubovas2012MNRASa}.

\subsection{Difference between AGN fossils and SF-driven outflows} \label{sec:sf_comp}

Galaxy-scale outflows are often found in galaxies without an AGN \citep[see, e.g.][for a review]{Rupke2018Galax}. In such cases, it is natural to assume that star formation (SF), encompassing supernovae, stellar winds and radiation pressure, is the process responsible for powering the outflow. On the other hand, the outflow may be a fossil of an earlier AGN episode. Generally, AGN outflows are more powerful than SF-driven ones \citep{Forster2019ApJ, Nelson2019MNRAS}, but fossil outflows have lost some of their energy, making them more similar to SF-driven counterparts. The potential confusion between the two types is confounded by the fact that AGN can occasionally trigger starbursts in the surrounding regions \citep{Silk2005MNRAS, Zubovas2013MNRASb, Zubovas2015MNRAS, Zubovas2016MNRASb}, so by the time the AGN outflow has become a fossil, there is ongoing star formation in the nuclear regions. Several lines of evidence help distinguish between the two possibilities of the outflow origin.

The first piece of evidence can be obtained by considering the evolutionary timescale of the outflow and the starburst, and the corresponding energy injection. A simple estimate of the outflow age, $t_{\rm age} \sim R_{\rm out} / v_{\rm out}$, is the upper limit for a fossil AGN outflow, since it has slowed down somewhat since the AGN switched off. Conversely, this estimate gives an approximately correct age for a SF-driven outflow. The maximum energy injected by a starburst during this time is \citep{Schneider2020ApJ}
\begin{equation}
    E_{\rm sb, max} \sim 10^{56} \frac{\dot{M}_*}{10\, \msun \, {\rm yr}^{-1}} \frac{t_{\rm age}}{{\rm Myr}} {\rm erg},
\end{equation}
where $\dot{M}_*$ is the star formation rate. We scale it to $10 \, \msun$~yr$^{-1}$ because this is the approximate star formation rate of a galaxy on the star-forming main sequence with a stellar mass $M_* \sim 10^{11} \, \msun$ \citep{CanoDiaz2016ApJ}, corresponding to a $10^8 \, \msun$ SMBH.  Conversely, an AGN injects
\begin{equation}
    E_{\rm AGN} \sim 0.05 L_{\rm AGN} t_{\rm q} \sim 2 \times 10^{58} l \, M_8 \frac{t_{\rm q}}{{\rm Myr}} {\rm erg},
\end{equation}
i.e. more than two orders of magnitude more energy. Even if the outflow energy decreases by a factor $\sim 30$ from its initial value, as seen in the bottom panel of Figure \ref{fig:mdot_pdot_edot}, the remaining energy is still almost an order of magnitude greater than what the starburst can inject over the same period. The actual values, of course, scale linearly with $\dot{M}_*$, but do not depend significantly on the age of the starburst, provided it exceeds the estimated age of the outflow.

The total mass and mass flow rate of the outflow are other properties that should be different between the two cases. SF-driven outflows have mass-loading factors, defined as $\dot{M}_{\rm out}/\dot{M}_*$, that decrease with stellar mass and have values $< 1$ as long as $M_* > 10^{10} \,\msun$ \citep{Chisholm2017MNRAS}. AGN-induced starbursts, on the other hand, should have star formation rates significantly lower than the outflow rate \citep{Zubovas2016MNRASb}, so the derived mass-loading factor would be $\gg 1$. The high loading factor persists even during the fossil phase of the outflow and can help identify a fossil AGN outflow (or, more precisely, a starburst that is a consequence of the outflow, rather than its cause). If the starburst is not caused by the AGN outflow but precedes it, we can use the energy estimate derived above to put constraints on the outflow properties. Assuming perfect coupling of the SF-injected energy to the outflow, the maximum outflowing mass is 
\begin{equation}
    M_{\rm sb, max} = \frac{2 E_{\rm sb,max}}{v_{\rm out}^2} \sim 10^7 \frac{\dot{M}_*}{10 \msun \, {\rm yr}^{-1}} \frac{t_{\rm age}}{{\rm Myr}} \left(\frac{10^8 \,{\rm km s}^{-1}}{v_{\rm out}}\right)^2 \, \msun.
\end{equation}
The ``perfect coupling'' implicitly requires the outflow to be spherical, so the maximum mass is reduced in proportion to the fraction of the solid angle subtended by the outflow. If the detected outflowing mass is much greater than this limit, it is unlikely that a starburst is responsible for launching the outflow.

Finally, multiphase data of outflow components can reveal the nature of the driving mechanism. In AGN-driven outflows, molecular gas comprises the majority - $\sim 90\%$ - of the outflowing mass, while in SF-driven outflows, the ratio of molecular to ionized gas is much closer to unity \citep[and references therein]{Venturi2021IAUS}. Once the AGN switches off, the outflow can cool rapidly, so the molecular gas fraction increases further \citep{Zubovas2014MNRASa, Richings2018MNRAS, Richings2018MNRASb, Costa2020MNRAS}, amplifying the difference from SF-driven ones.

\subsection{Impact on dark matter halos}

It is well known that rapid removal of gas from the centre of a galaxy can have a relaxing effect on its dark matter halo \citep{Pontzen2014Natur}. The effect was originally investigated in the context of dwarf galaxies undergoing supernova feedback \citep{Pontzen2012MNRAS, Governato2012MNRAS}, but was later extended to cover AGN feedback as well \citep{Martizzi2012MNRAS, Choi2018ApJ, vanderVlugt2019MNRAS, Maccio2020MNRAS}. An important aspect of the mechanism is the difference in timescales between gas removal and reaccretion. Although the dark matter halo responds to any change in the gravitational potential, slow removal of gas results in adiabatic change, which is reversible when gas falls back. It is only when gas is removed rapidly, on timescales much shorter than dynamical, that slow reaccretion does not restore the previous shape of the halo, resulting in long-term flattening of the inner density slope. 

Gas removal on sub-dynamical timescales corresponds to outflow velocities $v_{\rm out} \gg \sigma$. Fossil outflows in our simulations form only in regions where the radial velocity distribution of the driven outflow has a tail extending beyond $v_{\rm out} \simgt 2 \sigma$ (see Section \ref{sec:out_kinematics}). In addition, they live longer and expand to greater radii than outflows that stall and collapse very quickly after AGN switchoff. These properties make fossil outflows an important element in the long-term effect of AGN on the host galaxy's dark matter halo. Unfortunately, the effect is not going to be immediately apparent, so galaxies with fossil outflows should not necessarily have different dark matter properties from galaxies without them. A difference may be seen when comparing galaxies that are expected to have experienced several high-power AGN episodes with galaxies that have had multiple weaker ones, as the latter are less likely to have produced significant fossils.

\subsection{Realism of cooling rates in the simulations} \label{sec:discussion_cooling}

Once AGN driving switches off, the outflow keeps expanding due to residual inertia (snowplough effect) and the pressure of the hot shocked wind inside the outflow bubble. As we showed in section \ref{sec:thermodynamics}, the hot gas inside the bubble cools approximately adiabatically in the low-density simulation, but its cooling is dominated by radiative losses in the high-density one. Given that our numerical method does not represent the AGN wind with SPH particles, it is important to understand how much the properties of the hot gas in the bubble differ from those expected of the shocked wind and how much this affects the estimated cooling rate and, hence, outflow dynamics during the fossil phase.

We will use the M8L10 and M8fg002L02 simulations as an example. At $t = 2$~Myr, both of them have outflows with an approximate radius of $1$~kpc. These outflow bubbles should be filled with the shocked AGN wind that has been emanating for the whole $1$~Myr duration of the AGN phase. Assuming that the wind mass loss rate is equal to the SMBH accretion rate, the total wind mass injected over that time is
\begin{equation}
    M_{\rm wind} \simeq \frac{L_{\rm AGN} t_{\rm q}}{\eta c^2} = \frac{4 \pi G M_{\rm BH} l t_{\rm q}}{\kappa \eta c} = 2.2 \times 10^6 l \, \msun.
\end{equation}
The average wind particle density, assuming full ionization, i.e. $\mu = 0.63$, is
\begin{equation} \label{eq:nwind}
    n_{\rm wind} = \frac{M_{\rm wind}}{\mu m_{\rm p} V_{\rm out}} \simeq \frac{3 G M_{\rm BH} l t_{\rm q}}{\kappa \eta c \mu m_{\rm p} R_{\rm out}^3} \simeq 0.03 l \, {\rm cm}^{-3}.
\end{equation}
In fact, this estimate is a lower limit, because we assumed that the wind fills the whole sphere, rather than accounting for the dense filaments occupying a fraction of it. Looking at the phase diagrams (Figures \ref{fig:phase_diagrams} and \ref{fig:phase_diagrams_lowfg}), we see that the simulated hot gas has densities $1 \simlt n_{\rm hot} \simlt 30$~cm$^{-3}$ in M8L10 and $\sim 5$ times lower in M8fg002L02, with a negative correlation between density and temperature. These densities are a factor $30-10^3$ higher than given by eq. \ref{eq:nwind}. This happens mostly because our simulations do not have sufficient numerical resolution to resolve the very dilute wind phase. The energy of the AGN wind is transferred directly to these particles, but they heat up less than the wind would due to the higher total mass. The expected temperature of the shocked wind is of order $T_{\rm wind} \sim 10^{10} - 10^{11}$~K, some two-three orders of magnitude higher than the temperature reached in our simulations. This shows that the pressure in the simulated outflow bubble is approximately correct and confirms that the AGN wind energy injection works correctly. 

The main cooling process of the hot gas in our simulations is bremsstrahlung radiation; its luminosity is
\begin{equation}
    L_{\rm ff} \sim 1.42 \times 10^{-27} n^2 T^{1/2} \, {\rm erg} \, {\rm s}^{-1} \, {\rm cm}^3 \, {\rm K}^{-1/2}.
\end{equation}
Approximating the thermal energy of a particle by $E_{\rm th} \sim 3/2 k_{\rm B} T$, we can derive the cooling time
\begin{equation}
    t_{\rm cool} \sim \frac{3 k_{\rm B} T}{2 L_{\rm ff}} \sim 1.46 \times 10^{11} n^{-2} T^{1/2} \, {\rm s} \sim 4.6 \times 10^7 n_1^{-2} T_8^{1/2} \, {\rm yr},
\end{equation}
where we scale particle density to $n_1 \equiv n/1$~cm$^{-3}$ and the temperature to $T_8 \equiv T/10^8$~K in the last equality. Using the hot gas densities of the M8L10 simulation, we find the cooling time of the hottest gas is $t_{\rm cool, max} \sim 4.6 \times 10^7$~yr, while it drops to $t_{\rm cool, min} \sim 4.6 \times 10^4$~yr for gas at $T = 10^6$~K. The cooling times are some 25 times longer in the M8fg002L02 simulation, explaining why this gas is adiabatic. Other cooling processes included in our simulations should decrease the cooling time somewhat, but not significantly, because they are subdominant at these highest temperatures.

The cooling timescale of the shocked wind, assuming for the moment that it is a single-temperature plasma, can be calculated based on equations 5-8 in \cite{King2003ApJ}. Using values appropriate for the M8L10 simulation, we find $t_{\rm cool,w} \sim 10^7 R_{\rm kpc}^2$~yr. This timescale is inversely proportional to the AGN luminosity, so the expected timescale is $\sim 5 \times 10^7$~yr for the conditions of the M8fg002L02 simulation.

The analytically estimated cooling timescales and those calculated using the properties of our simulations are comparable. The gas with $T \sim 10^6$~K cools down much faster, but it contains a comparatively small fraction of the wind energy. We can conclude that the hot gas cooling in our simulations is tracked approximately correctly and does not introduce significant errors into our results. However, there are two caveats to this conclusion. First of all, the shocked wind is more likely to be a two-temperature plasma \citep{Faucher2012MNRASb}, with a cooling timescale several orders of magnitude longer. On the other hand, the ISM is nonuniform, so the expanding outflow bubble overtakes multiple dense clouds. They may evaporate very rapidly, on timescales $\ll 1$ Myr (see eq. 22 in \citealt{Cowie1977ApJ}; adopting $T = 10^{10}$~K and cloud density $n = 10^5$~cm$^{-3}$, we get $t_{\rm evap} \sim 3.3 r_{\rm pc}^2$~yr, where $r_{\rm pc}$ is cloud radius in parsecs). The cloud material adds to the mass of the wind, reducing its temperature and, hence, the cooling time. These two effects cancel each other to some extent.

\section{Summary and conclusion} \label{sec:concl}

We carried out a suite of idealised simulations intended to track the evolution of galactic outflows after the driving AGN switches off. Simulation parameters encompass different SMBH masses (together with corresponding changes to bulge mass and velocity dispersion), gas densities and driving AGN luminosities. Each driving AGN phase lasts 1 Myr and we follow the evolution of the system for up to 8 Myr after the AGN switches off. Our main results are the following:

\begin{itemize}
    \item Fossil outflows form in simulations with $l \simgt 7 f_{\rm g}$, where $l \equiv L_{\rm AGN}/L_{\rm Edd}$ is the Eddington ratio and $f_{\rm g} \equiv \rho_{\rm g} / \rho_{\rm tot}$ is the gas fraction. In simulations with lower AGN luminosity, outflows may form during the AGN phase, but they stall and collapse in less than a tenth of the AGN episode duration.
    
    \item Fossil outflows in gas-rich systems tend to last a few times longer than the duration of the driving episode; in gas-poor systems, they last far longer, eventually becoming undetectable due to dilution rather than actually stalling and falling back.

    \item As a result, fossil outflows should be rare in very gas-rich galaxies, but rather common in more typical systems in the local Universe. Their number should exceed that of driven AGN outflows by a factor of a few.

    \item At high redshift ($z \sim 2$), fossil outflows are probably somewhat more common than in the local Universe, mostly due to the typical bulge mass being lower by 1-2 orders of magnitude for a given SMBH mass.

    \item As fossil outflows expand, they rapidly become far more lopsided than driven outflows, because the outflow often stalls and collapses in some directions while expanding in others.

    \item Additionally, fossil outflows detach from the nucleus; the region around the nucleus fills up with backflowing gas that may trigger a subsequent AGN episode.

    \item Fossil outflows are dominated by the molecular phase (as opposed to ionized) more than driven ones.

    \item Fossil outflows tend to have a higher mass flow rate for a given velocity than driven ones, mostly due to having a larger radius and hence encompassing more material.
\end{itemize}

Identifying fossil outflows in real galaxies may be challenging, since their integrated properties are similar to those of both driven AGN outflows and star formation-driven outflows. Spatially resolved and/or multiphase observations should help in this regard. Furthermore, a large number of fossil outflows may be hiding in the data of inactive galaxies. Understanding the presence and properties of fossil outflows in real galaxies will help us investigate the activity histories of galaxies over the past several Myr, creating a more complete picture of AGN and host galaxy coevolution.

Our results are only a first step in trying to understand the full diversity of fossil AGN outflows. The wide range of galaxy morphologies and gas distributions, as well as AGN light curves, almost certainly leads to an immense variety of fossil outflow properties. Including additional physical processes, such as feedback from the pre-existing and newly forming stellar populations, complicates the picture further. Nevertheless, we hope this paper provides a foundation for more detailed studies to build upon.

\section*{Acknowledgements}

We thank Manuela Bischetti and Tiago Costa for valuable suggestions and illuminating discussions during the preparation of this manuscript. This research was funded by the Research Council Lithuania grant no. S-MIP-20-43. The simulations were performed on the supercomputer GALAX of the Center for Physical Sciences and Technology, Lithuania.

\section*{Data availability}

No new observational data was taken for the preparation of this manuscript. Simulation data is available from the corresponding author upon reasonable request.




\bibliographystyle{mnras}
\bibliography{zubovas} 




\appendix

\section{Resolution tests} \label{app:resolution}

\begin{figure*}
	\includegraphics[width=0.33\textwidth]{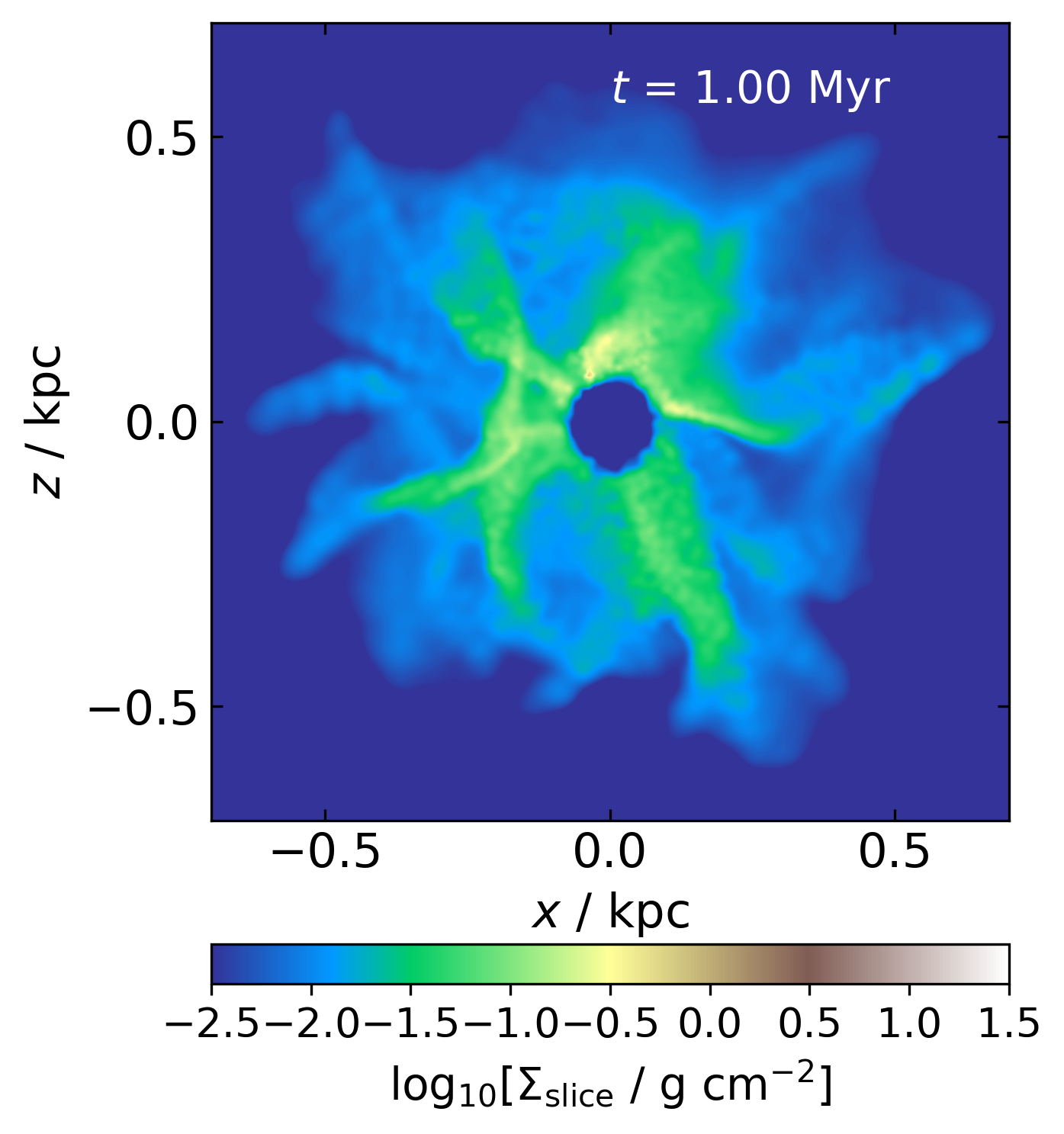}
	\includegraphics[width=0.33\textwidth]{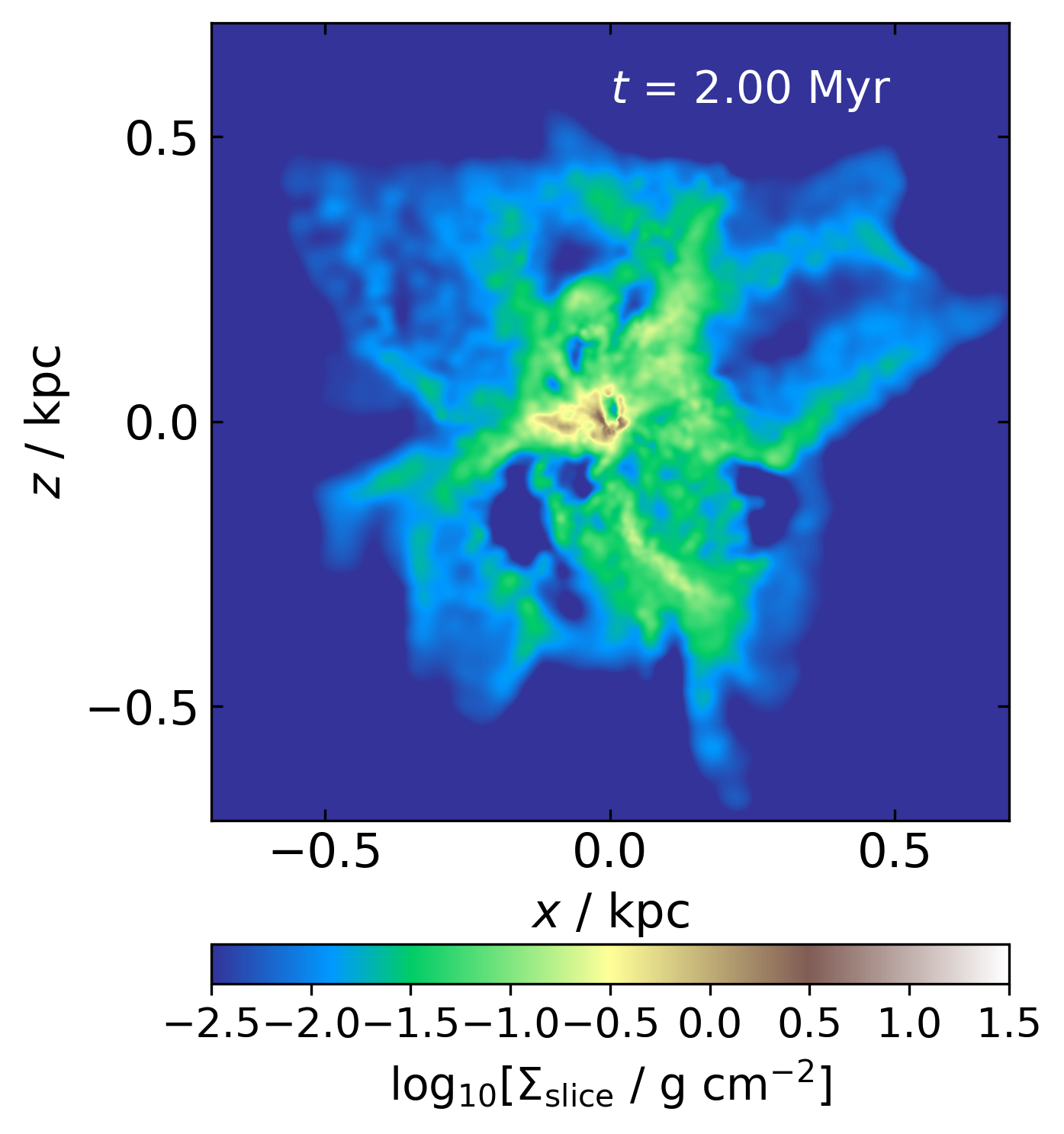}
	\includegraphics[width=0.33\textwidth]{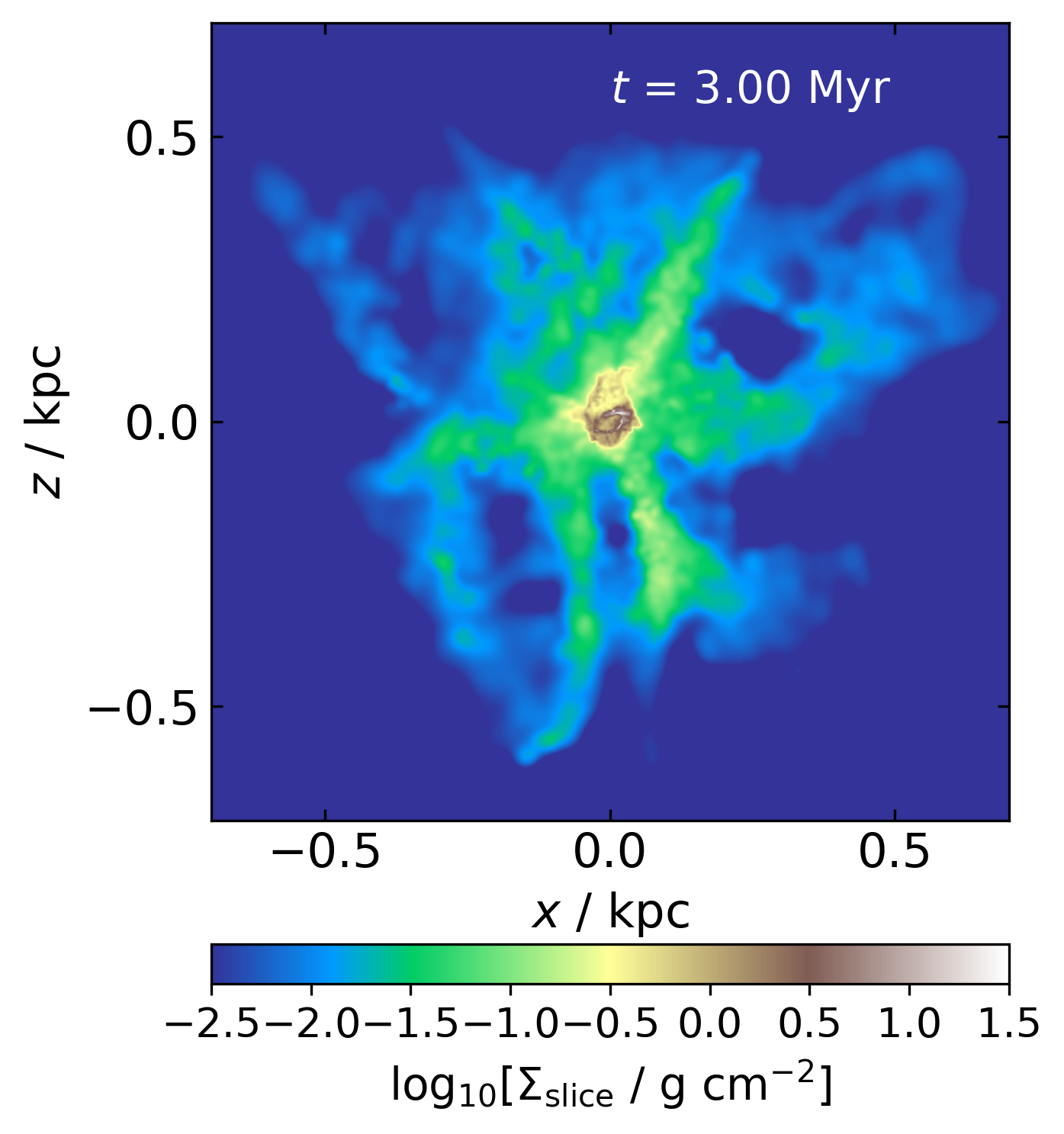}
    \caption{Density evolution of the M7L07 low-resolution simulation, to be compared with Figure \ref{fig:evolution_M7L07}. {\em Left}: density slice at $t = 1$~Myr, just before the AGN switches on. {\em Middle}: density slice at $t = 2$~Myr, when the AGN switches off. {\em Right}: density slice at $t = 3$~Myr.}
    \label{fig:evolution_M7L07_lr}
\end{figure*}

\begin{figure}
	\includegraphics[width=\columnwidth]{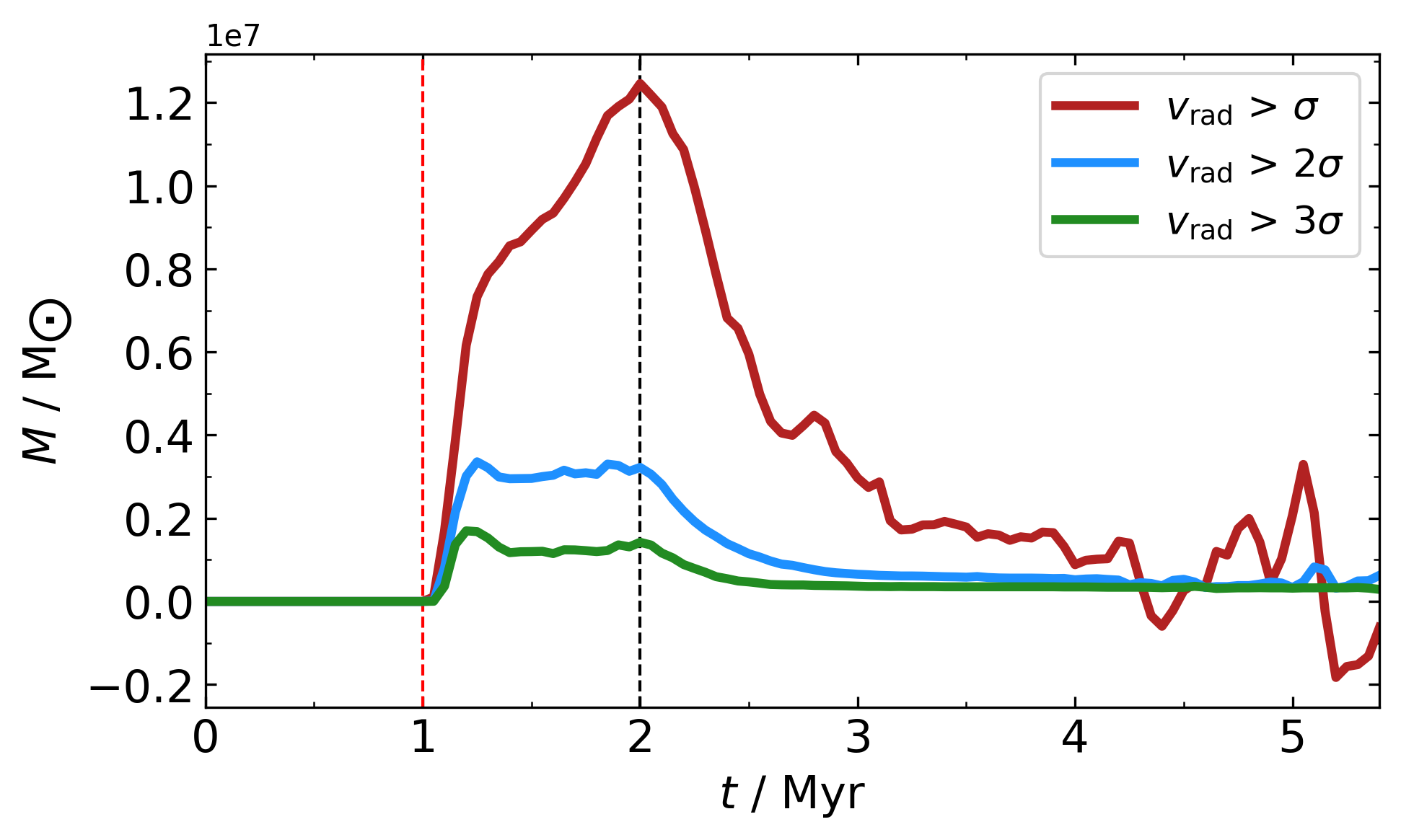}
    \caption{Total outflowing mass in the M7L07-lowres simulation. Lines show the difference in mass moving with radial velocity above $\sigma$ (brown), $2\sigma$ (blue) and $3\sigma$ (green) in the M7L07-lowres and M7L07-control-lowres simulations, against time. This is equivalent to the top panel of Figure \ref{fig:mass_time}.}
    \label{fig:mass_time_M7L07_lr}
\end{figure}

\begin{figure}
	\includegraphics[width=\columnwidth]{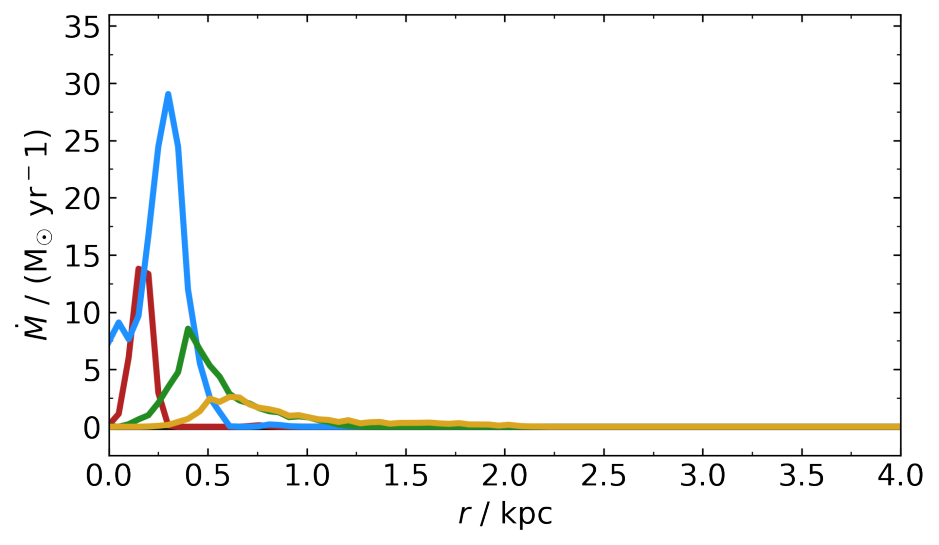}
	\includegraphics[width=\columnwidth]{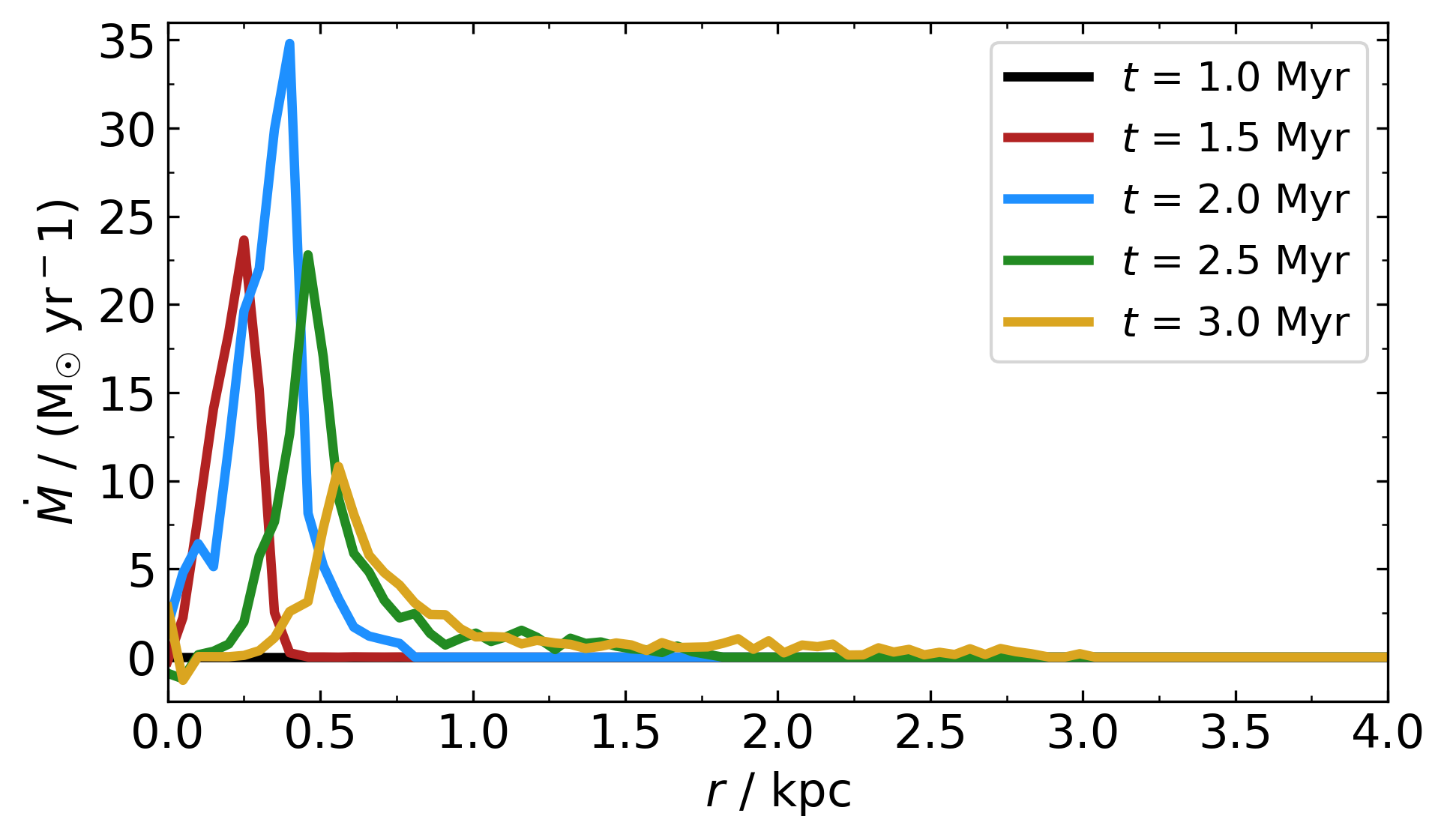}
    \caption{Radial distribution of mass outflow rate in the M7L10 (top) and M7L10-lowres (bottom) simulations. Line colours are the same as in Figure \ref{fig:mdot_pdot_edot}.}
    \label{fig:mdot_radial_M7L10_lr}
\end{figure}

\begin{figure*}
	\includegraphics[width=0.33\textwidth]{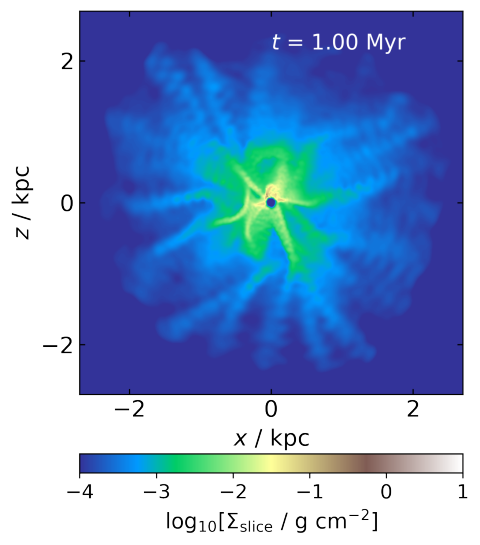}
	\includegraphics[width=0.33\textwidth]{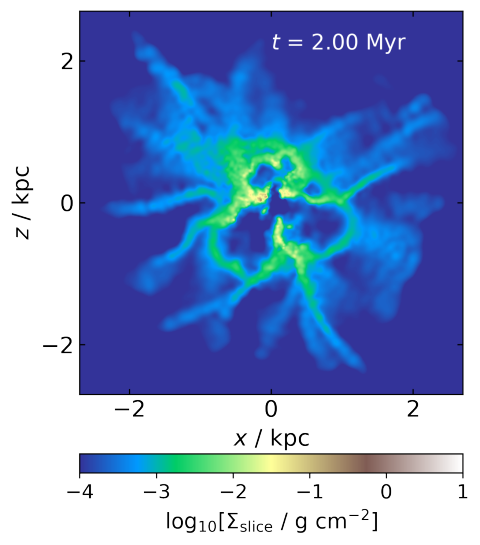}
	\includegraphics[width=0.33\textwidth]{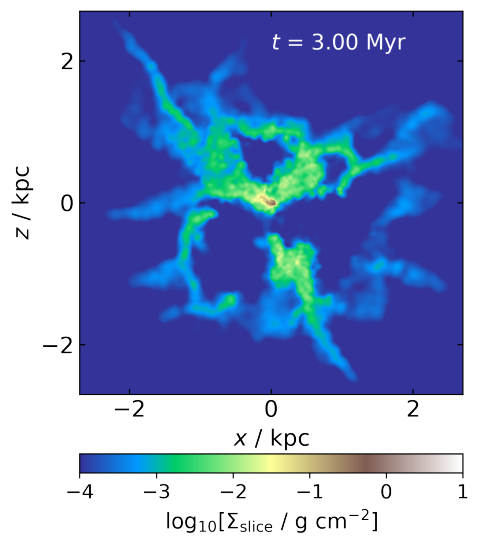}
    \caption{Density evolution of the M8L02fg002 low-resolution simulation, to be compared with Figure \ref{fig:evolution_M8fg002L02}. {\em Left}: density slice at $t = 1$~Myr, just before the AGN switches on. {\em Middle}: density slice at $t = 2$~Myr, when the AGN switches off. {\em Right}: density slice at $t = 3$~Myr.}
    \label{fig:evolution_M8fg002L02_lr}
\end{figure*}

\begin{figure}
	\includegraphics[width=\columnwidth]{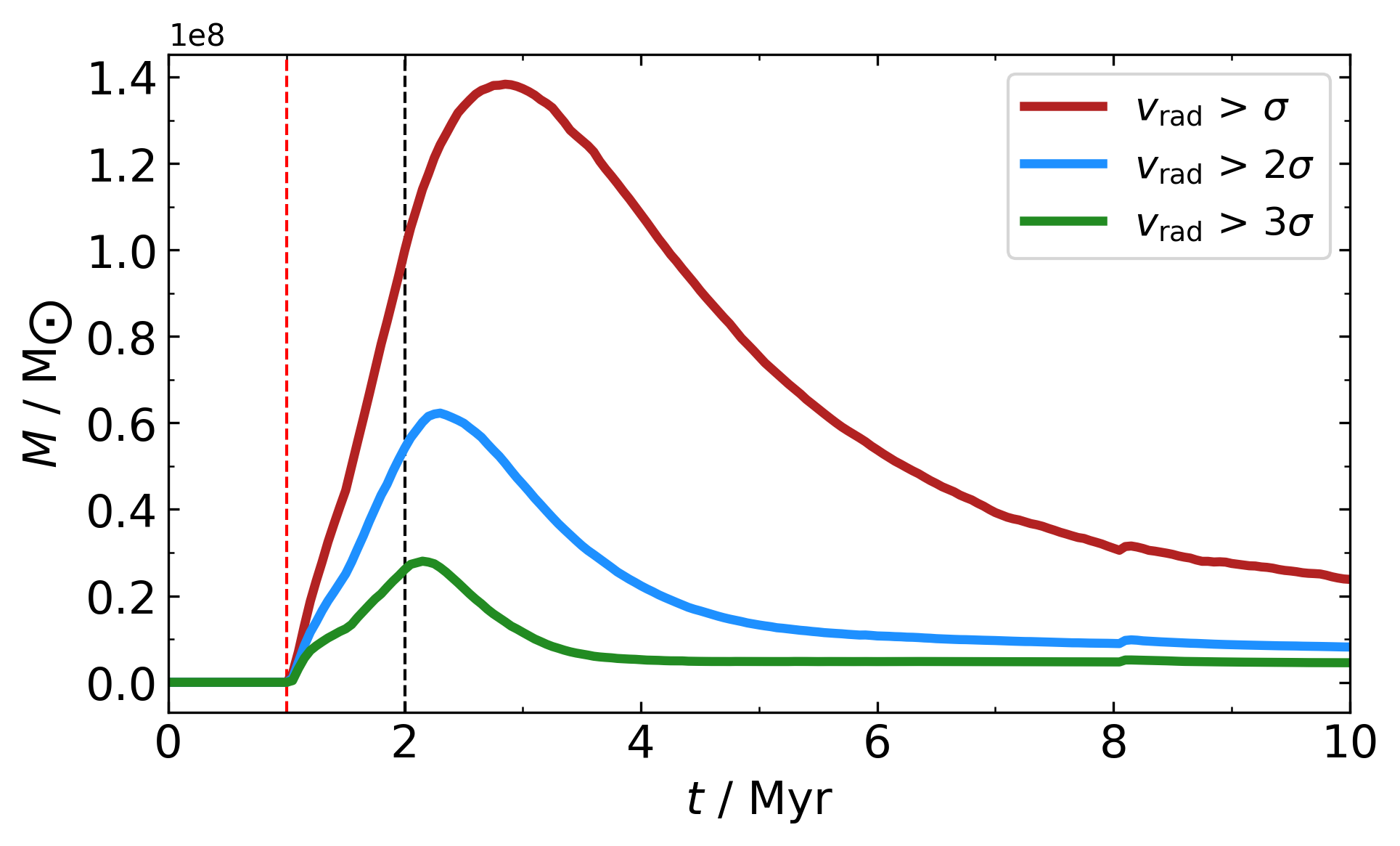}
    \caption{Total outflowing mass in the M8fg002L02-lowres simulation. Lines show the difference in mass moving with radial velocity above $\sigma$ (brown), $2\sigma$ (blue) and $3\sigma$ (green) in the M8fg002L02-lowres and M8fg002L02-control-lowres simulations, against time. This is equivalent to Figure \ref{fig:mass_time_fg002}.}
    \label{fig:mass_time_M8fg002L02_lr}
\end{figure}

\begin{figure*}
	\includegraphics[width=\textwidth]{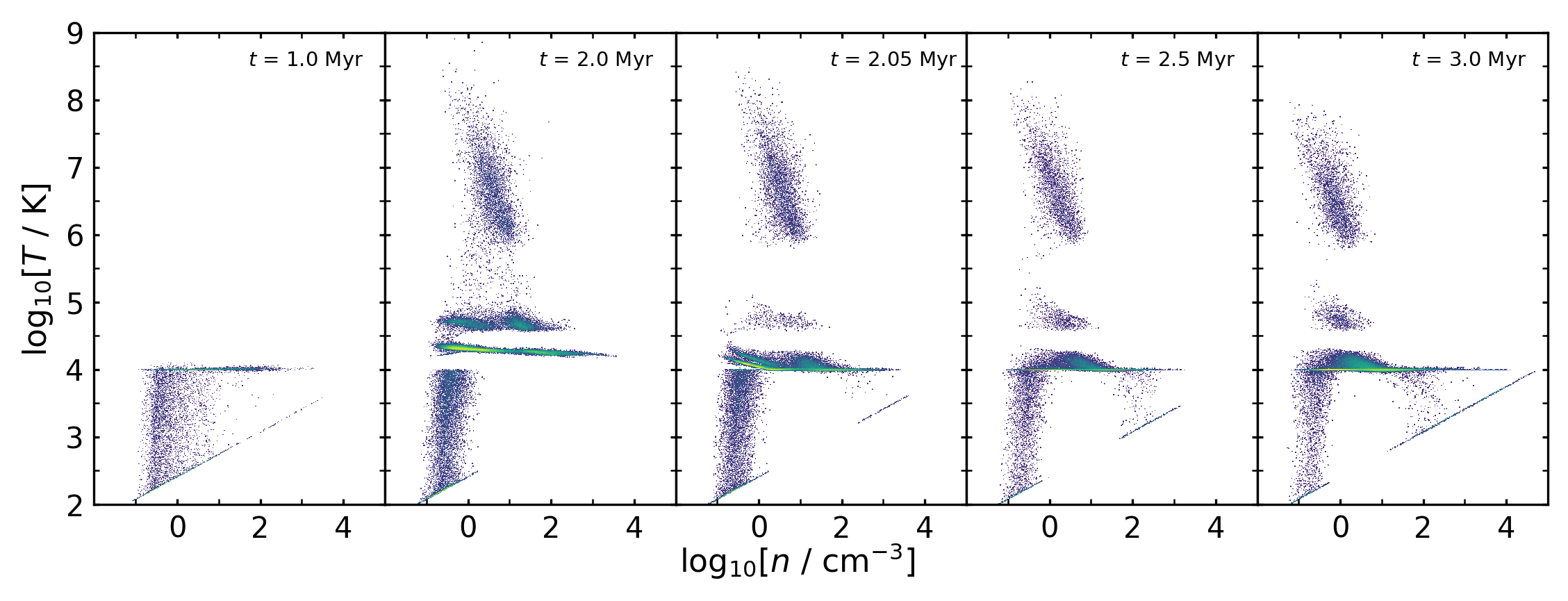}
    \caption{Phase diagrams of gas in the M8fg002L02 simulation. Panels show snapshots at $t = 1$, 2, 2.05, 2.5 and 3~Myr. Colour represents gas mass in each pixel. This is equivalent to Figure \ref{fig:phase_diagrams_lowfg}.}
    \label{fig:phase_diagrams_M8fg002L02_lr}
\end{figure*}

The four groups of simulations we run, with different SMBH masses and different gas densities, all have $N \sim 10^6$ gas particles in their initial conditions. This leads to the particle masses, and hence the mass resolution, being significantly different: from $82 \, \msun$ in M7fg002 simulations to $3700 \, \msun$ in M8. The reason for using constant particle numbers was the availability of computing resources, however, the question remains of whether different mass resolution leads to significant differences in results beyond those caused by the different physical conditions of the simulations. In order to check this, we reran several simulations at lower resolution (designated with an appendix `-lowres' below). In particular, we ran simulations M7-control-lowres, M7L07-lowres, M7L10-lowres, M8fg002-control-lowres, M8fg002L014-lowres and M8fg002L02-lowres with particle masses $m_{\rm SPH} = 3700 \, \msun$, the same as in the M8 simulations, and also simulations M7fg002-control-lowres, M7fg002L014-lowres and M7fg002L02-lowres with particle masses $m_{\rm SPH} = 410 \, \msun$, the same as in the M7 simulations. Overall, the differences between simulations run at different resolutions are minor and do not affect our conclusions. We present them in more detail below.

We first compare the M7-lowres simulations with their high-resolution counterparts. Figure \ref{fig:evolution_M7L07_lr} shows density maps of M7L07-lowres at $t = 1, 2$ and $3$~Myr, same as Figure \ref{fig:evolution_M7L07}. Comparing the two figures, we see some obvious resolution-dependent differences, such as wider radial filaments at $t = 1$~Myr and a rather large disc-like structure in the centre at $t = 3$~Myr. However, the main properties of outflow shape and evolution remain the same: a single bubble is produced by $t = 2$~Myr (although it extends toward the bottom left rather than directly downward), but all evidence of outflow is gone by $t = 3$~Myr.

Outflowing mass against time in M7L07-lowres (Figure \ref{fig:mass_time_M7L07_lr}) shows greater differences from the high-resolution counterpart (top panel of Figure \ref{fig:mass_time}). The peak outflowing mass is higher in the low-resolution simulation by a factor of three ($1.2\times10^7\, \msun$ compared to $4 \times 10^6 \, \msun$ for the $v > \sigma$ threshold; the ratios are similar for higher thresholds). The peak mass also occurs later, at $t = 2$~Myr, rather than at $t = 1.7$~Myr, and the decay of outflowing gas mass is slower, lasting $\sim 1-2$~Myr after the AGN switches off instead of being essentially complete within $0.5$~Myr; the timescale difference is smaller in the case of higher velocity thresholds. The reason for these differences is the shortest effective wavelength of turbulence: the simulation having fewer particles leads to a coarser filamentary structure before the AGN switches on, so the hot gas bubble is more efficiently trapped in the centre and able to push against more gas for longer.

Although the differences are notable, we believe they do not impact our overall conclusions. We tentatively identify a trend that with higher resolution, outflows are less massive and dissipate more quickly after the AGN switches off. If this holds true, with even higher resolution than our main simulations, the L07 models (both in the M7 and M8 groups) would have even weaker outflows and the non-existence of fossil outflows would be maintained. The higher luminosity simulations (L10 and L12) may end up with weaker fossil outflows if the resolution were increased further (cf. Fig. \ref{fig:fossil_duration}), but that seems unlikely, because the outflowing mass evolution in the M7L10-lowres simulation is very similar to that of M7L10, suggesting the two are converged.

The tendency of higher-resolution simulations to produce weaker fossil outflows can also be seen when comparing the radial profiles of outflow properties. In Figure \ref{fig:mdot_radial_M7L10_lr}, we show the radial distributions of outflowing mass in simulations M7L10 (top) and M7L10-lowres (bottom); this figure is comparable to the top panel of Fig. \ref{fig:mdot_pdot_edot}, except that it shows a different simulation. Qualitatively, the two sets of radial profiles are very similar: the outflow becomes more massive as it expands to $R_{\rm out} \sim 0.3-0.4$~kpc by $t = 2$~Myr, after which time it starts to decay. The peak mass outflow rate at $t = 2$~Myr (and in the whole simulation), $\dot{M}_{\rm max} \sim 30 \, \msun$~yr$^{-1}$, is similar in both simulations. At other times, however, the high-resolution simulation produces outflows that are less massive and less spread out than its low-resolution counterpart. The reason behind these differences is the higher density contrast in the high-resolution simulation: the AGN wind is able to escape from the bubble more efficiently through low-density gaps, leaving less energy to inflate the outflow bubble.

Now, we turn to the M8fg002-lowres simulations and their high-resolution counterparts. Figure \ref{fig:evolution_M8fg002L02_lr} shows three density maps, at $t=1$, $2$ and $3$~Myr from left to right, of the simulation M8fg002L02-lowres. These are equivalent to the maps shown in Figure \ref{fig:evolution_M8fg002L02} except for the five times lower mass resolution. Much like in the case of M7L07-lowres, the morphological evolution of the low-resolution simulation is generally similar to its high-resolution counterpart, except for a few obvious resolution-dependent differences. The azimuthal density contrasts produced by turbulence are weaker at lower resolution, and there is a disc-like structure forming in the centre by $t=3$~Myr. However, the outflow bubbles at $t=2$~Myr are remarkably similar between the two simulations, with a three-lobed shape in both cases and the greatest bubble extent of just under $1$~kpc. By $t=3$~Myr, the outflow bubbles in the low-resolution simulation are somewhat more confined, with only one bubble, expanding towards the right, showing clear signs of breaking out of the bulge. The plot of outflowing mass against time (Figure \ref{fig:mass_time_M8fg002L02_lr}) shows the same tendency: the curves have almost identical shape to those of the high-resolution simulation (Figure \ref{fig:mass_time_fg002}), but the total mass is $\sim 30\%$ lower at $t=2$~Myr. The difference increases to $\sim 50\%$ by $t=10$~Myr. However, importantly, the outflow breaks out of the bulge and persists for longer than $10$~Myr in both the high- and low-resolution simulations.

The differences can be explained by considering the phase diagrams of the M8fg002L02-lowres simulation (Figure \ref{fig:phase_diagrams_M8fg002L02_lr}) and comparing them with the high-resolution equivalent (Figure \ref{fig:phase_diagrams_lowfg}). Qualitatively, the two sets of phase diagrams are very similar, with the major groups of gas particles appearing in both. However, the  fraction (and, equivalently, total mass) of highest-temperature gas particles is significantly smaller in the low-resolution simulation. At $t= 2$~Myr, $\sim 13.1\%$ of gas has $T> 10^6$~K in the low-resolution simulation, compared with $\sim 17.7\%$ in the high-resolution one. The difference becomes starker by $t=3$~Myr, where the fraction has dropped to $\sim 9.0\%$ in the low-resolution simulation but has increased to $\sim 20.7\%$ in the high-resolution one. The main reason for this difference is the particle mass. In both simulations, the energy injected into the gas is the same, but with more massive particles, the low-resolution simulation develops a smaller population of gas with extremely high temperature (this can be seen when comparing the topmost parts of the phase diagrams at $t = 2$~Myr). This extremely hot gas is responsible for heating up the neighbouring gas particles and maintaining their population in the high-resolution simulation. The large particle masses dilute this effect at low resolution. As a result, the fossil outflow is weaker. The mean gas temperature, however, evolves virtually identically in the high- and low-resolution simulations.

It is interesting to note that the trend of outflow size/persistence with resolution is opposite when considering the simulation groups with different SMBH masses and with different gas densities. This happens because of the different relative importance of gas shell porosity and cooling. If the escape of AGN feedback energy through gaps in the dense gas distribution is more important, lower resolution leads to stronger outflows due to the gas density being more uniform. Conversely, if cooling is a more important channel of energy loss, then higher resolution allows for more gas to be heated to very high temperatures with long cooling times, leading to more powerful outflows. In our low-density simulations, the outflows tend to break out of the bulge in both the high- and low-resolution versions, so it is the cooling rate that determines the properties of the fossil outflow; conversely, in the high-density simulations, the gas shell is thick enough to significantly confine the outflow, so its porosity is the crucial property when it comes to outflow persistence. The two effects cancel each other to some extent, mitigating the effects of numerical resolution on outflow properties in galaxy-scale simulations.

\bsp	
\label{lastpage}
\end{document}